\documentclass[useAMS,usenatbib,usegraphicx]{mn2e}
\usepackage{amsmath}
\usepackage{url}
\usepackage{fixltx2e}

\title[Investigating the spectroscopic, magnetic and circumstellar variability of HD\,57682]{Investigating the spectroscopic, magnetic and circumstellar variability of the O9 subgiant star HD\,57682\thanks{Based on observations obtained at the Canada-France-Hawaii Telescope (CFHT) which is operated by the National Research Council of Canada, the Institut National des Sciences de l'Univers of the Centre National de la Recherche Scientifique of France,  and the University of Hawaii.}}
\author[Grunhut et al.]{J.H. Grunhut$^{1,2}$\thanks{E-mail: Jason.Grunhut@rmc.ca}, G.A. Wade$^2$, J.O. Sundqvist$^3$, A. ud-Doula$^4$, C. Neiner$^5$,
\newauthor 
 R. Ignace$^{6}$, W.L.F. Marcolino$^7$, Th. Rivinius$^8$, A. Fullerton$^9$, L. Kaper$^{10}$,    
\newauthor
B. Mauclaire$^{11}$,C. Buil$^{12}$, T. Garrel$^{13}$, J. Ribeiro$^{14}$, S. Ubaud$^{15}$,
\newauthor
and the MiMeS Collaboration\\
$^1$Department of Physics, Engineering Physics \& Astronomy, Queen’s University, Kingston, Ontario, Canada, K7L 3N6\\
$^2$Department of Physics, Royal Military College of Canada, P.O. Box 17000, Station Forces, Kingston, Ontario, Canada, K7K 7B4\\
$^3$University of Delaware, Bartol Research Institute, Newark, Delaware 19716, USA\\
$^4$Penn State Worthington Scranton, 120 Ridge View Drive, Dunmore, PA USA, 18512\\
$^5$LESIA, Observatoire de Paris, CNRS UMR 8109, UPMC, Universit\'e Paris Diderot; 5 place Jules Janssen, 92190 Meudon, France\\
$^{6}$Department of Physics and Astronomy, East Tennessee State University, Johnson City, TN 37614, USA\\
$^7$Universidade Federal do Rio de Janeiro, Observat\'{o}rio do Valongo Ladeira Pedro Ant\^{o}nio, 43, CEP 20080-090, Rio de Janeiro, Brasil\\
$^8$ESO - European Organisation for Astronomical Research in the Southern Hemisphere, Casilla 19001, Santiago 19, Chile\\
$^9$Space Telescope Science Institute, 3700 San Martin Drive, Baltimore, MD 21218, USA\\
$^{10}$Astronomical Institute ``Anton Pannekoek", University of Amsterdam, Science Park 904, 1098 XH Amsterdam, The Netherlands\\
$^{11}$Observatoire du Val de l'Arc, route de Peynier, 13530 Trets, France\\
$^{12}$Castanet Tolosan Observatory, 6 place Cl\'{e}mence Isaure, 31320 Castanet Tolosan, France\\
$^{13}$Observatoire de Juvignac, 19, avenue du Hameau du Golf, 34990 Juvignac, France\\
$^{14}$Obervatorio do Instituto Geografico do Exercito, R. Venezuela 29, 3 Esq., 1500-618 Lisboa, Portugal\\
$^{15}$Observatoire des Quatres chemins, Calade, 06410 Biot, France\\
}

\begin{document}
\date{\today}
\pagerange{\pageref{firstpage}--\pageref{lastpage}} \pubyear{2012}
\maketitle
\label{firstpage}
\begin{abstract}
The O9IV star HD\,57682, discovered to be magnetic within the context of the MiMeS survey in 2009, is one of only eight convincingly detected magnetic O-type stars. Among this select group, it stands out due to its sharp-lined photospheric spectrum. Since its discovery, the MiMeS Collaboration has continued to obtain spectroscopic and magnetic observations in order to refine our knowledge of its magnetic field strength and geometry, rotational period, and spectral properties and variability. In this paper we report new ESPaDOnS spectropolarimetric observations of HD\,57682, which are combined with previously published ESPaDOnS data and archival H$\alpha$ spectroscopy. This dataset is used to determine the rotational period ($63.5708\pm0.0057$\,d), refine the longitudinal magnetic field variation and magnetic geometry (dipole surface field strength of $880\pm50$\,G and magnetic obliquity of $79\pm4\degr$ as measured from the magnetic longitudinal field variations, assuming an inclination of 60$\degr$), and examine the phase variation of various lines. In particular, we demonstrate that the H$\alpha$ equivalent width undergoes a double-wave variation during a single rotation of the star, consistent with the derived magnetic geometry. We group the variable lines into two classes: those that, like H$\alpha$, exhibit non-sinusoidal variability, often with multiple maxima during the rotation cycle, and those that vary essentially sinusoidally. Based on our modelling of the H$\alpha$ emission, we show that the variability is consistent with emission being generated from an optically thick, flattened distribution of magnetically-confined plasma that is roughly distributed about the magnetic equator.  Finally, we discuss our findings in the magnetospheric framework proposed in our earlier study.
\end{abstract}


\section{Introduction}
HD\,57682 is an 09 subgiant, and one of only eight O-type stars convincingly detected to host magnetic fields: the zero-age main sequence (ZAMS) O7V star $\theta^1$ Ori C=HD\,37022 \citep{don02, wade06}, the more evolved Of?p stars HD\,108 \citep{martins10}, HD\,148937 \citep{hub08, wade12a}, HD\,191612 \citep{don06, wade11}, NGC 1624-2 \citep{wade12b} and CPD -28 2561 \citep{hub11b, wade12c}, the recently detected Carina nebula star Tr16-22 \citep{naze12}, and HD\,57682 \citep{grun09}. In addition, a small number of other O-type stars have been tentatively reported to be magnetic in modern literature \citep[e.g.][]{bouret08, hub08,hub11a, hub11b}. These stars have either been found through independent observation and/or analysis to be non-magnetic \citep[e.g.][]{fullerton11, bagnulo12}, or have yet to be independently re-observed or re-analysed. These small numbers are both a reflection of the rarity of O-type stars with detectable magnetic fields, and the challenge of detecting such fields when present.

Based on seven spectropolarimetric observations of HD\,57682, \citet{grun09} characterised its physical and wind properties, deriving, in particular, a mass of $17^{+19}_{-9}\,M_\odot$, a radius of $7.0^{+2.4}_{-1.8}\,R_\odot$, a mass-loss rate $\log \dot M=-8.85$\,$M_\odot$\,yr$^{-1}$ and wind terminal velocity $v_\infty=1200^{+500}_{-200}$\,km\,s$^{-1}$. Their spectroscopic and magnetic data indicated that the star was variable, likely periodically on a time-scale of a few weeks. Using the Bayesian statistical method of \citet{petit12}, \citet{grun09} estimated the surface dipole strength to be $1680^{+135}_{-355}$\,G. This magnetic field strength, combined with the inferred physical and wind parameters, indicated a wind magnetic confinement parameter of $1.4\times 10^4$ \citep{uddoula02}, leading the authors to suggest that the observed line profile variations were a consequence of dense wind plasma confined in closed magnetic loops above the stellar surface.


The present study seeks to refine and elaborate the preliminary results reported by \citet{grun09}. In Sect.~\ref{obs_sect} we discuss the new and archival observations used in our analysis, including extraction of Least-Squares Deconvolved (LSD) mean profiles and measurement of the longitudinal magnetic field. In Sect.~\ref{period_sect} we perform a period analysis of the H$\alpha$ and longitudinal field data and demonstrate that a single period of $\sim$63\,d is required to reproduce the periodic variations of the measurements, which span over 15 years. In Sect.~\ref{rot_sect} we discuss the implications of the inferred rotation period for the projected rotational velocity of the star, and, in particular, the estimate of 15~km\,s$^{-1}$ proposed by \citet{grun09}. In Sect.~\ref{mag_field_sect} we constrain the magnetic field properties using the longitudinal field measurements and direct fits to the observed LSD Stokes $V$ profiles. In Sect.~\ref{spec_var_sect} we present an extensive overview of the variability of line profiles of H, He and metals. In Sect.~\ref{magneto_sect} we revisit the magnetospheric framework proposed by \citet{grun09}, and in Sect.~\ref{disc_sect} we discuss our results and present our conclusions.

\section{Observations}\label{obs_sect}
\subsection{Polarimetric data and measurements}
\begin{figure*}
\centering
\includegraphics[width=7in]{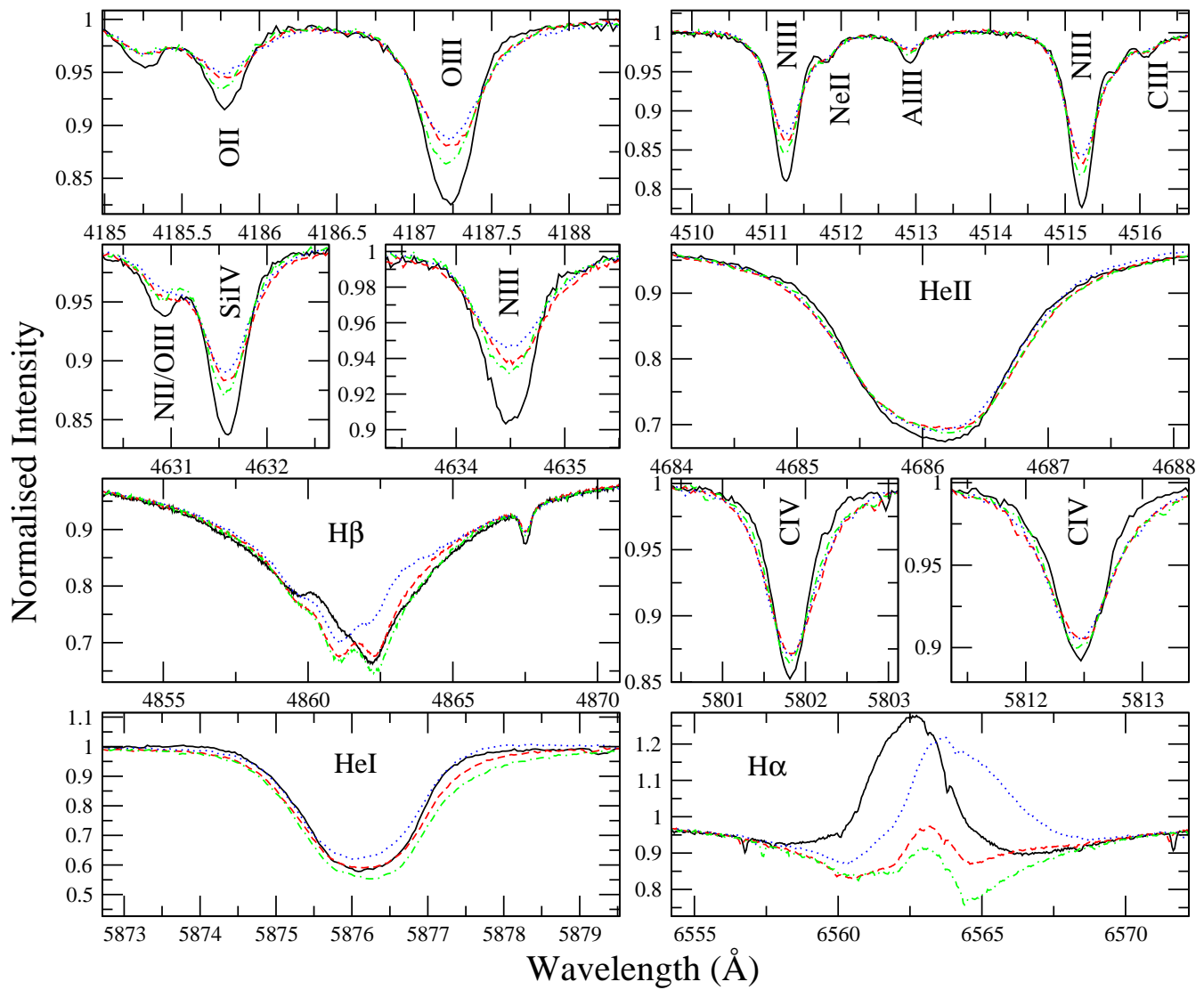}
\caption{Selected regions of visible spectra from ESPaDOnS. Spectra at different rotational phases are shown to highlight the observed variability in almost all lines (0.02 - solid black, 0.30 - dashed red, 0.45 - dotted blue, 0.72 - dash-dotted green; see Sect.~\ref{spec_var_sect} for further details). The ions with the most significant contribution to each line are also labelled.}
\label{spec_comp}
\end{figure*}

A total of 38 polarimetric observations were collected with the high-resolution ($R\sim68\,000$) ESPaDOnS spectropolarimeter mounted on the Canada-France-Hawaii telescope (CFHT). The observations were obtained in the context of the Magnetism in Massive Stars (MiMeS) CFHT Large Program on 20 different nights over a period of two years from 2008 December to 2010 December. Each observation consists of a sequence of four sub-exposures that were processed using the {\sc upena} pipeline running {\sc libre-esprit}, as described by \citet{don97}. The final reduced products consist of an unpolarised Stokes~$I$ and circularly polarised Stokes $V$ spectrum. The individual sub-exposures are also combined in such a way that the polarisation should cancel out, resulting in a diagnostic null spectrum that is used to test for spurious signals in the data. During most nights two polarimetric sequences were obtained and the un-normalised spectra were combined to produce 20 distinct observations. Seven of our twenty observations were previously discussed by \citet{grun09} and we present 13 new observations here, the details of which are listed in Table~\ref{obs_tab1}. In Fig.~\ref{spec_comp} we show spectra for selected spectral regions at a few different rotational phases to highlight the variability observed in almost all spectral lines (see Sect.~\ref{period_sect} for further details).

\begin{figure}
\centering
\includegraphics[width=3.1in]{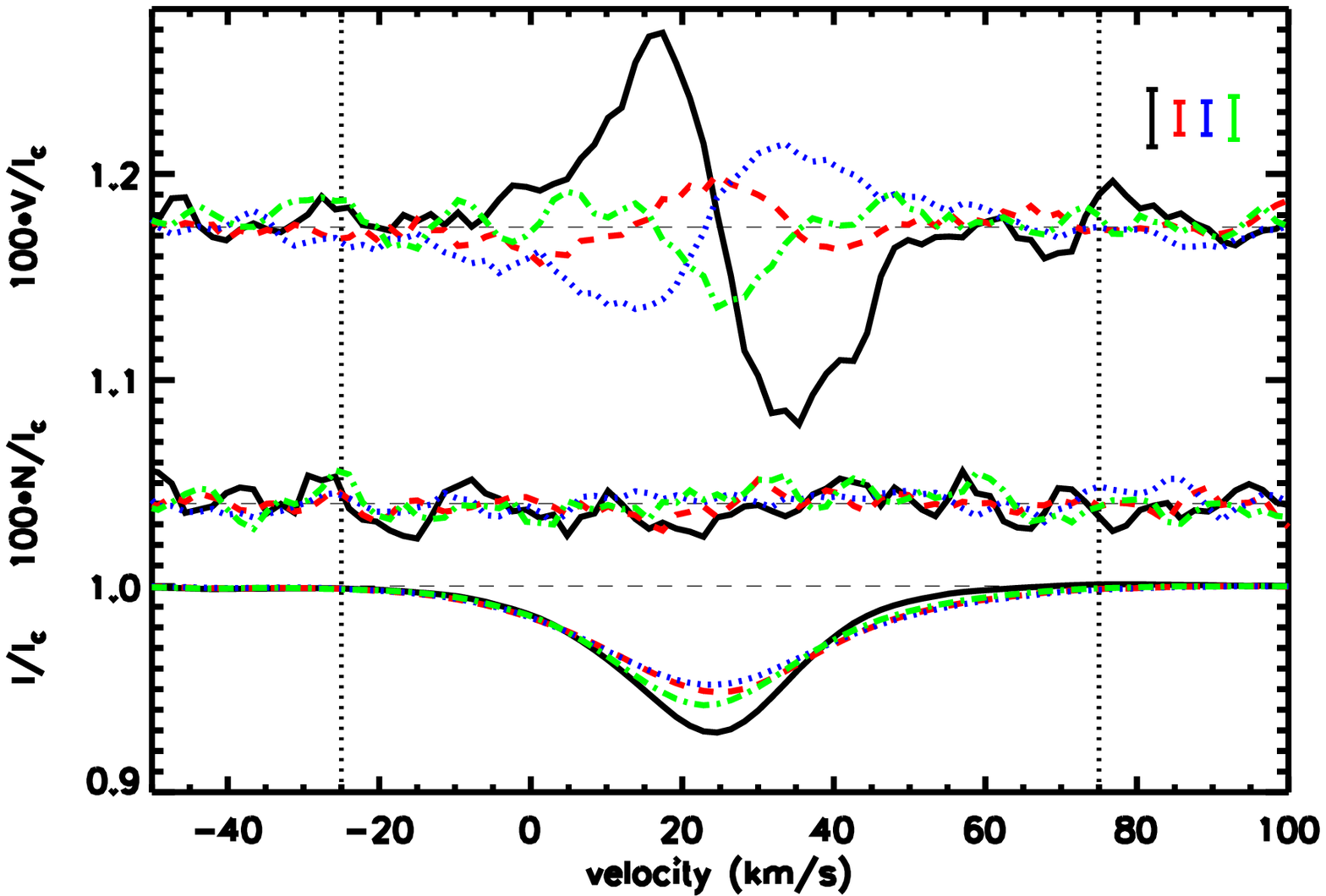}
\caption{Mean LSD Stokes $V$ (top), diagnostic null (middle) and Stokes $I$ profiles (bottom) of HD\,57682 from observations obtained at four different rotational phases (see Sect.~\ref{period_sect}; 0.02 - solid black, 0.30 - dashed red, 0.45 - dotted blue, 0.72 - dash-dotted green). The $V$ and $N$ profiles are expanded by the indicated factor and shifted upwards, and smoothed with 3 pixel boxcar for display purposes. A clear Zeeman signature is detected at most phases in the Stokes $V$ profiles, while the null profile shows no signal. The integration limits used to measure the longitudinal field are indicated by the dotted lines and the mean 1$\sigma$ pixel uncertainty for each profile is also shown.}
\label{lsd_fig}
\end{figure}

From each observation we extracted LSD profiles using the technique as described by \citet{don97}. The measurements reported here differ in details from those reported by \citet{grun09} as we have adopted an updated line mask with significantly more lines, but the measurements are fully consistent within their respective uncertainties. Our new mask has an average Land{\' e} factor and average wavelength of 1.180 and 4682\,\AA\ versus 1.135 and 4649\,\AA\ from the original mask. The new mask was derived from the Vienna Atomic Line Database \citep[VALD; ][]{pisk95} for a $T_{\rm eff}=35\,000$\,K, $\log(g)=4.0$ model atmosphere. Only lines with intrinsic depths greater than 1 percent of the continuum were included in the line list over the full ESPaDOnS spectral range (370 to 1050\,nm), resulting in 1648 He and metallic lines. The line list was further reduced by removing all intrinsically broad He and H lines, all lines that are blended with these lines, in addition to all lines blended with atmospheric telluric lines. This resulted in a final list of 571 metallic lines used to extract the final LSD profiles computed on a 1.8\,km\,s$^{-1}$ velocity grid, as illustrated in Fig.~\ref{lsd_fig}. The detection probability for each LSD spectrum was computed according to the criteria of \citet{don97}. Each observation resulted in a definite detection (DD) of an excess signal in Stokes~$V$ within the line profile (False Alarm Probability (FAP) $<10^{-5}$), except for the observations obtained on the nights of 2008-12-05 and  2009-05-07, which were non-detections (ND). 

We calculated the longitudinal magnetic field ($B_\ell$) and null measurements ($N_\ell$) from each LSD profile using the first-order moment method discussed by \citet{rees79}, using an integration range between -25 and 75\,km\,s$^{-1}$ (see Table~\ref{obs_tab1}). This range was adopted to encompass the full profile range of the Stokes~$V$ signature. The resulting $B_\ell$ measurements range from -170  to -350\,G with a median uncertainty of 20\,G. 

\begin{table*}
\centering
\caption{Journal of polarimetric observations listing the date, the heliocentric Julian date (2\,450\,000+), the number of sub-exposures and the exposure time per individual sub-exposure, the phase according to Eq.~\ref{ephemeris}, the peak signal-to-noise (S/N) ratio (per 1.8 km\,s$^{-1}$ velocity bin) in the $V$-band, the mean S/N in LSD Stokes $V$ profile, the evaluation of the detection level of the Stokes $V$ Zeeman signature according to \citet{don97} (DD=definite detection, ND=no detection), and the derived longitudinal field and longitudinal field detection significance $z$ from both $V$ and $N$. The first seven observations were already discussed by \citet{grun09}.}
\begin{tabular}{ccccrrcrrrr}
\hline
\ 	&	HJD 	& $t_{\rm exp}$ &	\ & \multicolumn{1}{c}{PK} & \multicolumn{1}{c}{LSD} & Det.& \multicolumn{2}{c}{~~~$V$}	& 	\multicolumn{2}{c}{~~~$N$}	\\								
\multicolumn{1}{c}{Date}	& (2\,450\,000+)		& (s)	& Phase & \multicolumn{1}{c}{S/N}	&  \multicolumn{1}{c}{S/N} & Flag & $B_\ell\pm\sigma_B$ (G) & \multicolumn{1}{c}{$z$}& $N_\ell\pm\sigma_N$ (G) & \multicolumn{1}{c}{$z$}	\\
\hline
2008-12-05  &   4806.0798  & $1\times4\times500$  &  0.9408 &   311	 & 2186  & ND &$354  \pm 93$ & 3.8  	&$-59 \pm93$&0.6 \\
2008-12-06  &   4807.1081  & $1\times4\times500$  &  0.9570 &   949	 & 8339  & DD &$248  \pm 24$ & 10.3  &$-11 \pm24$&0.5 \\
2009-05-04  &   4955.7675  & $1\times4\times540$  &  0.2955 &   1463	 & 12769 & DD &$-42  \pm 17$ & 2.5   &$-19 \pm17$&1.1 \\
2009-05-05  &   4956.7498  & $2\times4\times600$  &  0.3109 &   1388	 & 12192 & DD &$-64  \pm 18$ & 3.6   &$-23 \pm18$&1.3 \\
2009-05-07  &   4958.7805  & $2\times4\times540$  &  0.3429 &   653	 & 5546  & ND &$-78  \pm 39$ & 2.0   &$-17 \pm39$&0.4 \\
2009-05-08  &   4959.7489  & $1\times4\times540$  &  0.3581 &   1094	 & 9515  & DD &$-153 \pm 23$ & 6.7   &$9   \pm23$&0.4 \\
2009-05-09  &   4960.7480  & $2\times4\times540$  &  0.3738 &   900	 & 7634  & DD &$-126 \pm 29$ & 4.4   &$-7  \pm29$&0.2 \\
\hline                                                                                                      
2009-12-31  &   5197.0957  & $2\times4\times540$  &  0.0917 &   1566	 & 14309 & DD &$233  \pm 14$ & 16.2  &$-7  \pm14$&0.5 \\
2010-01-03  &   5201.0427  & $2\times4\times540$  &  0.1538 &   1110	 & 10354 & DD &$133  \pm 20$ & 6.6   &$-1  \pm20$&0.0 \\
2010-01-23  &   5219.9545  & $2\times4\times540$  &  0.4512 &   1376	 & 12569 & DD &$-169 \pm 18$ & 9.6   &$-13 \pm18$&0.8 \\
2010-01-29  &   5225.9836  & $2\times4\times540$  &  0.5461 &   1610	 & 14758 & DD &$-174 \pm 15$ & 11.5  &$12  \pm15$&0.8 \\
2010-01-31  &   5228.0015  & $2\times4\times540$  &  0.5779 &   1179	 & 10817 & DD &$-172 \pm 20$ & 8.5   &$-33 \pm20$&1.6 \\
2010-02-01  &   5228.9032  & $2\times4\times540$  &  0.5920 &   1250	 & 8088  & DD &$-162 \pm 27$ & 6.0   &$-8  \pm27$&0.3 \\
2010-02-24  &   5251.9131  & $2\times4\times414$  &  0.9540 &   1198	 & 10940 & DD &$234  \pm 18$ & 12.7  &$34  \pm18$&1.9 \\
2010-02-28  &   5255.9891  & $2\times4\times414$  &  0.0181 &   932	 & 7208  & DD &$190  \pm 28$ & 6.8   &$-33 \pm28$&1.2 \\
2010-03-04  &   5259.9420  & $2\times4\times414$  &  0.0803 &   1163	 & 10477 & DD &$208  \pm 20$ & 10.6  &$-10 \pm20$&0.5 \\
2010-03-08  &   5263.9096  & $2\times4\times414$  &  0.1427 &   878	 & 7921  & DD &$187  \pm 26$ & 7.1   &$-4  \pm26$&0.2 \\
2010-11-28  &   5529.0352  & $2\times4\times415$  &  0.3133 &   1224	 & 10980 & DD &$-60  \pm 19$ & 3.1   &$-5  \pm20$&0.3 \\
2010-12-24  &   5555.0639  & $2\times4\times415$  &  0.7227 &   1083	 & 9567   & DD &$-13  \pm 22$ & 0.6   &$-27 \pm22$&1.2 \\
2010-12-30  &	5561.0731  &	$2\times4\times415$  &	0.8172 &   995	 & 8850  & DD &$89   \pm 23$ & 3.8   &$20 \pm 23$&0.9 \\
\hline
\hline
\end{tabular}
\label{obs_tab1}
\end{table*}

\subsection{Spectroscopic data and measurements}
In addition to the polarimetric spectra, we also utilised archival spectroscopic observations to analyse the line profile variability (LPV) of H$\alpha$. Seven spectra were obtained with the Coud{\' e} Auxiliary Telescope (CAT) using the Coud{\' e} Echelle Spectrograph (CES, $R\sim45\,000$) at the European Southern Observatory (ESO), La Silla, Chile in 1996 February. These observations were obtained as part of a large monitoring campaign of bright O-type stars, as described by \citet{kaper98}. Eight spectra were obtained with the Ultraviolet and Visual echelle Spectrograph (UVES, $R\sim50,000$) mounted on ESO's Very Large Telescope (VLT) on 2002-12-13. These spectra were combined to yield a single observation. Five spectra were obtained with the Fibre-fed Extended Range Optical Spectrograph (FEROS, $R\sim48,000$) mounted on ESO's 2.2-m telescope at La Silla: four observations between 2004-02-07 and 2004-02-10 (on three separate nights), and another observation on 2004-12-22. Between 2009-10-07 and 2009-10-09 ten spectra were obtained with the Echelle spectrograph ($R\sim40,000$) on the 2.5-m Du Pont telescope at the Las Campanas Observatory (LCO). The nightly spectra were combined to produce three distinct observations. Lastly, we also used spectra from the BeSS database obtained between 2008 November and 2011 April. Fifty-one spectra were collected, primarily with a LHIRES3 spectrograph ($R\sim15\,000$) using various telescopes with diameters between 0.2 to 0.3-m. A log of these spectroscopic observations is listed in Table~\ref{obs_tab2}.

From the spectroscopic datasets we measured the equivalent width (EW) variations of H$\alpha$ (in addition to several additional spectral lines present in the ESPaDOnS dataset - see Sect.~\ref{spec_var_sect} for further details), following the same procedure as described by \citet{wade12a}. The H$\alpha$ EW measurements are listed in Table~\ref{online_tab}.

We also utilized archival {\it IUE} spectra, the details of which are described by \citet{grun09}.

\begin{table}
\centering
\caption{Journal of spectroscopic observations listing the identification of the dataset, the typical resolving power of the instrument, the epoch of the observations, the number of spectra obtained within the given dataset, the resulting number of distinct observations, and the median per pixel signal-to-noise (S/N) ratio in the H$\alpha$ region for the dataset.}
\begin{tabular}{cccccc}
\hline
\ & Res. & \ & Num & Num & Median \\
Name & Power & Epoch & Spec & Obs & S/N \\
\hline
CES & 45\,000 & 1996 & 7 & 7 & 540 \\
UVES & 50\,000 & 2002 & 8 & 1 & 160 \\
FEROS & 48\,000 & 2004 & 5 & 5 & 170\\
LCO & 40\,000 & 2009 & 10 & 3 & 360\\
BeSS & 15\,000 & 2008-2011 & 51 & 51 & 70 \\
\hline
\hline
\end{tabular}
\label{obs_tab2}
\end{table}

\begin{table}
\centering
\caption{Spectroscopic H$\alpha$ EW measurements. Included is the instrument or observatory name, the heliocentric Julian date, and the measured H$\alpha$ equivalent width (EW) and its corresponding 1$\sigma$ uncertainty. Shown here is only a sample of the table. The full table can be found in Appendix \ref{apndx1}.}
\begin{tabular}{cccc}
\hline
Dataset & HJD & H$\alpha$ EW\,(\AA) & $\sigma_{\rm EW}$\,(\AA) \\
\hline
CES	&	2450125.6216	&	1.270	&	0.014	\\
CES	&	2450126.6291	&	1.108	&	0.007	\\
CES	&	2450127.6296	&	0.926	&	0.008	\\
CES	&	2450128.6284	&	0.735	&	0.012	\\
CES	&	2450129.6287	&	0.569	&	0.011	\\
CES	&	2450130.6194	&	0.454	&	0.010	\\
\hline
\end{tabular}
\label{online_tab}
\end{table}

\subsection{Photometric data}\label{obs_phot_sect}
Photometric observations were also utilised to study the optical variability (see Sect.~\ref{period_sect}) and to characterise the stellar and circumstellar emission properties (see Sect.~\ref{grtr_circ}). 

The optical photometric time series corresponds to 85 measurements from Hipparcos \citep{esa97} collected on 64 different nights between 1990-04-03 and 1993-06-06. 

Additional photometric observations used to construct the spectral energy distribution (SED) consists of optical data obtained from the Centre de Donn\'{e}es astronomiques de Strasbourg (CDS), and infrared measurements from the Two Micron All Sky Survey (2MASS), the Deep Near Infrared Survey (DENIS), the Infrared Astronomical Satellite (IRAS), the AKARI satellite and the Wide-field Infrared Survey Explorer (WISE), which were all obtained from NASA's Infrared Science Archive (IRSA). Additionally, we also used photometry from \citet{the86}.

The optical and infrared magnitudes were converted to flux measurements at the zero-point of the associated photometric filters using conventional methods.

\section{Period Analysis}\label{period_sect}
\begin{figure}
\centering
\includegraphics[width=3.3in]{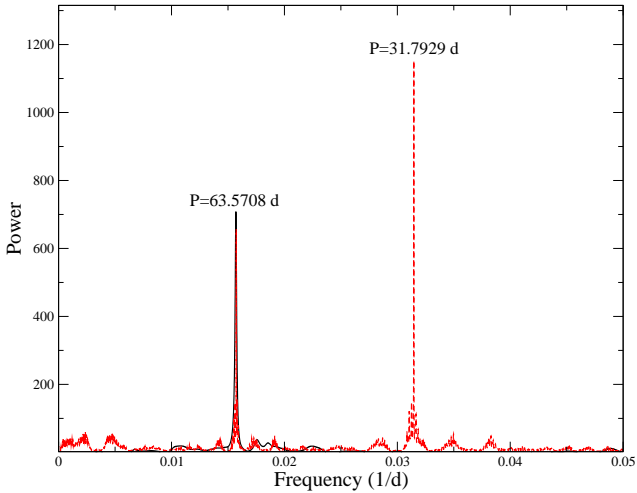}
\caption{Periodograms obtained from the $B_\ell$ measurements (solid black) and from the H$\alpha$ EW variations including contributions from the first harmonic (dashed red). Note the strong power present in the H$\alpha$ periodogram at $P=31.7927$\,d but the lack of power from the $B_\ell$ measurements at this period. However, a consistent period is found at $P=63.5708$\,d from both datasets.}
\label{period_fig}
\end{figure}

\citet{grun09} attempted to infer the rotational period ($P_{\rm rot}$) of HD\,57682 based on the limited temporal sampling of their dataset. With the larger dataset presented in this paper, we re-evaluate the periodicity and rotation period of HD\,57682.

Using a Lomb-Scargle technique \citep{press96}, we analysed the longitudinal magnetic field variations $B_\ell$ and the EW measurements of H$\alpha$. The periodograms obtained from these analyses are displayed in Fig.~\ref{period_fig}. The results obtained from the $B_\ell$ measurements show a clear peak in the periodogram at a period of $63.52^{+0.18}_{-0.17}$\,d, where the uncertainties represent the 1$\sigma$ limits. When phased with this period, the $B_\ell$ measurements show a clear sinusoidal variation. However, the periodogram obtained from the H$\alpha$ measurements shows a clear peak occurring at $31.7929^{+0.0029}_{-0.0014}$\,d, roughly half the period that is obtained from the $B_\ell$ measurements. If we phase the H$\alpha$ EW measurements to this period, we obtain a roughly sinusoidal variation, but with a relatively broad dispersion. On the other hand, the $B_\ell$ measurements cannot be reasonably phased with this period. We proceeded to obtain a periodogram of the H$\alpha$ data using the multi-harmonic fitting technique of \citet{sc96}, including contributions from the first harmonic only. The resulting periodogram shows two peaks, one occurring at about 31\,d and the other occurring at $63.5708\pm0.0057$\,d. This second peak is consistent with the period found from the magnetic measurements. Therefore, adopting this period as the stellar rotation period and the phase of maximum positive $B_\ell$ as ${\rm HJD}_0$ we derive the following ephemeris:
\begin{equation}\label{ephemeris}
{\rm HJD}^{max}_{B_\ell}=2453347.71(35) + 63.5708(57)\cdot E,
\end{equation}
where the uncertainties (1$\sigma$ limits) in the last digits are indicated in brackets. Unless otherwise stated, all further data are phased according to this ephemeris. 

\begin{figure*}
\centering
\includegraphics[width=6.7in]{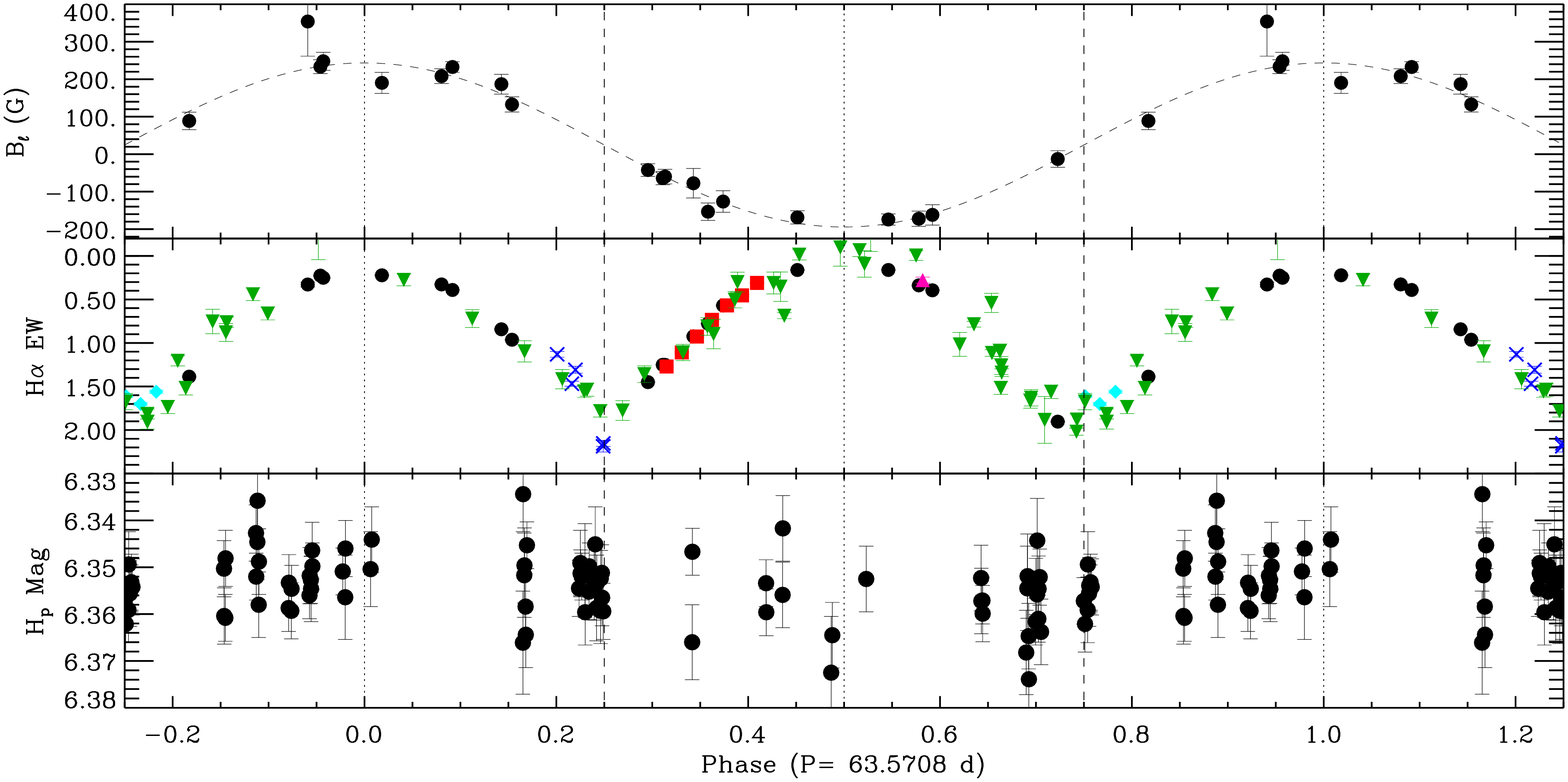}
\caption{Phased observational data using the ephemeris of Eq.~\ref{ephemeris}. {\bf Upper:} Longitudinal magnetic field variations measured from LSD profiles of the ESPaDOnS spectra. {\bf Middle:} H$\alpha$ equivalent width variations measured from ESPaDOnS (black circles), CES (red squares), FEROS (blue Xs), Las Campanas Observatory (turquoise diamonds), UVES (pink up-facing triangles), and BeSS (green down-facing triangles) datasets. {\bf Lower:} Hipparcos photometry. The dashed curve in the upper frame represents a least-squares sinusoidal fit to the data.}
\label{phot_mag_comp}
\end{figure*}

In Fig.~\ref{phot_mag_comp} we show the phased $B_\ell$ measurements, H$\alpha$ EW variations, and Hipparcos photometry of HD\,57682. The $B_\ell$ measurements follow a simple, sinusoidal variation with maximum positive $B_\ell$ occurring at phase 0.0 and maximum negative $B_\ell$ occurring at phase 0.5. The H$\alpha$ EW measurements show a double peaked variation with maximum emission occurring at phases 0.0, and 0.5, and minimum emission occurring at phases 0.25 and 0.75. This is further discussed in Sect.~\ref{magneto_sect}. The Hipparcos measurements do not show any significant variations when phased with the rotation period. Furthermore, the periodogram shows no significant power at this or any other period and the scatter of the photometric variations and their associated uncertainty is fully consistent with no variability.

\section{Rotational Broadening}\label{rot_sect}
\citet{grun09} measured the projected rotational velocity ($v\sin i$) using the Fourier transform method \citep[e.g.][]{gray81, jankov95, simon07}. They compared the positions of the first nodes in the Fourier spectrum to a theoretical Gaussian profile convolved with a rotationally broadened profile corresponding to a particular $v\sin i$, and found a $v\sin i=15\pm3$\,km\,s$^{-1}$. 

If we assume rigid rotation and take the radius as inferred by \citet{grun09} ($R_\star=7.0^{+2.4}_{-1.8}\,R_\odot$) and the period obtained in this work ($P=63.5708$\,d), we find a maximum allowed $v\sin i$ (i.e. for $\sin i=1$) $\sim$7.5\,km\,s$^{-1}$, which is inconsistent with the results obtained from the Fourier analysis. This implies that contributions from non-rotational broadening, as noted by \citet{grun09}, significantly affect the Fourier spectrum resulting in the incorrect interpretation of the first node representing the rotational broadening. 

\begin{figure}
\centering
\includegraphics[width=3.2in]{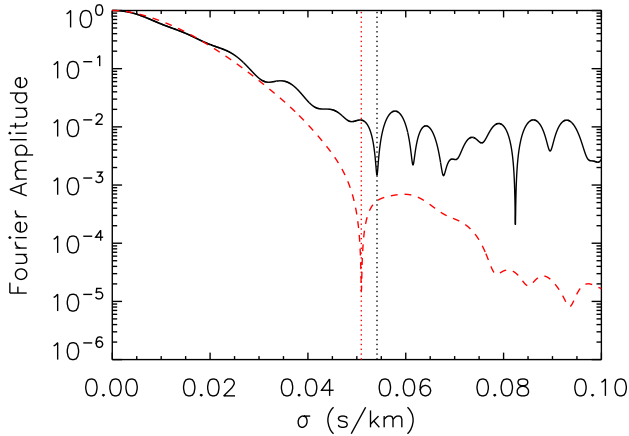}
\caption{Fourier spectrum of the C\,{\sc iv} $\lambda5801$ line (black) compared to a theoretical line profile with $v\sin i=5$\,km\,s$^{-1}$ with an additional macroturbulent velocity of 20\,km\,s$^{-1}$ (red), which we find necessary to provide a good fit to the observed line profile. The difference in the positions of the first nodes (as indicated by the vertical lines) are used to determine the best-fit $v\sin i=4.2$\,km\,s$^{-1}$ for this observation.}
\label{four_fig}
\end{figure}

In our endeavour to constrain the rotational broadening, we re-analysed the spectra, particularly focusing on those obtained at phases where the lines appeared their narrowest and therefore were least affected by macroturbulence (see Sect.~\ref{spec_var_sect} for further details). We identified lines from a theoretical {\sc synth3} \citep{oleg} spectrum for which the resulting Fourier spectrum was minimally affected when significant contributions of macroturbulence were included. We then compared the Fourier spectrum obtained from a theoretical profile with 5\,km\,s$^{-1}$ rotational broadening to that obtained from the observed spectrum and measured the difference between the positions of the first nodes to obtain the $v\sin i$ of the spectral line. Using this procedure on several lines from the spectrum obtained on 2009 December 31 indicate a $v\sin i=6.1\pm1.9$\,km\,s$^{-1}$, consistent with the range implied by the rotation period and estimated radius of HD\,57682. However, we note that this period is sufficiently imprecise that it cannot be used to usefully constrain the inclination of the rotation axis. We also found that the C\,{\sc iv} $\lambda5801$ line was one of the lines least affected by the addition of macroturbulence and therefore also used this single line to determine a unique $v\sin i$, as illustrated in Fig.~\ref{four_fig}. From measurements of this line at several rotation phases we find a best-fit $v\sin i=4.6\pm0.6$\,km\,s$^{-1}$, where the uncertainties represent the 1$\sigma$ limits. Given the range of $v\sin i$ obtained from the C\,{\sc iv} $\lambda5801$, our adopted $R_\star$ and $P_{\rm rot}$, implies that $i\sim56^{+35}_{-23}\degr$, or that we can place a lower limit on $i$ such that $i>30\degr$, at a $1\sigma$ confidence. However, at $\sim$4.6\,km\,s$^{-1}$ we are approaching the spectral resolution of ESPaDOnS ($\sim$4.4\,km\,s$^{-1}$) and therefore one must be cautious to not over interpret these results.

\section{Magnetic Field Geometry}\label{mag_field_sect}
The magnetic field geometry of HD\,57682 was investigated assuming the field is well described by the Oblique Rotator Model (ORM), which is characterised by four parameters: the phase of closest approach of the magnetic pole to the line-of-sight $\phi_0$, the inclination of the stellar rotation axis $i$, the obliquity angle between the magnetic axis and the rotation axis $\beta$ and the dipole polar strength $B_d$. This model naturally predicts the simple sinusoidal variation of the longitudinal magnetic field measurements as shown in Fig.~\ref{phot_mag_comp}, implying that the longitudinal field curve of HD\,57682 is well described by a simple dipole topology of its global magnetic field.

We first attempted to infer the characteristics of the magnetic field based on the $B_\ell$ measurements. These measurements are nearly symmetric about 0 and we can therefore infer that either $i$ or $\beta$ (or both) are close to 90$\degr$. This is similar to the $B_\ell$ variation of $\beta$~Cephei \citep[e.g.][]{don01}, and similar to this star, $B_d$ and $\beta$ are well-constrained except as $i$ approaches 0 or 90\,$\degr$. As $i$ approaches 0$\degr$, $B_d$ becomes unconstrained, while as $i$ approaches 90$\degr$, $\beta$ becomes unconstrained. To infer the characteristics of the magnetic field, we carried out a $\chi^2$ minimisation, comparing the observed $B_\ell$ curve to a grid of computed longitudinal field curves to determine $B_d$ and $\beta$ for a fixed linear limb darkening coefficient of 0.35 \citep{claret2000} and various fixed inclinations.

\begin{figure*}
\centering
\includegraphics[width=6.9in]{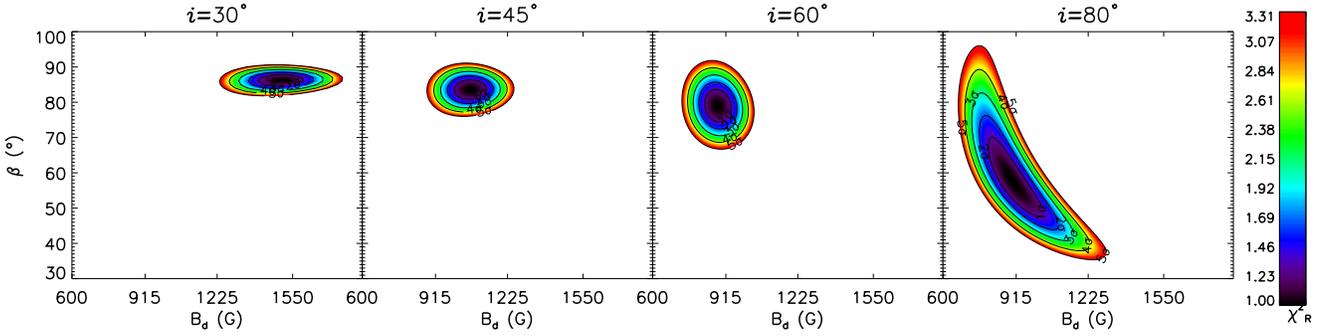}
\caption{Reduced $\chi^2$ contours of dipole field strength $B_d$ versus magnetic obliquity $\beta$ permitted by the longitudinal field variation of HD\,57682 for varying inclinations.}
\label{chisq_land}
\end{figure*}

The resulting $\chi^2$ distributions from our fits are shown in Fig.~\ref{chisq_land}. For an assumed $i=60\degr$ we find that $B_d=880\pm50$\,G and $\beta=79\pm4\degr$. If we take $i=30\degr$, we infer that $B_d=1500\pm90$\,G and $\beta=86\pm2\degr$, which places an upper limit on the dipole field strength if $i>30\degr$ as implied from the constraints derived in Sect.~\ref{rot_sect}. Also shown in Fig.~\ref{chisq_land} are $\chi^2$ distribution for other inclinations to illustrate how the distribution varies as a function of inclination.

We also attempted to constrain the magnetic field characteristics based on fits to the individual LSD Stokes~$V$ profiles by comparing our profiles to a large grid of synthetic profiles, also characterised by the ORM. The models were computed by performing a disc integration of local Stokes $V$ profiles assuming the weak field approximation and uniform surface abundance. The parameters of the Stokes $V$ profiles were chosen to provide the best fit to the width of the average of all the observed $I$ profiles, and the best fit to the depths of each individual observed $I$ profile. For each observation we found the parameters that provided the lowest $\chi^2$ value and inferred the maximum likelihood by combining these results in a Bayesian framework. 

We conducted our fits for varying inclinations and found that the maximum likelihood model for $i\sim80\degr$ provided the lowest combined total $\chi^2$, but this value was not significantly different from those obtained for other inclinations. However, as $i$ increased we did find increasing deviations between the best $B_\ell$-fit and profile-fit model parameters; at $i=30\degr$ we found a best $B_\ell$-fit $\beta=86\degr$ and a profile-fit $\beta=80\degr$; at $i=80\degr$ we found a best $B_\ell$-fit $\beta=56\degr$ but $\beta=34\degr$ from the profile fits. We therefore conclude that we are unable to constrain the inclination from fits to the LSD profiles and continue to adopt $i=60\degr$ for the remainder of this discussion as this is the inclination inferred from the $v\sin i$ measurements and provides similar magnetic parameters from both the $B_\ell$ and profile fits.

The maximum likelihood profile-fit model for $i=60\degr$ was found with $B_d=700\pm30$\,G, where the uncertainty corresponds to the 95.6 percent significance level, and $\beta=68\pm2\degr$, with the uncertainty corresponding to the 98.2 percent confidence limits. While the formal uncertainties are very low, we note that individual best-fit parameters to each observation range from $B_d\sim400-2000$\,G and $\beta\sim0-150\degr$. However, while the fits to $B_d$ and $\beta$ are not well constrained from fits to the individual observations at most phases, the maximum likelihood model is well constrained. 

\begin{figure*}
\includegraphics[width=6.8in]{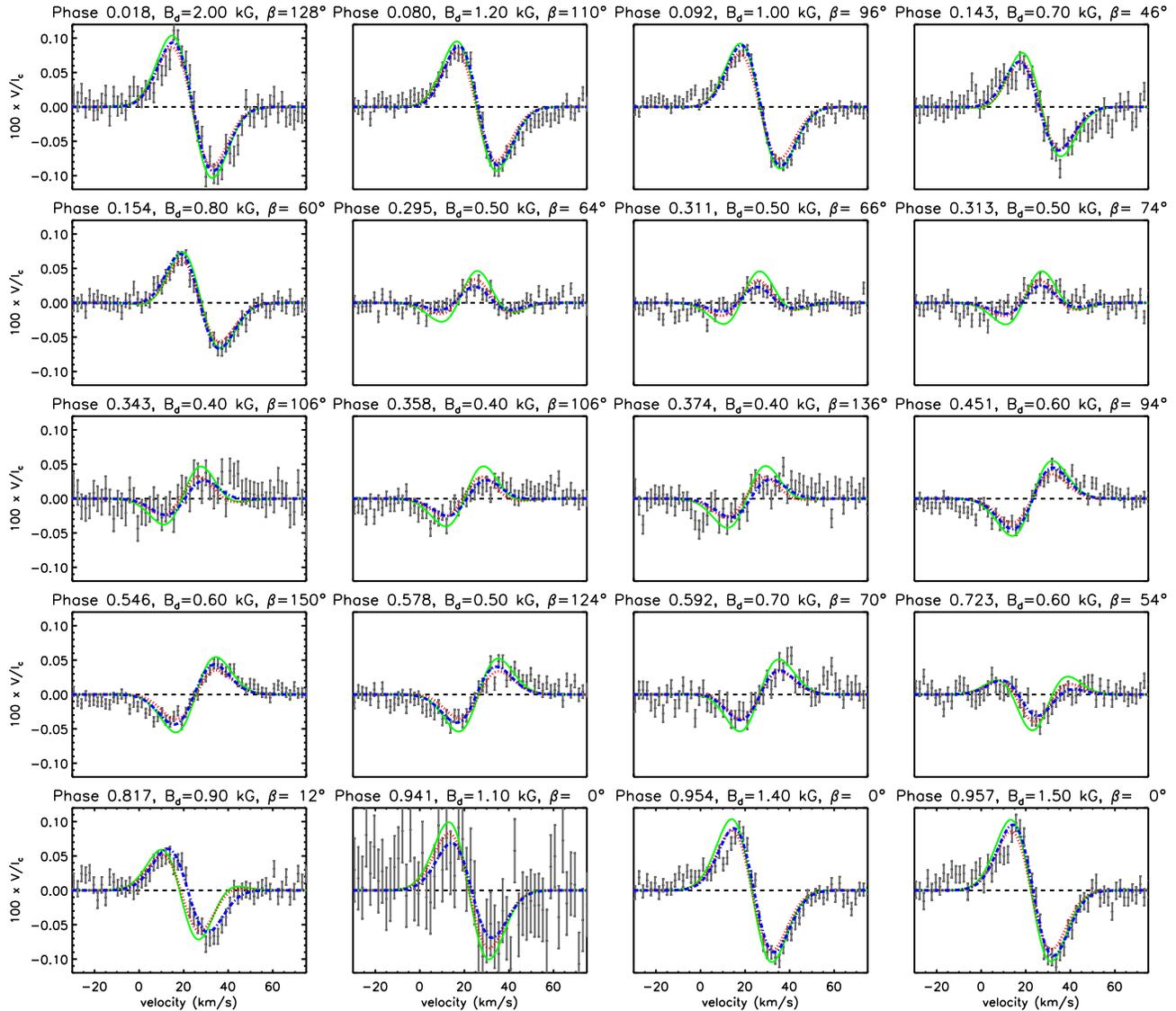}	  
\caption{Mean circularly polarised LSD Stokes $V$ profiles (grey circles). The error bars represent the 1$\sigma$ uncertainties of each pixel. Also shown are the individual best-fit model profiles for each phase (dash-dotted blue), the model that provides the global maximum likelihood (dotted red; $B_d=700$\,G, $\beta=68\degr$) and the model corresponding to the best-fit to the $B_\ell$ measurements ($B_d=880$\,G, $\beta=79\degr$; solid green), all for an assumed inclination of $60\degr$. The phase and best-fit parameters of the individual observations are also indicated.}
\label{lsd_figure}
\end{figure*}

In Fig.~\ref{lsd_figure} we compare the observed Stokes $V$ profiles (grey circles) with the best profile-fit model for the given observation (dash-dotted blue), the maximum likelihood LSD model (dotted red) and the model corresponding to the best-fit parameters that were obtained from the $B_\ell$ measurements for $i=60\degr$ ($B_d=880$\,G and $\beta=79\degr$; solid green). The quality of the fits of the best-fit and maximum likelihood models are similar, which shows that a single dipole configuration can reasonably reproduce the observed Stokes $V$ profiles. However, systematic differences, particularly in the wings of the profiles, are evident between the models and the observations, resulting from the fact that our simple model cannot reproduce the extended wings in the observed $I$ or $V$ profiles. It is also evident that the model corresponding to the best-fit to the $B_\ell$ curve does a poorer job of fitting the observed Stokes $V$ profiles at several phases, possibly indicating that there may be an issue with the way we chose to model the line profiles.

We investigated this potential issue by re-conducting our fits, but neglecting the varying line depths for each observation as was done in the original procedure. Instead we adopted a line depth as determined from the average of all our observations. From the new fits for $i=60\degr$ we found a best-fit $B_d\sim900$\,G, which is consistent with the results of the $B_\ell$ measurements. However, we find a larger disagreement between the inferred magnetic obliquity as $\beta$ was found to be $\sim$60$\degr$. We also note that the overall combined total $\chi^2$ value was slightly worse from these fits.  

Another possibility is that the global magnetic topology contains significant contributions from higher order multipoles. The $B_\ell$ curve is rather insensitive to these higher order moments, in contrast
to the velocity-resolved Stokes $V$ profiles. We investigated this possibility by applying our fitting procedure to a timeseries of synthetic profiles calculated assuming our original dipole, supplemented by a significant aligned quadrupolar field moment. This model yields a longitudinal field curve essentially indistinguishable from that produced by the pure dipole. Our tests indicate that by trying to fit a pure dipole model to the dipole+quadrupole profiles, the procedure would systematically infer dipole field strengths that were strongly incompatible with the results obtained from the $B_\ell$ curve. Specifically, our tests showed a systematic over-estimation of the dipole field strength at most phases. This trend is inconsistent with the measurements presented in Fig.~\ref{lsd_figure} and we therefore conclude that our polarimetric observations do not suggest any significant detectable contributions from higher order multipoles.

\section{Line Variability}\label{spec_var_sect}
\subsection{Equivalent width and radial velocity measurements}
HD\,57682 stands out from the sample of magnetic O-type stars due to its sharp spectral lines and the apparently limited contamination of its spectrum by wind and circumstellar plasma emission. Detection and measurement of the line profile variations \citep[e.g.][]{stahl96} can therefore be performed more sensitively for a much larger sample of lines than for any other magnetic O-type star. It therefore represents a uniquely suited target for understanding the general influence of magnetised winds of O-type stars on their spectrum and spectral variability. 

To quantify the variability, we used the EW variations for numerous spectral lines. We also measured the radial velocities of each spectral line by fitting a Gaussian to the cores of the profiles to determine the central wavelength at each phase. A number of other methods were investigated, but we ultimately settled on fitting the cores using a Gaussian as it provided less scatter in the measurements. However, since H$\alpha$ displayed too much emission to carry out this fitting, we quantified the radial velocity of H$\alpha$ by measuring the centre of gravity (or centroid) of the line profile.

Our observations reveal that HD\,57682 displays a rich variety of line profile variability. In addition to the double-wave variations of the H$\alpha$ line discussed in Sect.~\ref{period_sect}, we observe variations in essentially every line present in the stellar spectrum, including other H Balmer lines, lines of neutral and ionised He, and lines of metals: C\,{\sc iii} and {\sc iv}, N\,{\sc iii}, O\,{\sc ii}, Mg\,{\sc ii}, and Si\,{\sc iii} and {\sc iv}. The EW and radial velocities of all lines vary coherently when phased according to the rotational ephemeris (Eq.~\ref{ephemeris}).

\begin{figure*}
\centering
\includegraphics[width=3.46in]{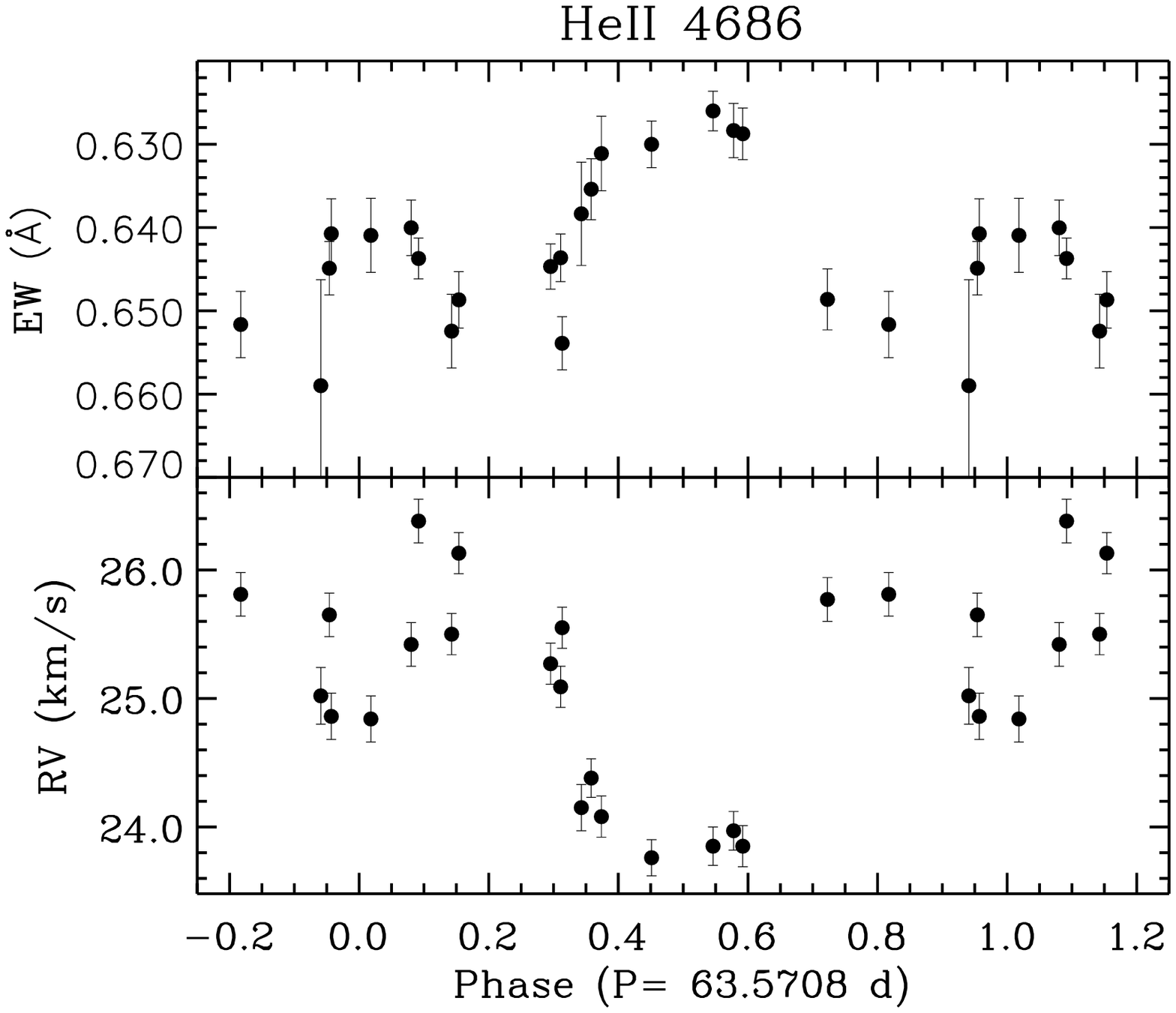}
\includegraphics[width=3.46in]{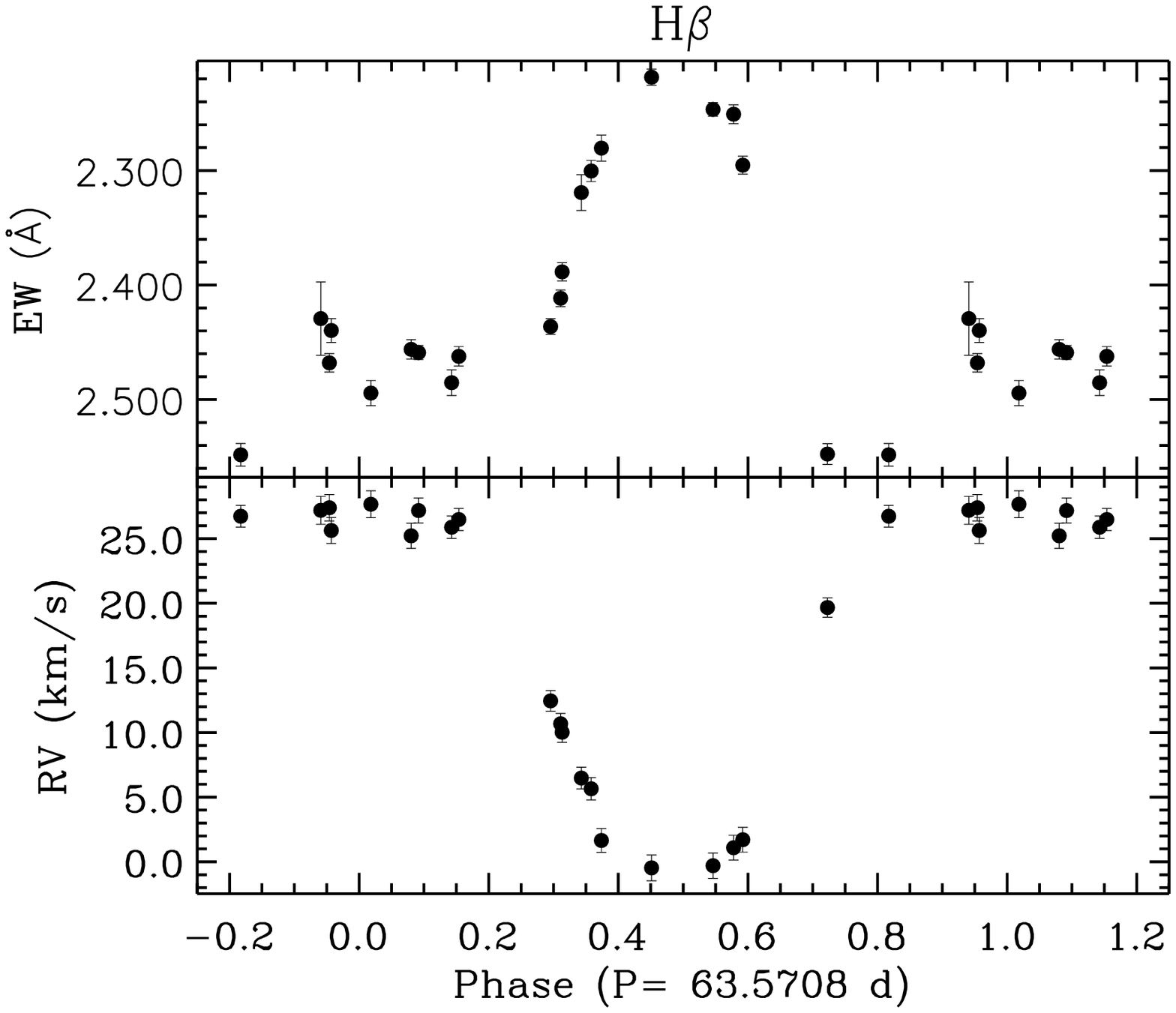}\\
\includegraphics[width=3.46in]{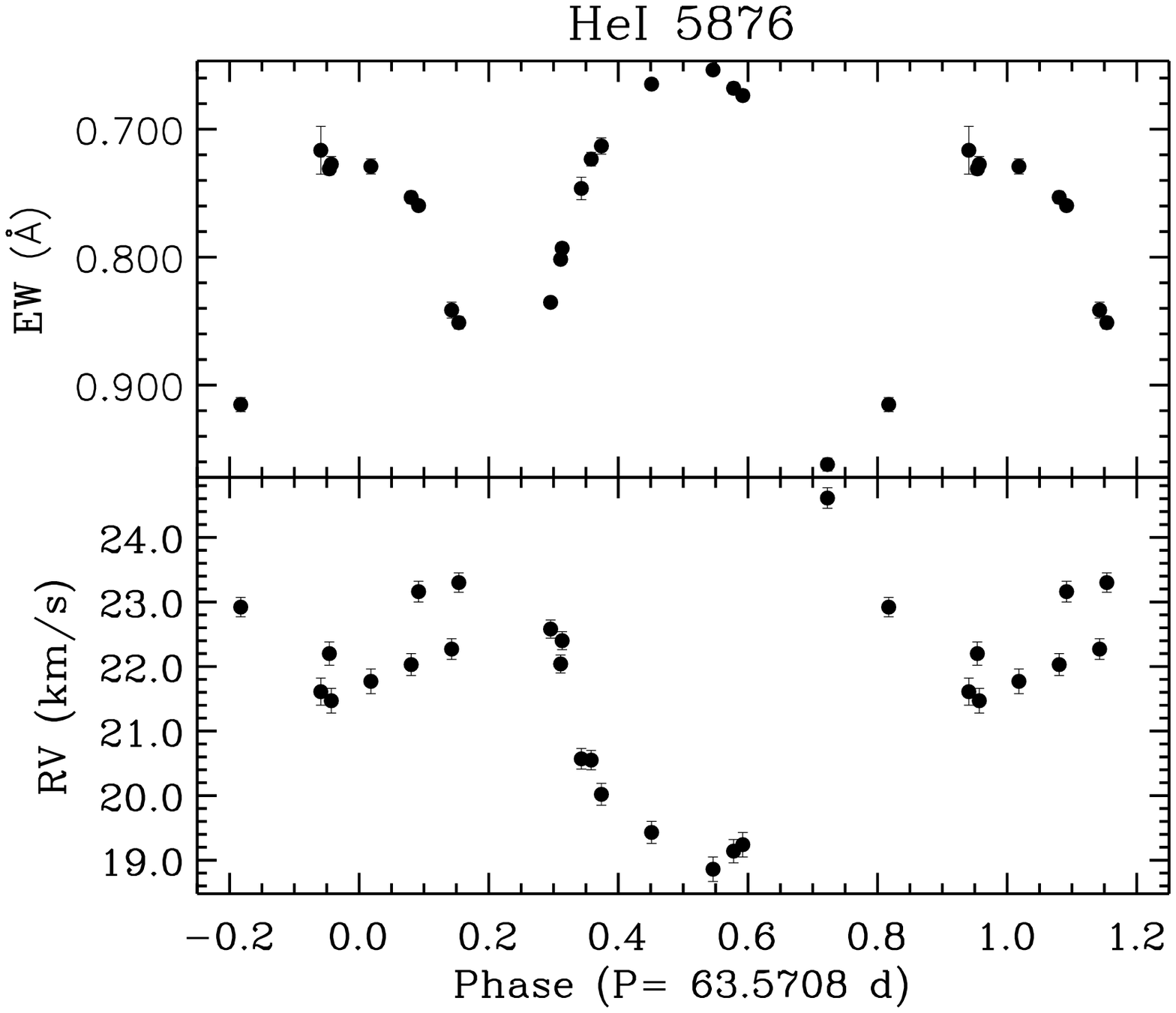}
\includegraphics[width=3.46in]{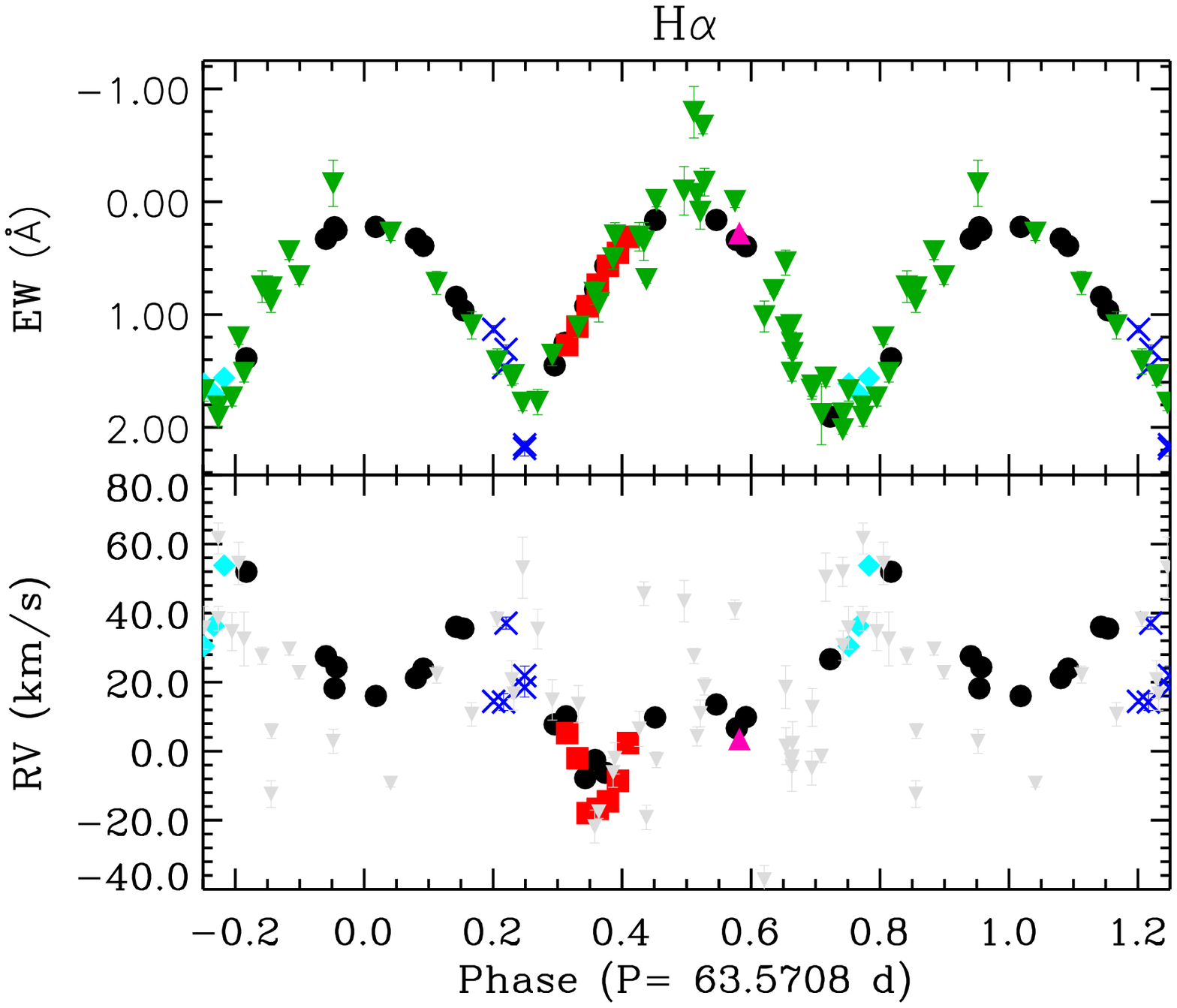}
\caption{Rotationally phased EW measurements (top panels) and radial velocity measurements (bottom panels) for selected spectral absorption lines with clearly non-sinusoidal EW variations. H$\alpha$ measurements from ESPaDOnS (black circles), CES (red squares), FEROS (blue Xs), Las Campanas Observatory (turquoise diamonds), UVES (pink up-facing triangles), and BeSS (green down-facing triangles) datasets. The BeSS H$\alpha$ radial velocity measurements are shown as grey triangles. The large scatter of these measurements with respect to the other datasets likely results from their poor S/N and low spectral resolution.}
\label{nonsin_ew}
\end{figure*}

The line variations can be approximately grouped into two categories. The first represents lines (mostly metals and some neutral He lines) with EW measurements that vary sinusoidally, as displayed in Fig.~\ref{sin_ew}. The second includes those, like H$\alpha$, that display EW phase variations that are distinctly non-sinusoidal (as shown in Fig.~\ref{nonsin_ew}), and likely directly reflect, to varying degrees, the variable projection of the flattened equatorial magnetospheric plasma density enhancement, as further discussed in Sect.~\ref{magneto_sect}. These groups are discussed in more detail below.

Except for C\,{\sc iv} $\lambda 5801, 5812$, all lines with sinusoidally varying EW measurements show minimum emission (or maximum absorption) at about phase 0.0 and maximum emission (or minimum absorption) at phase 0.5. At phase 0.5, these lines are shallower and broader than at phase 0.0, as illustrated in Fig.~\ref{spec_comp}. The difference in line width and depth is quite significant. Most lines have an extended red wing at phase 0.5 that disappears at phase 0.0.

The H$\alpha$ line shows variable emission with two maxima per rotation cycle. The phases of the emission extrema, 0.0 and 0.5, correspond to the phases of longitudinal magnetic field extrema and therefore closest approach of the magnetic poles to the line-of-sight. In addition, the H$\alpha$ EW curve exhibits two emission minima at phases 0.25 and 0.75, corresponding to the phases where the magnetic equator crosses the line-of-sight. These phenomena are usually interpreted \citep[e.g.][]{don01} as the consequence of the variable projection of a flattened distribution of magnetospheric plasma trapped in closed loops near the magnetic equatorial plane, and possibly the occultation of the stellar disc by this plasma at phases 0.25 and 0.75. Only three other lines - He\,{\sc i} $\lambda5876$, He{\sc ii}\,$\lambda4686$ and H$\beta$ (see Fig.~\ref{nonsin_ew}) - show clear evidence of similar reversals at phases 0.25 and 0.75, although a marginal contribution is suspected for He\,{\sc i} $\lambda 4921$ and He\,{\sc i} $\lambda 4471$. The emission level in H$\alpha$ at phase 0.5 is roughly 10 percent larger than at phase 0.0. Likewise, the emission minimum at phase 0.75 appears 20 percent less than the corresponding emission minimum at phase 0.25. However, these measurements are primarily measured from BeSS spectra with limited S/N.

In almost all the spectral lines that show sinusoidally varying phased EW measurements, we find approximately sinusoidally varying radial velocity measurements. These variations reach a maximum radial velocity between phases 0.2-0.3 and show a peak-to-peak amplitude of about 2\,km\,s$^{-1}$, with a mean radial velocity of $\sim$22\,km\,s$^{-1}$. This is not the case for H$\gamma$, which reaches a maximum radial velocity at phase 0.0 and minimum at phase 0.5, and shows a peak-to-peak difference of $\sim$8\,km\,s$^{-1}$. Other than H$\beta$, we see non-sinusoidal variations in the radial velocity measurements for the lines that show clear, non-sinusoidal EW variations, which often mirror the EW variations with maximum radial velocity occurring at minimum emission and minimum radial velocity occurring at maximum emission. The radial velocity measurements for H$\beta$ phase in a similar manner with H$\gamma$ - they are nearly sinusoidal, reach maximum radial velocity at phase 0.0 and appear flat-topped.

\begin{figure*}
\centering
\includegraphics[width=2.3in]{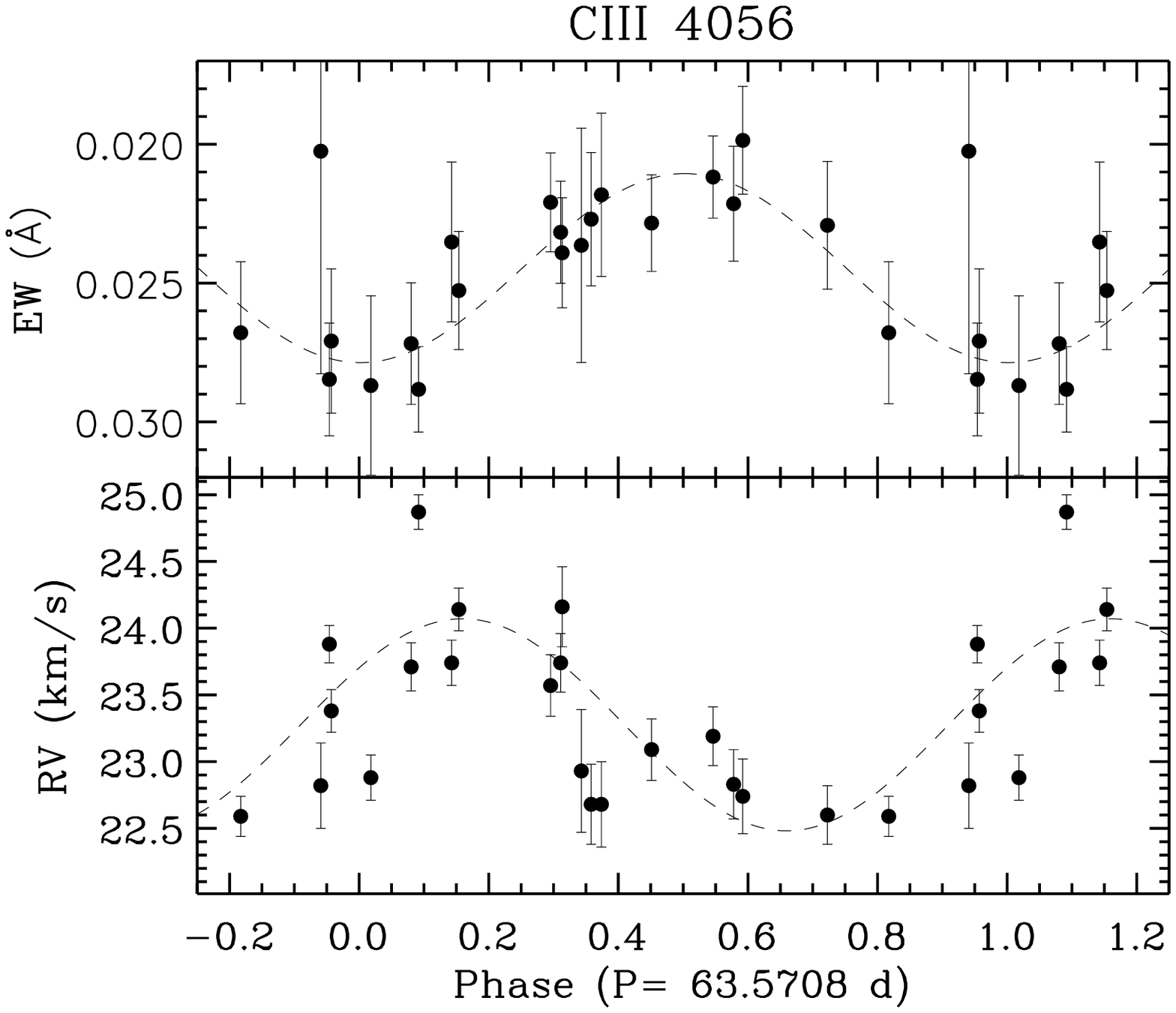}
\includegraphics[width=2.3in]{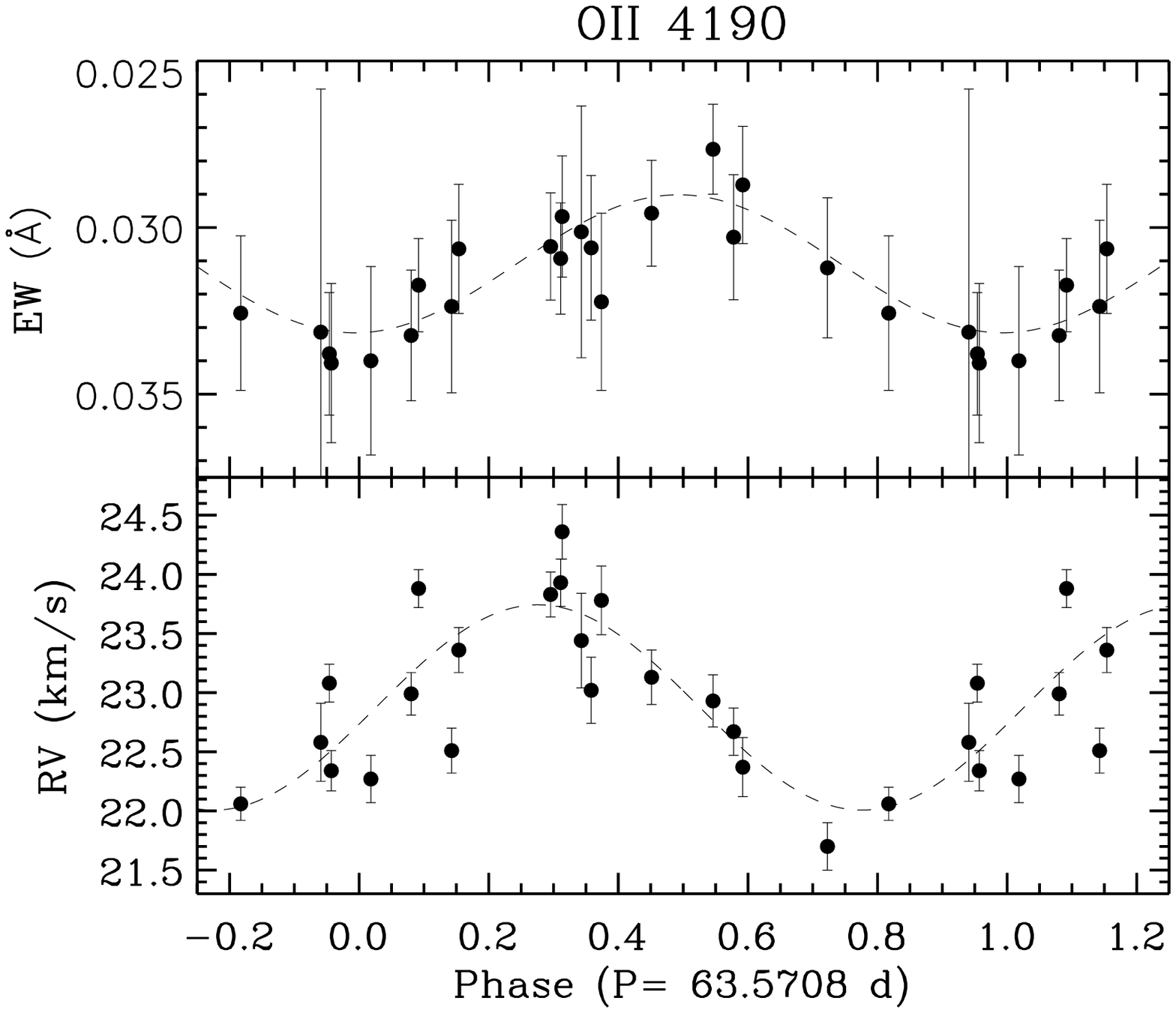}
\includegraphics[width=2.3in]{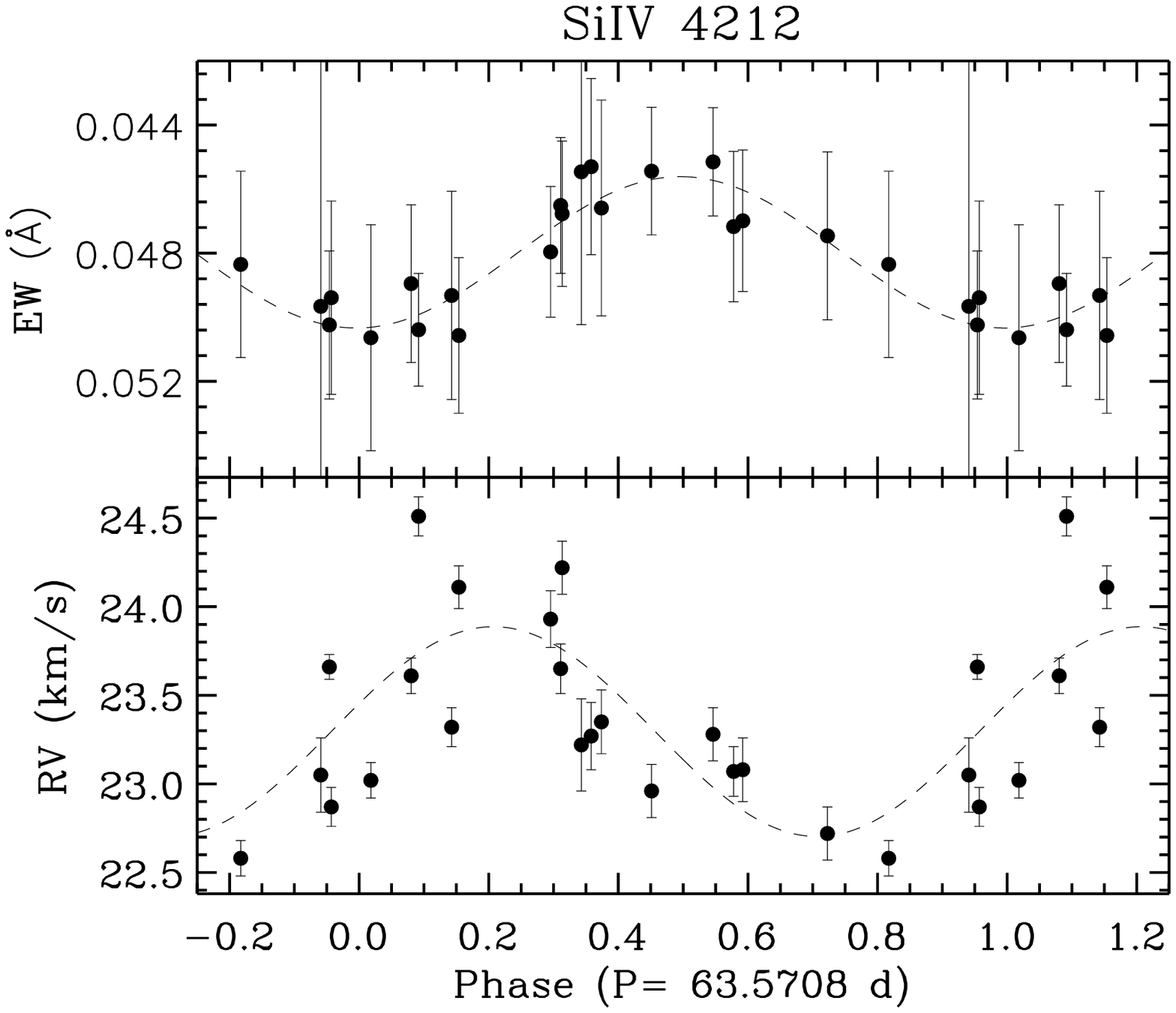}\\
\includegraphics[width=2.3in]{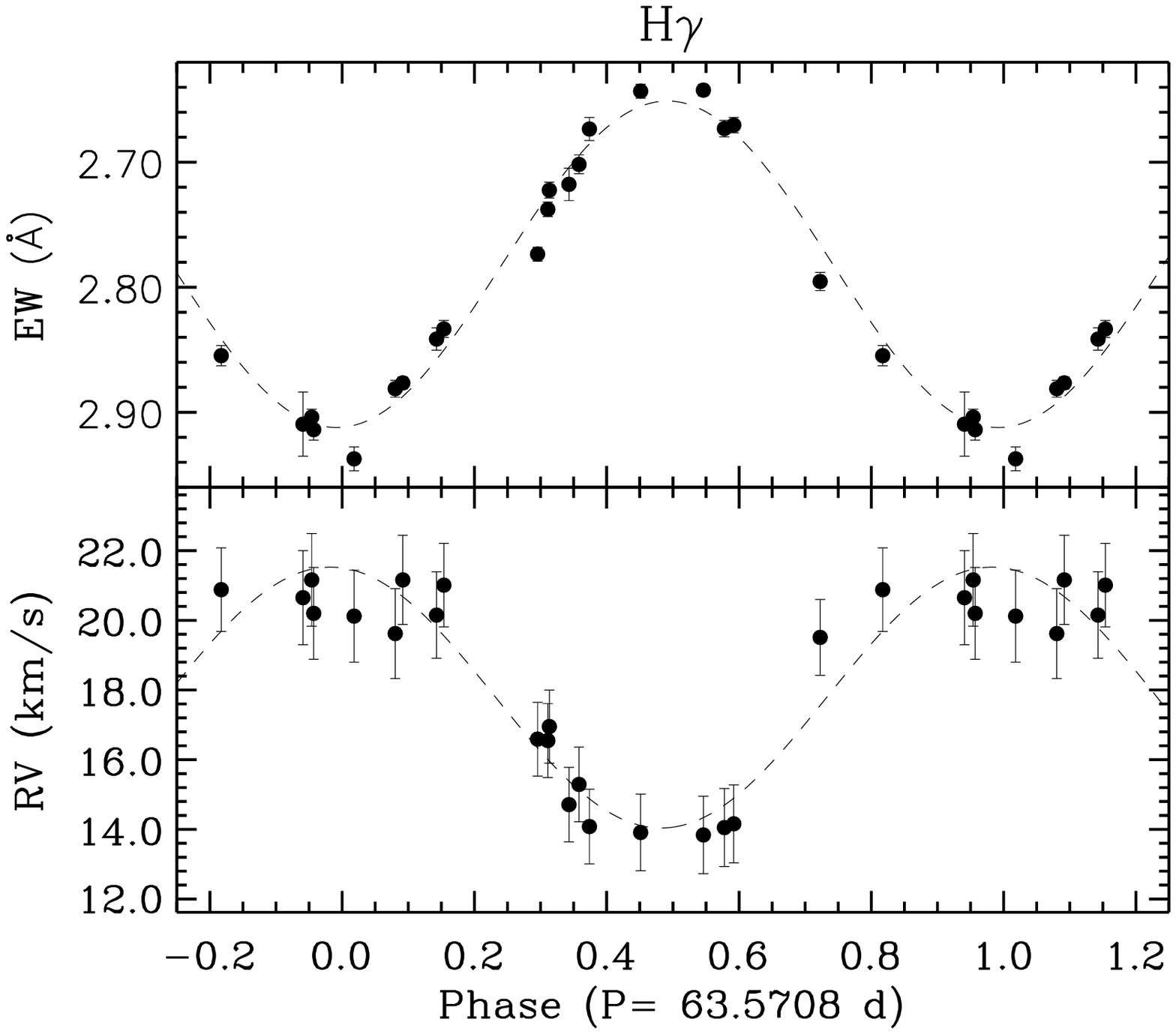}
\includegraphics[width=2.3in]{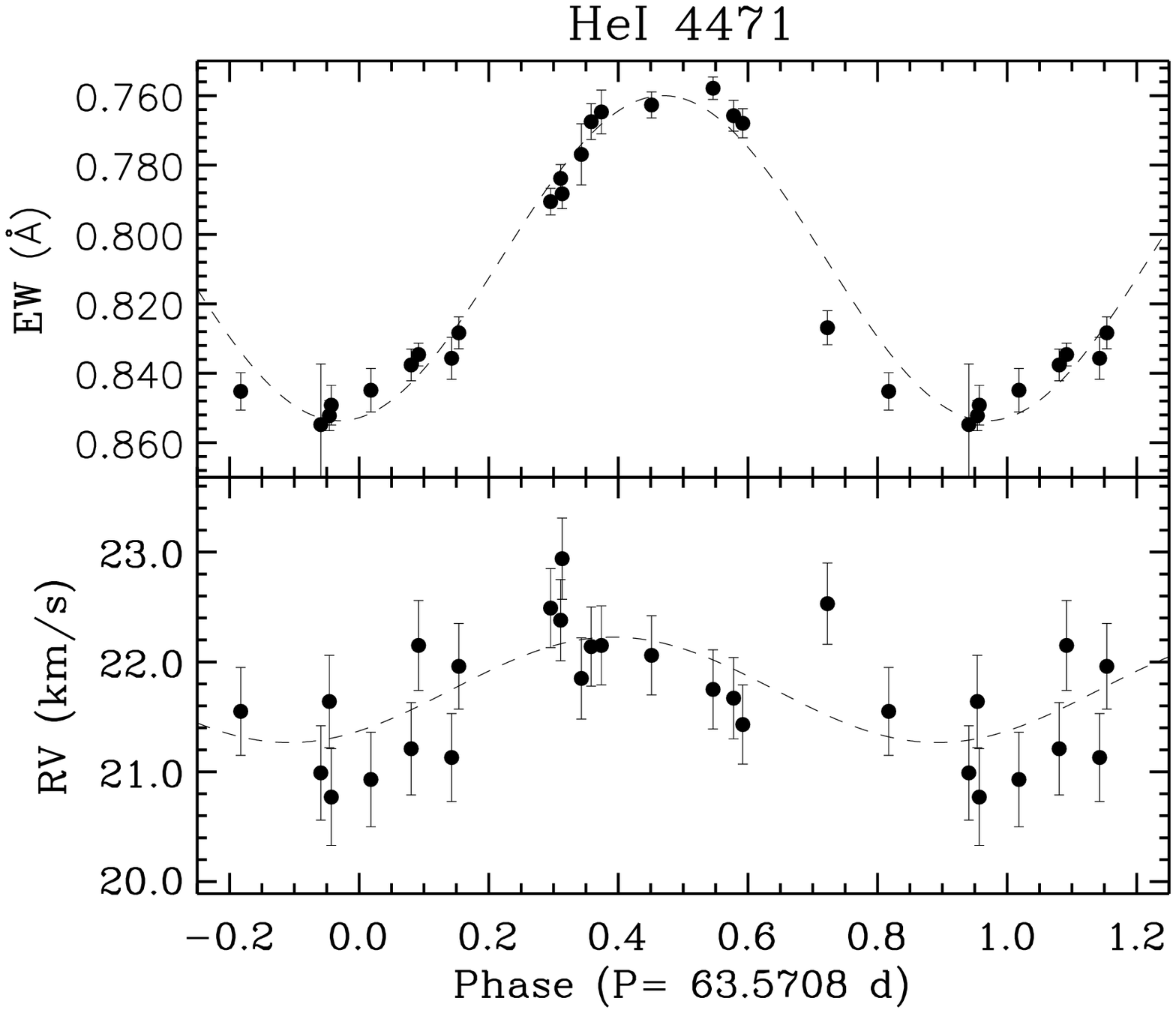}
\includegraphics[width=2.3in]{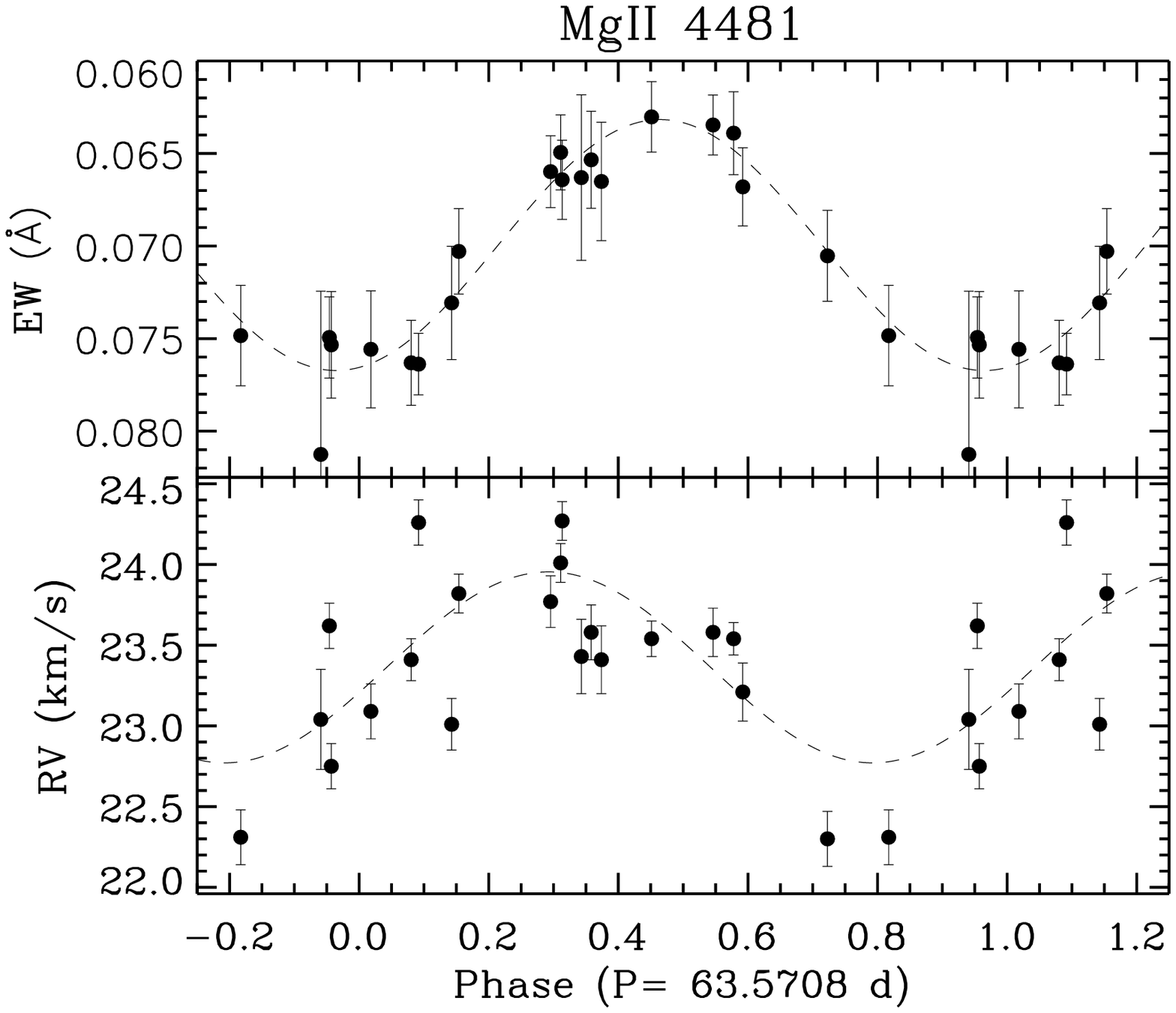}\\
\includegraphics[width=2.3in]{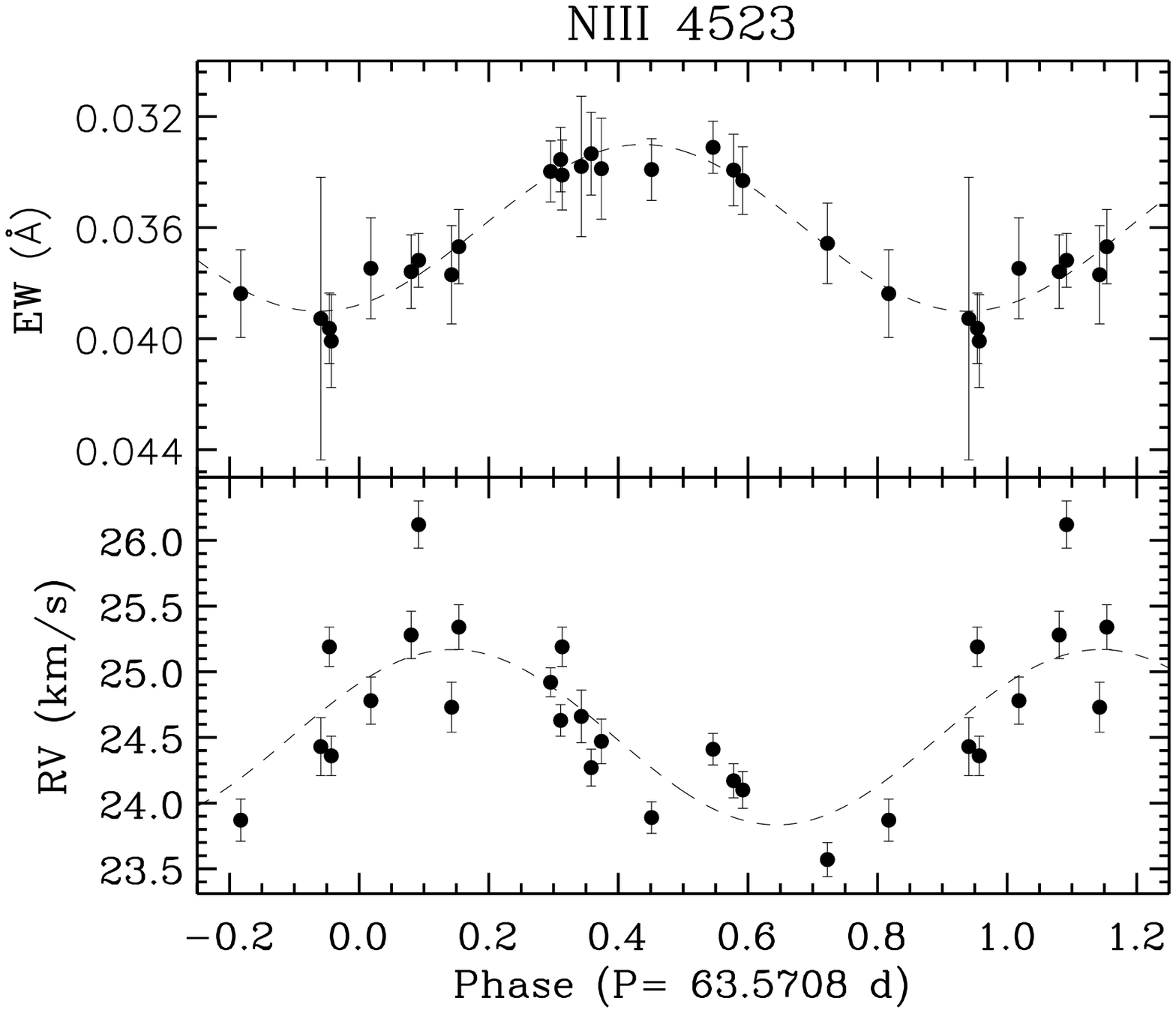}
\includegraphics[width=2.3in]{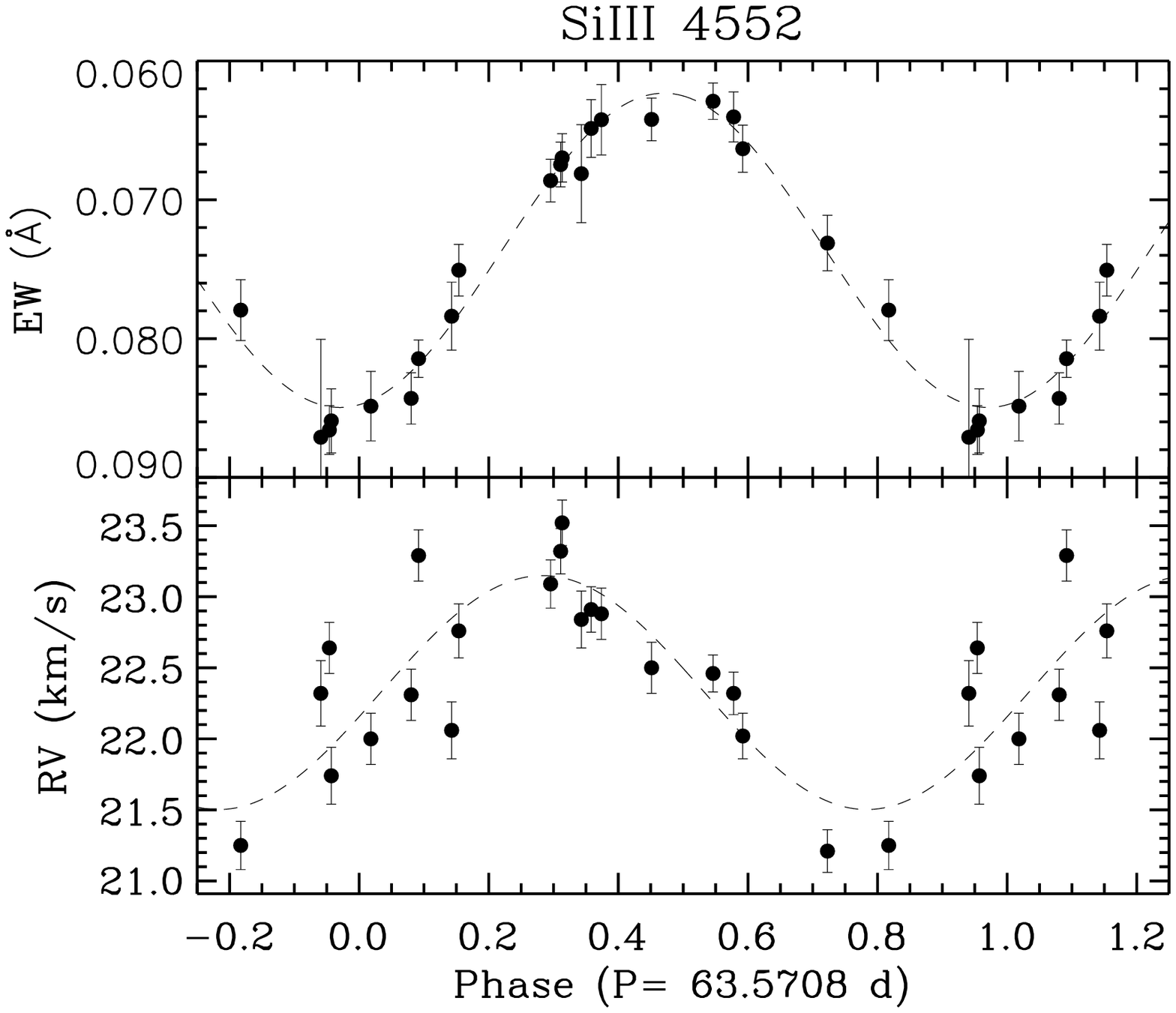}
\includegraphics[width=2.3in]{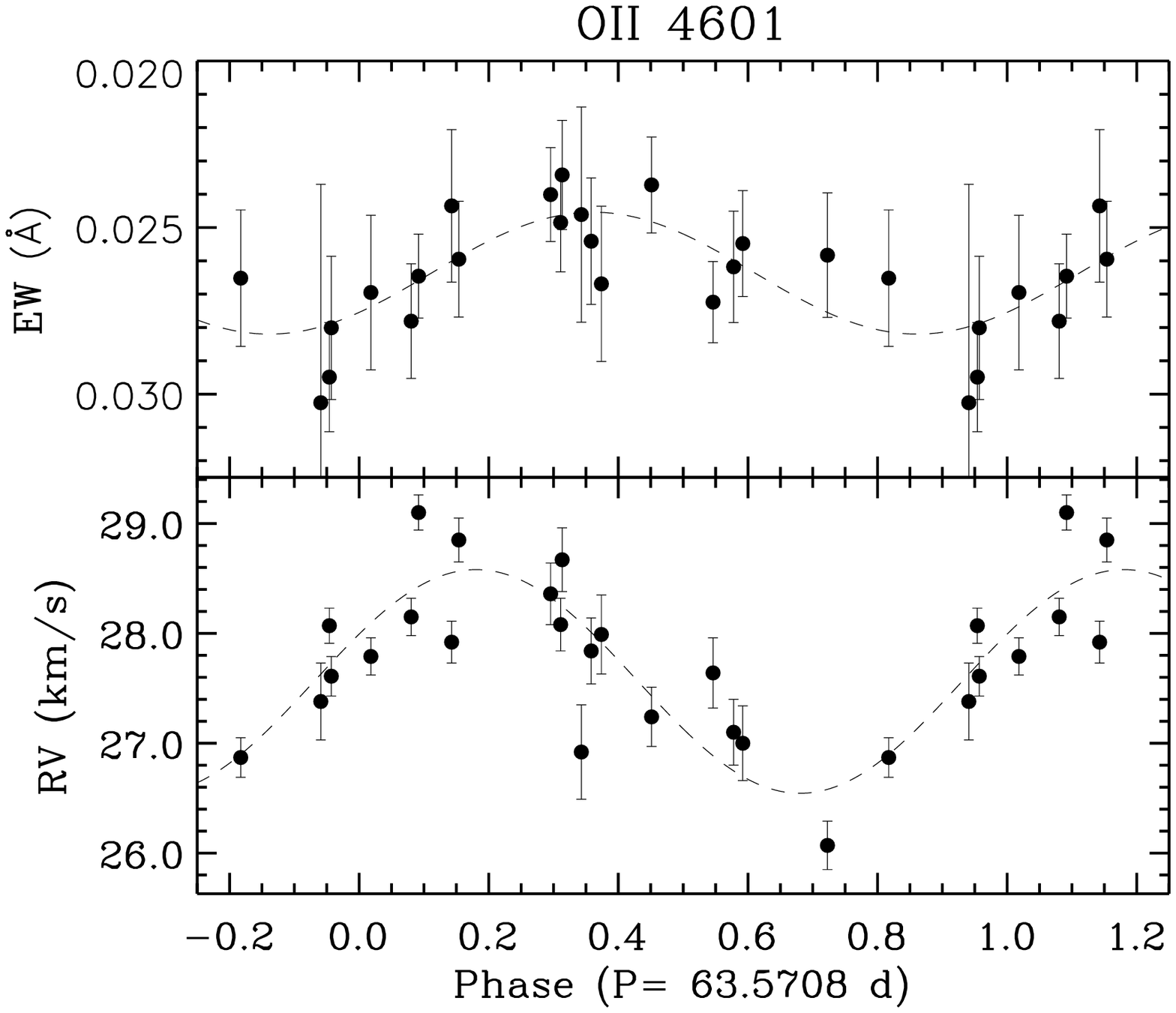}
\caption{Rotationally phased EW measurements (top panels) and radial velocity measurements (bottom panels) for selected spectral absorption lines that appear to have sinusoidally varying EW measurements. Note that while the EW variations of H$\gamma$ are consistent with the other lines, the radial velocity variations are out of phase with the rest of the lines. Least-squares sinusoidal fits to the data are also shown (dashed curves).}
\label{sin_ew}
\end{figure*}

\begin{figure*}
\centering
\includegraphics[width=2.3in]{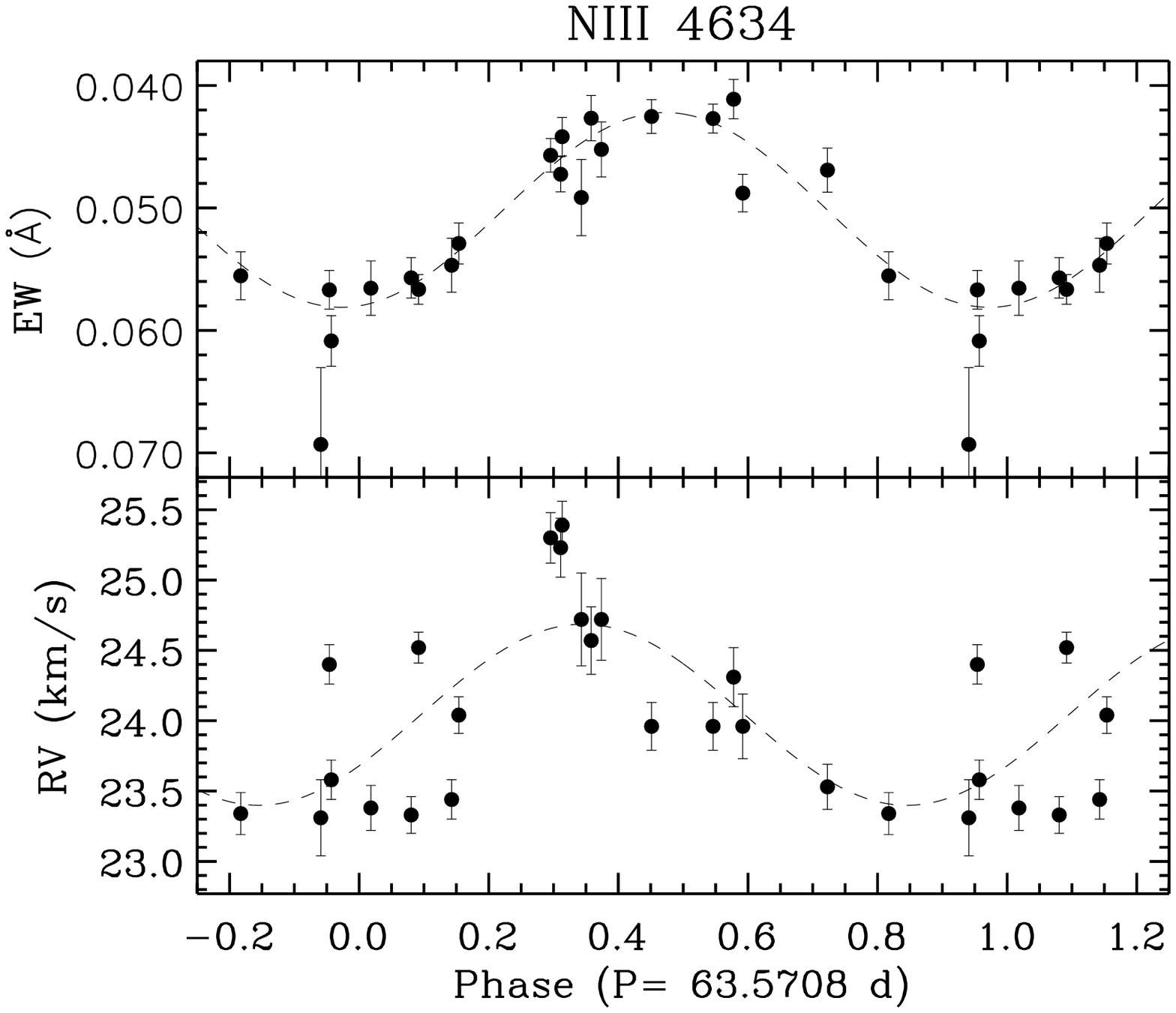}
\includegraphics[width=2.3in]{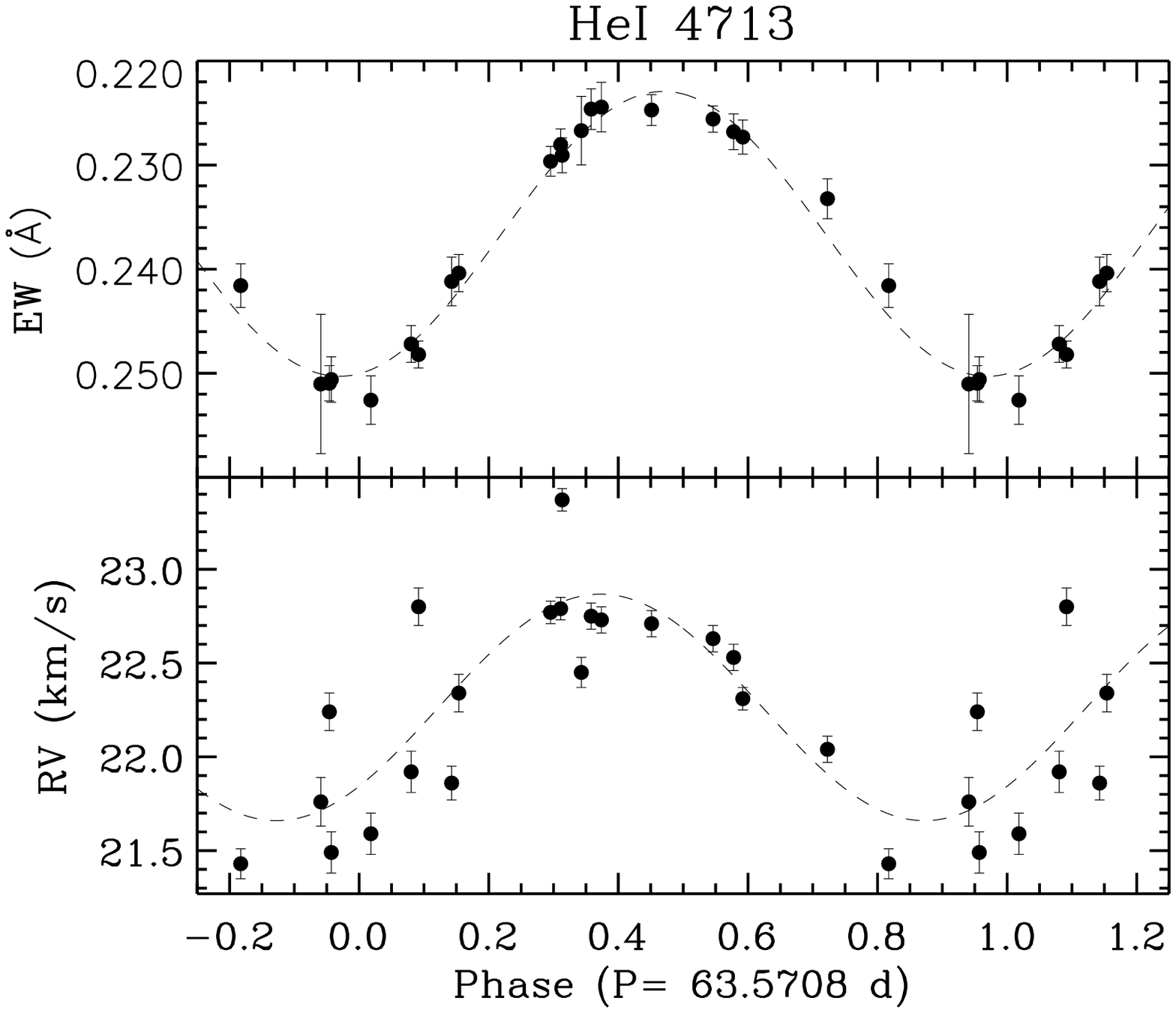}
\includegraphics[width=2.3in]{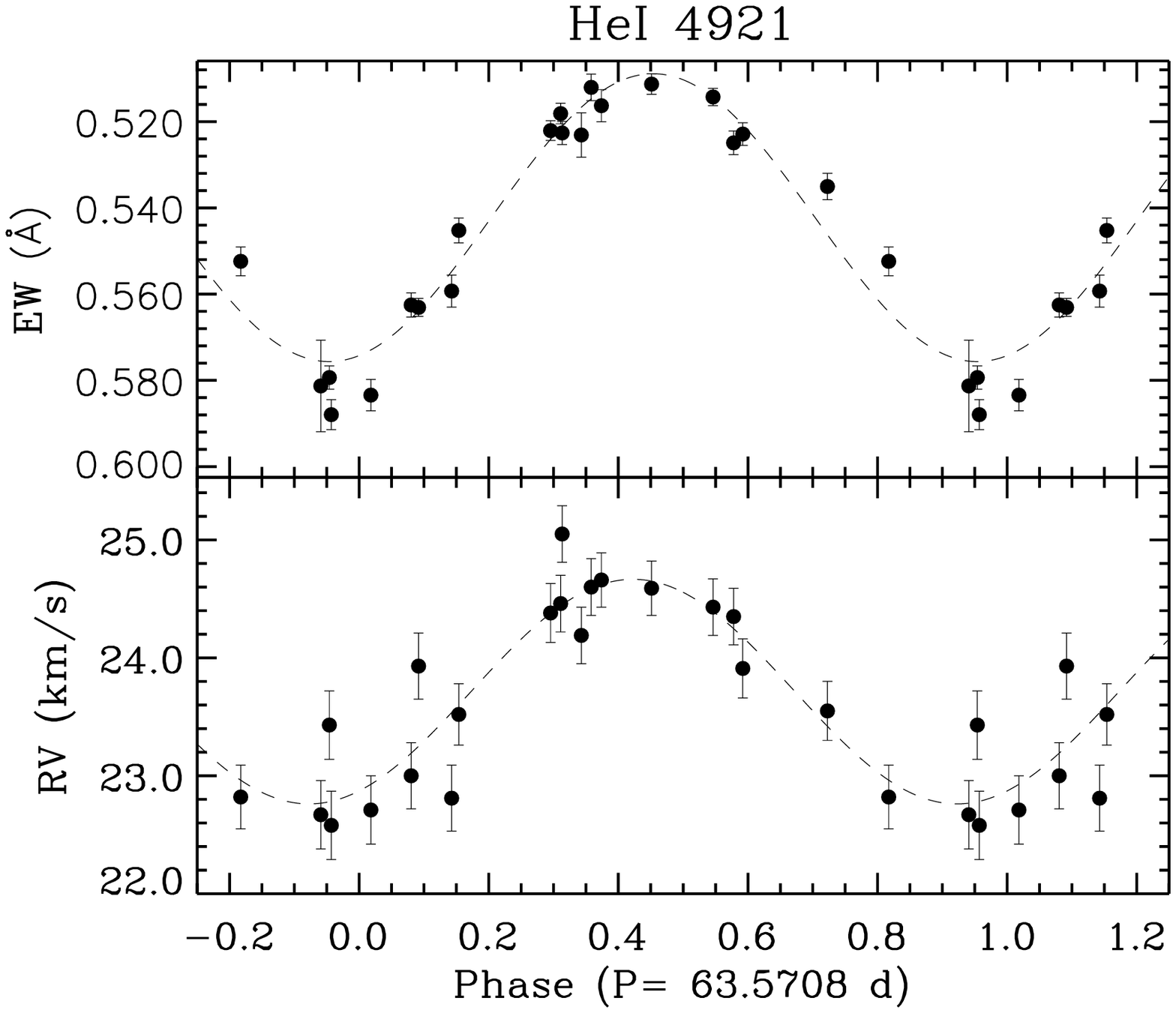}\\
\includegraphics[width=2.3in]{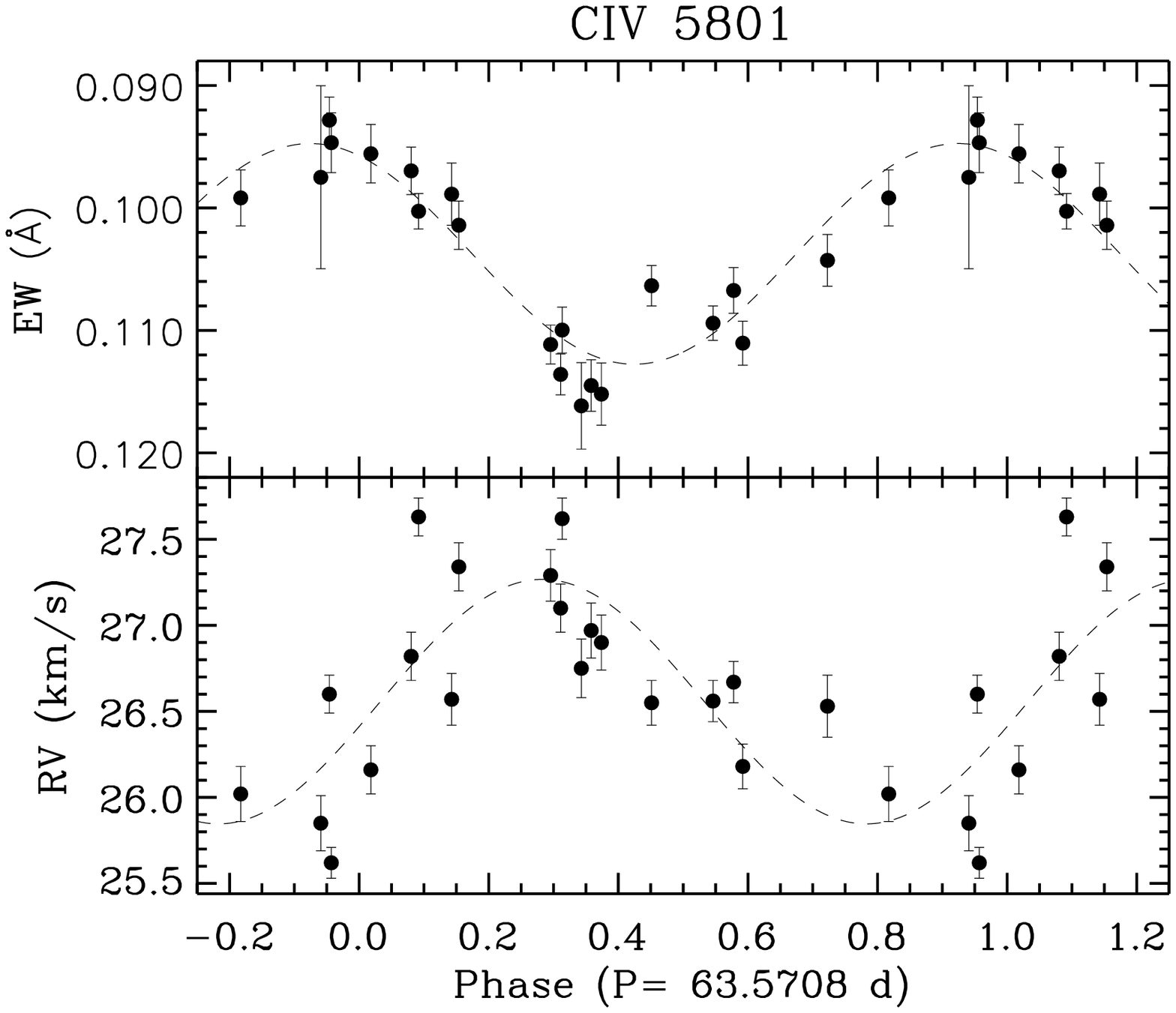}
\includegraphics[width=2.3in]{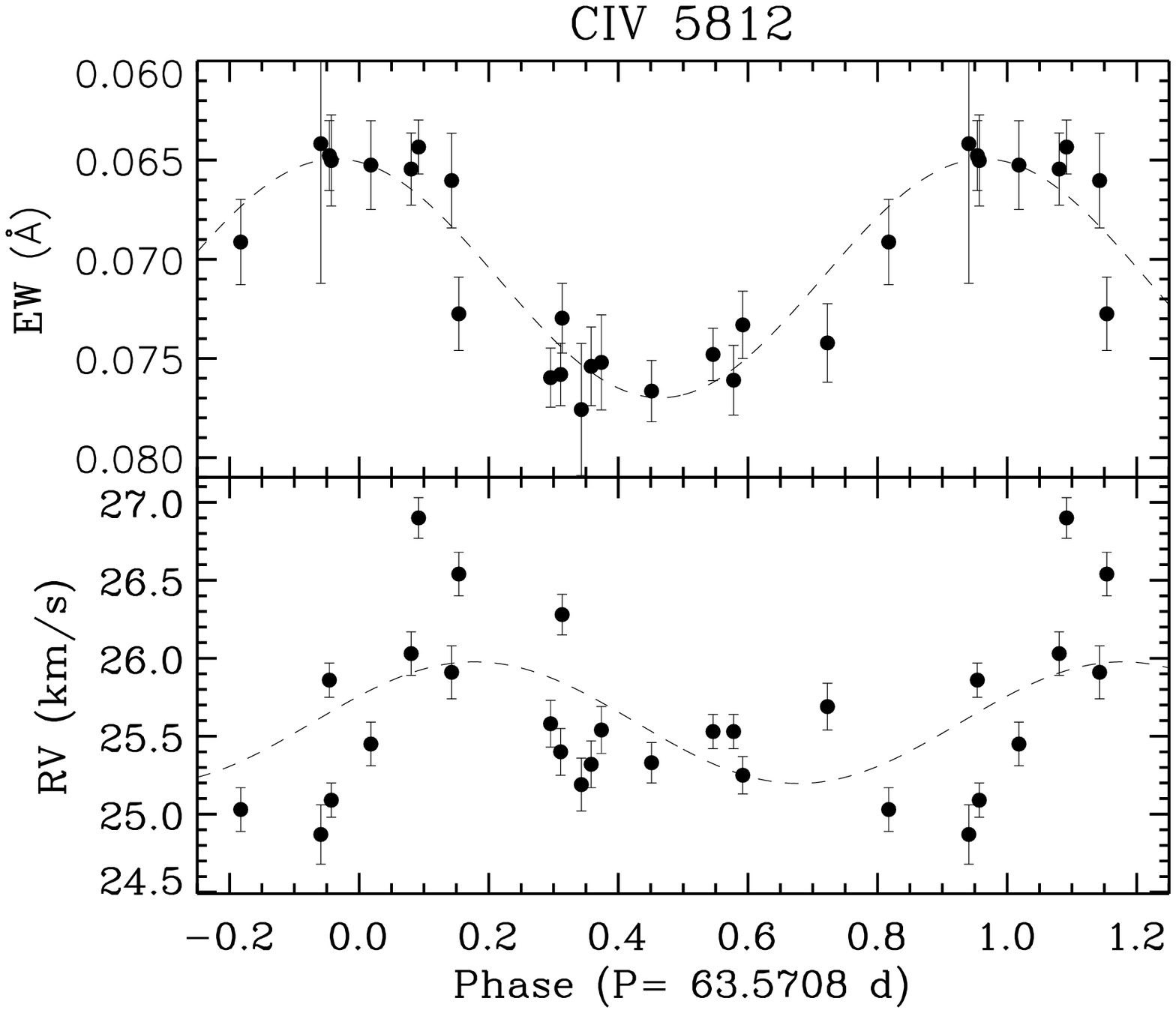}
\includegraphics[width=2.3in]{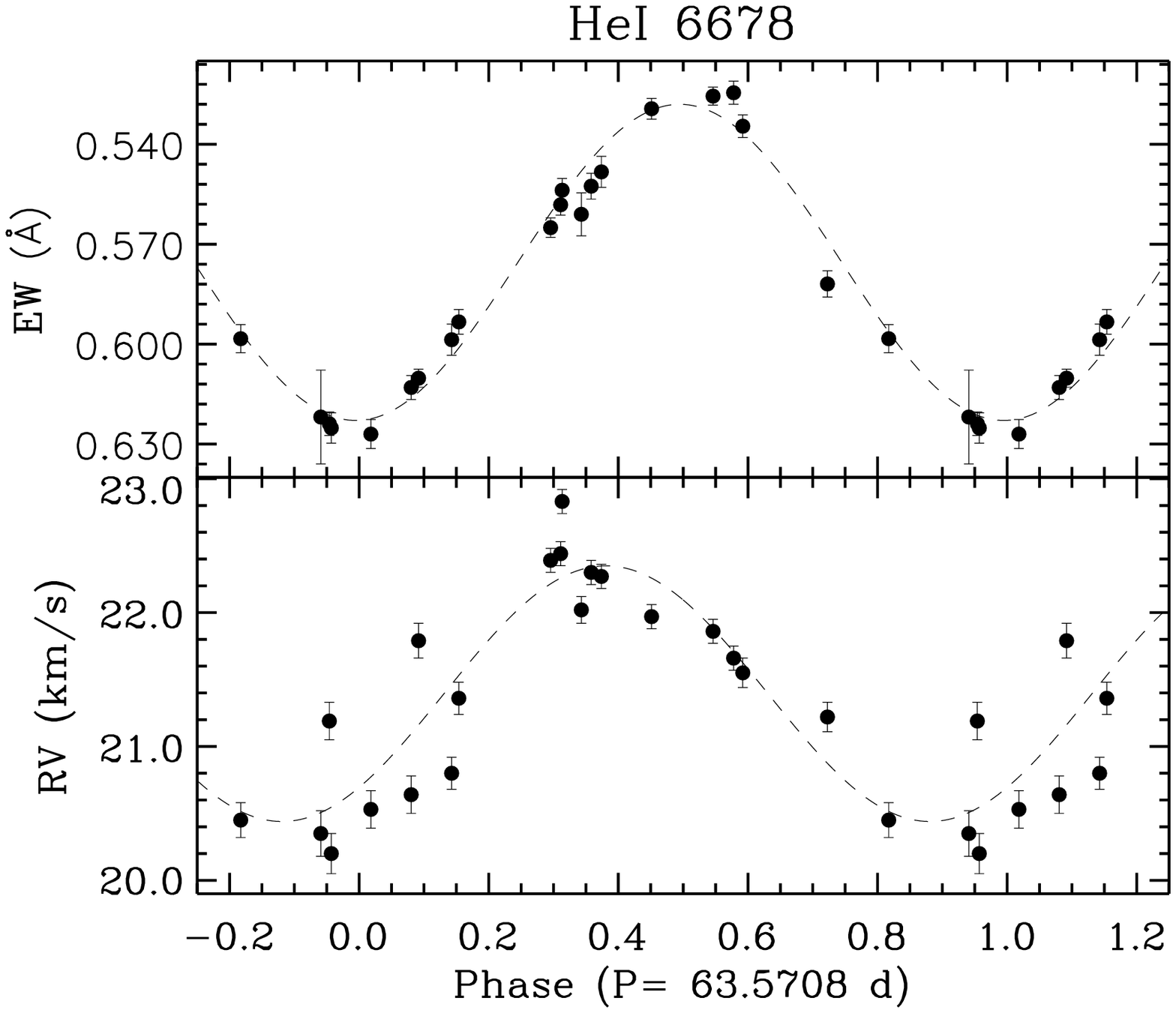}
\contcaption{}
\end{figure*}

\subsection{Line profile variations}
In Figs.~\ref{sin_dyn_fig} and \ref{nonsin_dyn_fig} we plot the phased line profile variations, or dynamic spectra. In these figures, we have subtracted the profile obtained on 24 December 2010 to highlight the variability. This spectrum was obtained at phase 0.72 and corresponds to the observation where the magnetic equator is passing our line-of-sight and, as previously discussed, when the flattened distribution of magnetospheric plasma is viewed nearly edge-on, and therefore provides the lowest contribution of emission. Our dynamic spectra indicate that a single pseudo-emission feature (a feature that appears in emission relative to the 24 December 2010 profile) occurs once per cycle, and is centred at phase 0.5 in the dynamic spectra of the lines with sinusoidally varying EW measurements (Fig.~\ref{sin_dyn_fig}). In almost all these lines, we find that the pseudo-emission is mostly confined to the inner core region and extends further out in the blue wing than the red wing. During phases of maximum emission we also find increased pseudo-absorption in the red wing, except for He\,{\sc i} $\lambda$4471, which does not appear to show additional pseudo-absorption, but does show strong evidence of pseudo-emission during phases of maximum absorption, which is centred at phase 0.0. Other lines show varying amounts of pseudo-emission during this phase as well, but this feature is most prominent in the He lines shown in Fig.~\ref{sin_dyn_fig}.

The dynamic spectra for the C\,{\sc iv} lines appear different. Maximum emission still occurs once per cycle, centred at phase 0.5, but the emission feature is mainly constrained to the inner core region and does not extend into the blue wing as it does for other lines. During phases of maximum emission, there appears to be an increase in pseudo-absorption in both wings, but the absorption in the red wing appears stronger. A similar dynamic spectrum is also found for the He\,{\sc ii} $\lambda4686$ line, as shown in the left panel of Fig.~\ref{nonsin_dyn_fig}.  In almost all the lines shown in Figs.~\ref{sin_dyn_fig} and \ref{nonsin_dyn_fig} we find that the features all appear at low velocities with the centres of the features increasing in velocity with increasing phase. However, this does not appear to be the case with the He lines, which appear more symmetric, except for the extended pseudo-absorption into the blue wing, which is likely caused by the forbidden transitions in these He lines. 

In Fig.~\ref{halpha_dyn_fig} we show the dynamic spectra of the other lines with non-sinusoidally varying EW variations. A NLTE {\sc tlusty} \citep{lanz03} synthetic profile was subtracted from the Balmer lines in this figure to highlight the circumstellar emission contribution. As was found in the EW variations, we see two emission features per rotation cycle, one centred around phase 0.0 and the other centred around phase 0.5. In the He\,{\sc i} $\lambda5876$, H$\gamma$ and H$\beta$ line we find the lines to be mainly in absorption and the emission feature at phase 0.0 to be narrower and weaker in emission than the feature at phase 0.5. In the He\,{\sc i} line shown, both emission features appear redshifted with respect to the systemic velocity of HD\,57682; the narrower feature is found to have a central velocity of $\sim$75\,km\,s$^{-1}$ while the broader feature has a central velocity of $\sim$65\,km\,s$^{-1}$. There also appears to be an additional weak emission feature that is blueshifted with respect to the systemic velocity occurring at phase 0.0. H$\beta$ and H$\gamma$ appear similar to the He\,{\sc I} $\lambda5876$ line. Both show a strong and broad emission feature at phase 0.5, and a weaker, narrower emission feature at phase 0.0. This is not the case for H$\alpha$, which shows both emission features with nearly the same width and same emission level; it is evident that the strength of the weaker emission feature at phase 0.0 is decreasing for higher Balmer lines. Unlike the He\,{\sc i} $\lambda5876$ line, the emission features of the Balmer lines have their central velocities alternating about the systemic velocity; the feature at phase 0.0 appears blueshifted, while the feature at phase 0.5 is redshifted. 

\begin{figure*}
\centering
\includegraphics[width=2.3in]{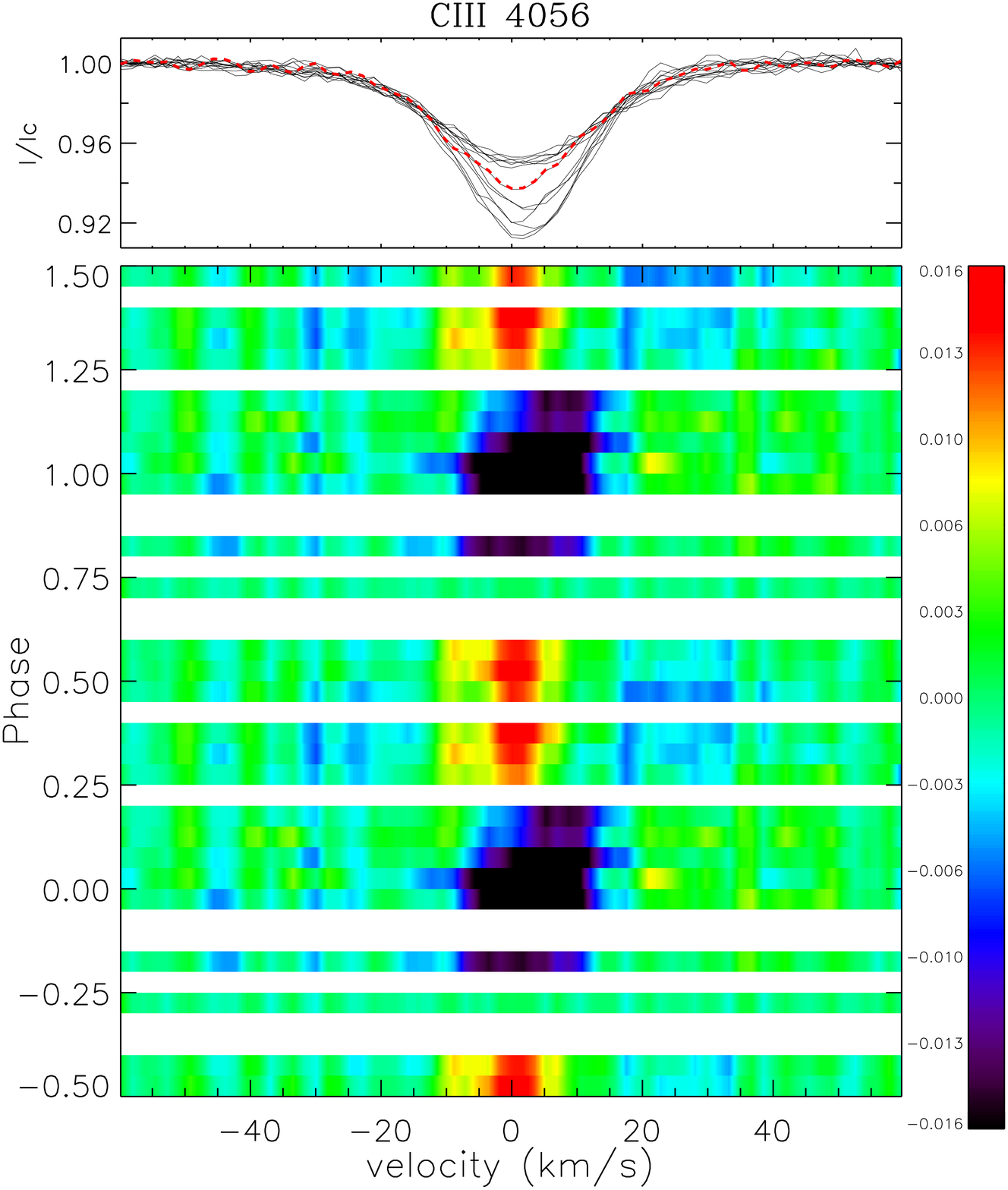}
\includegraphics[width=2.3in]{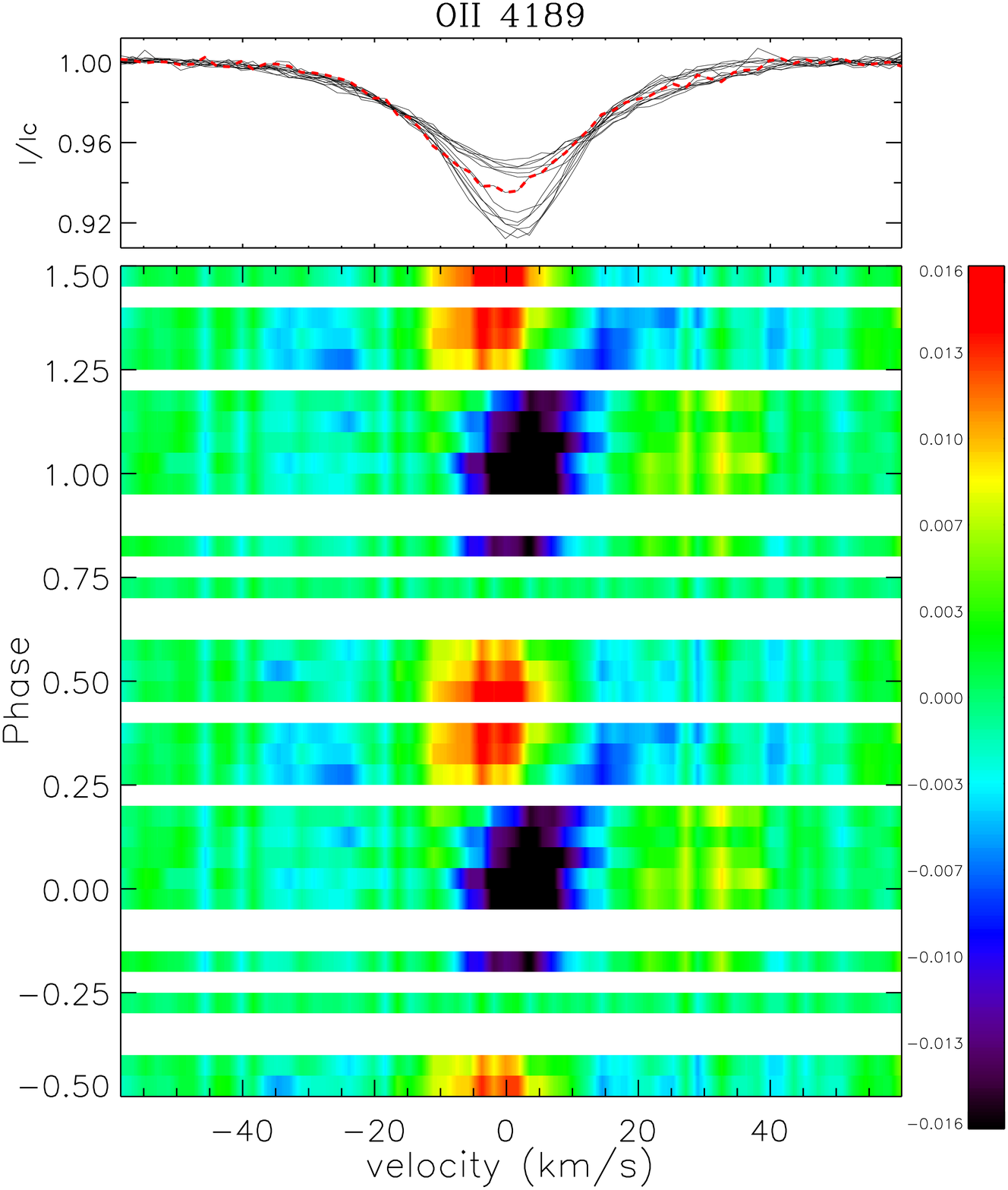}
\includegraphics[width=2.3in]{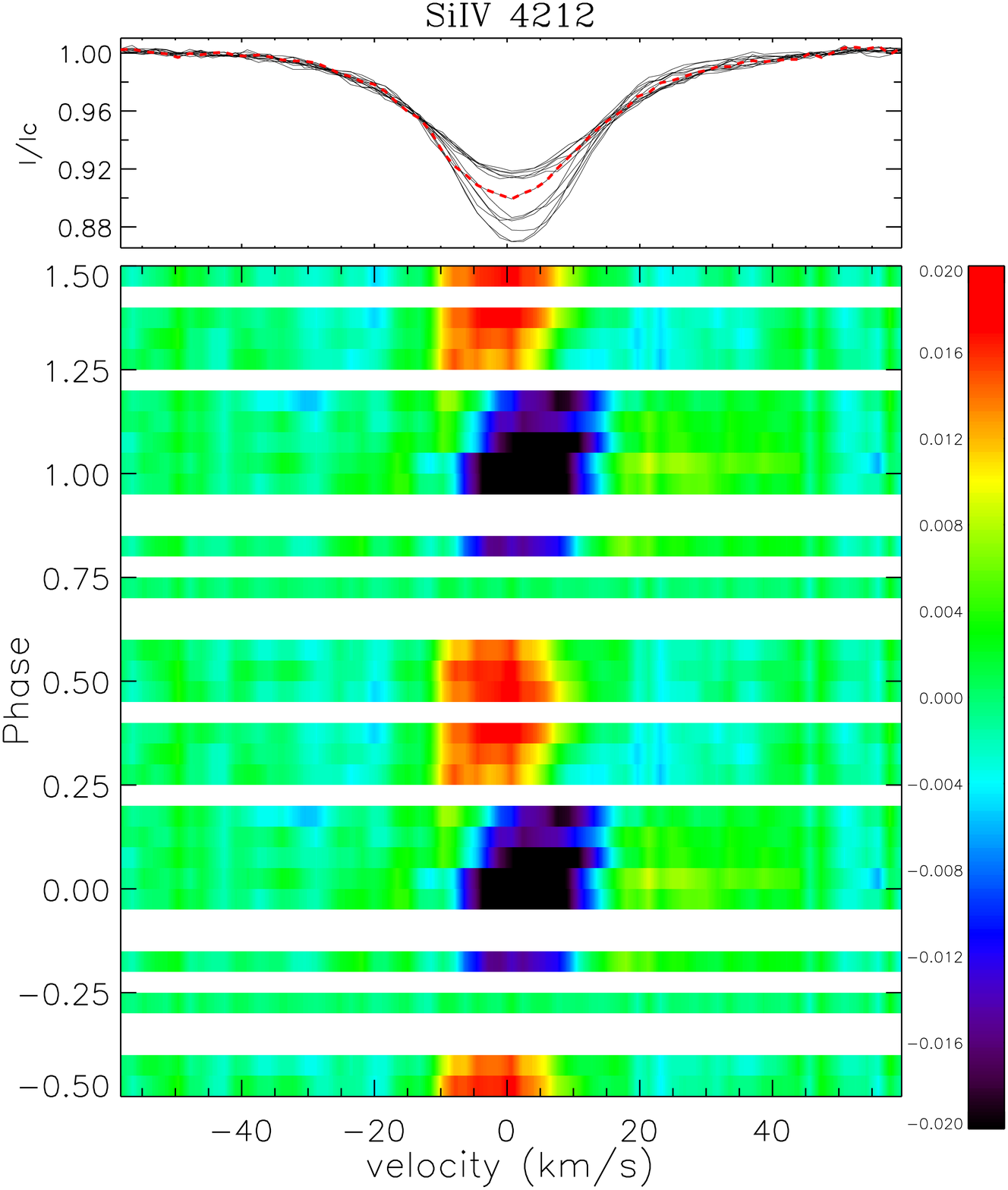}\\
\includegraphics[width=2.3in]{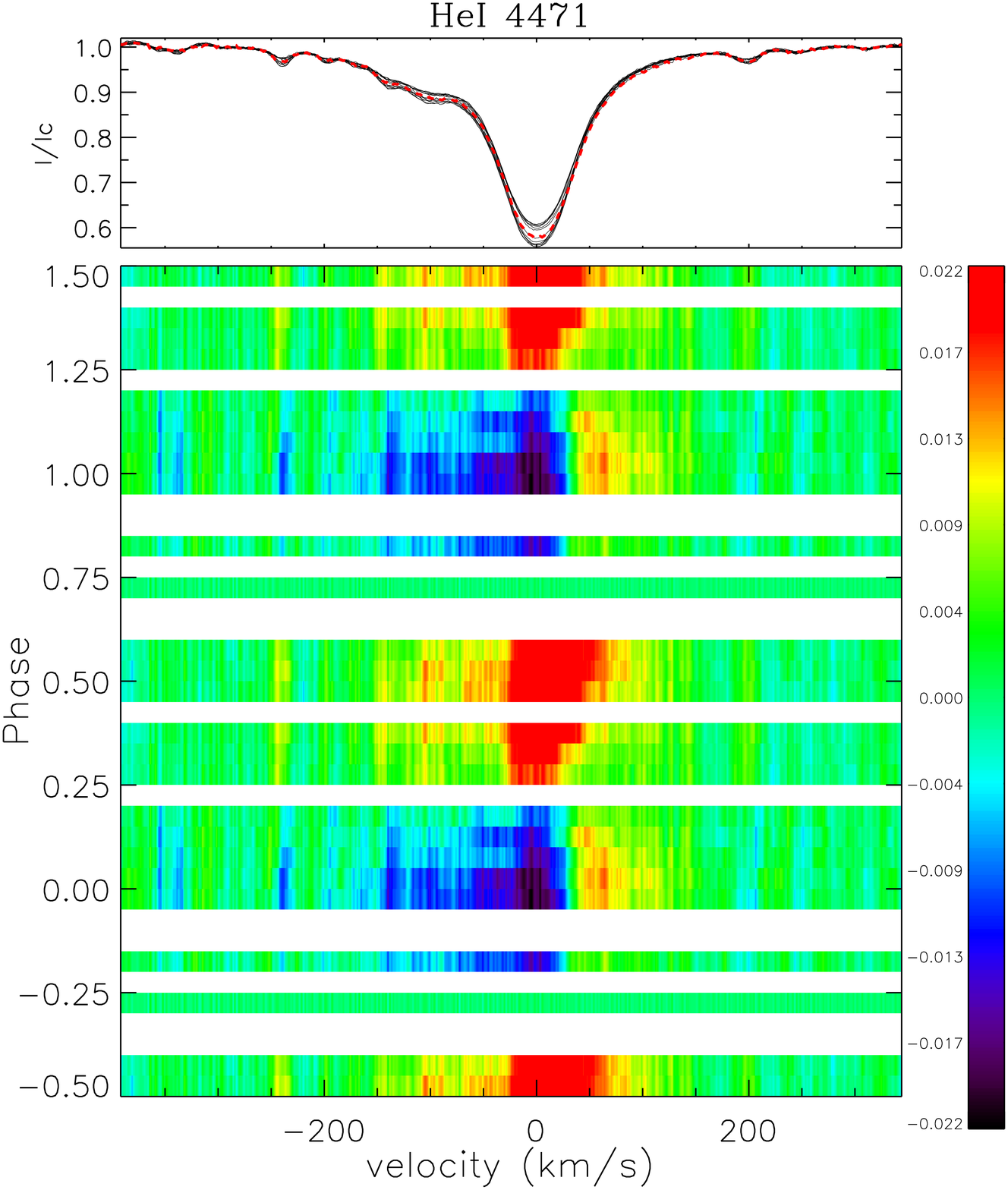}
\includegraphics[width=2.3in]{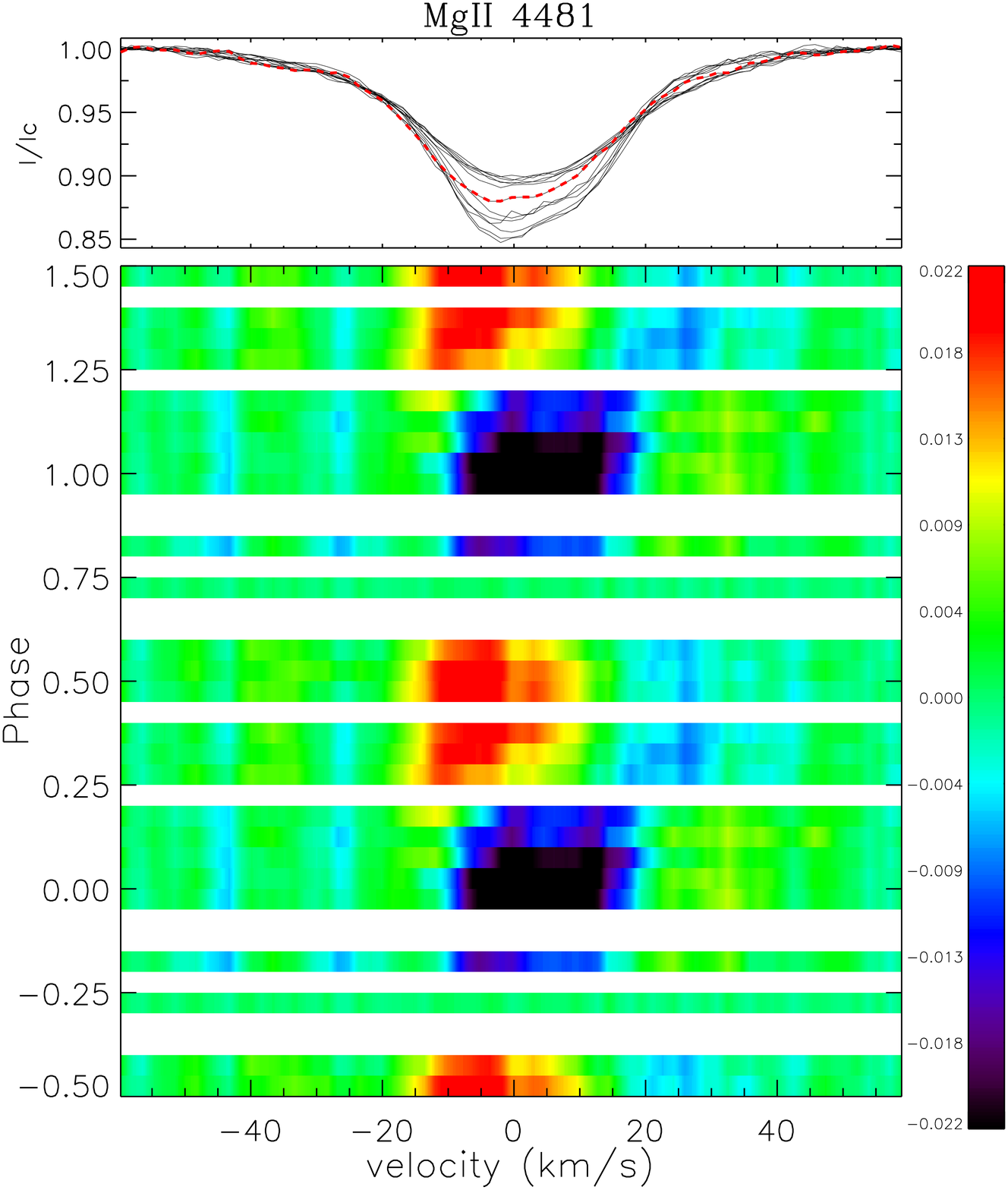}
\includegraphics[width=2.3in]{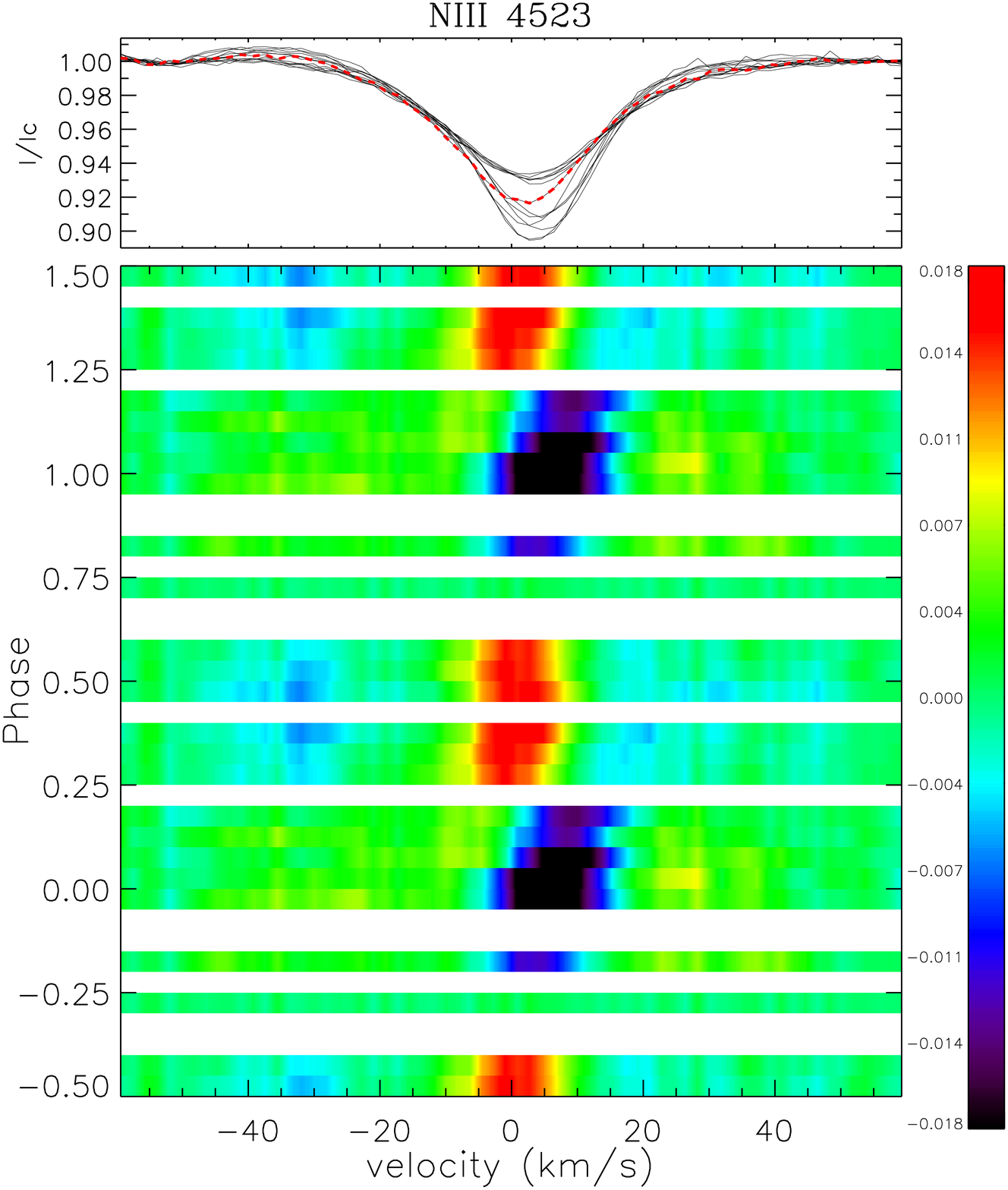}\\
\includegraphics[width=2.3in]{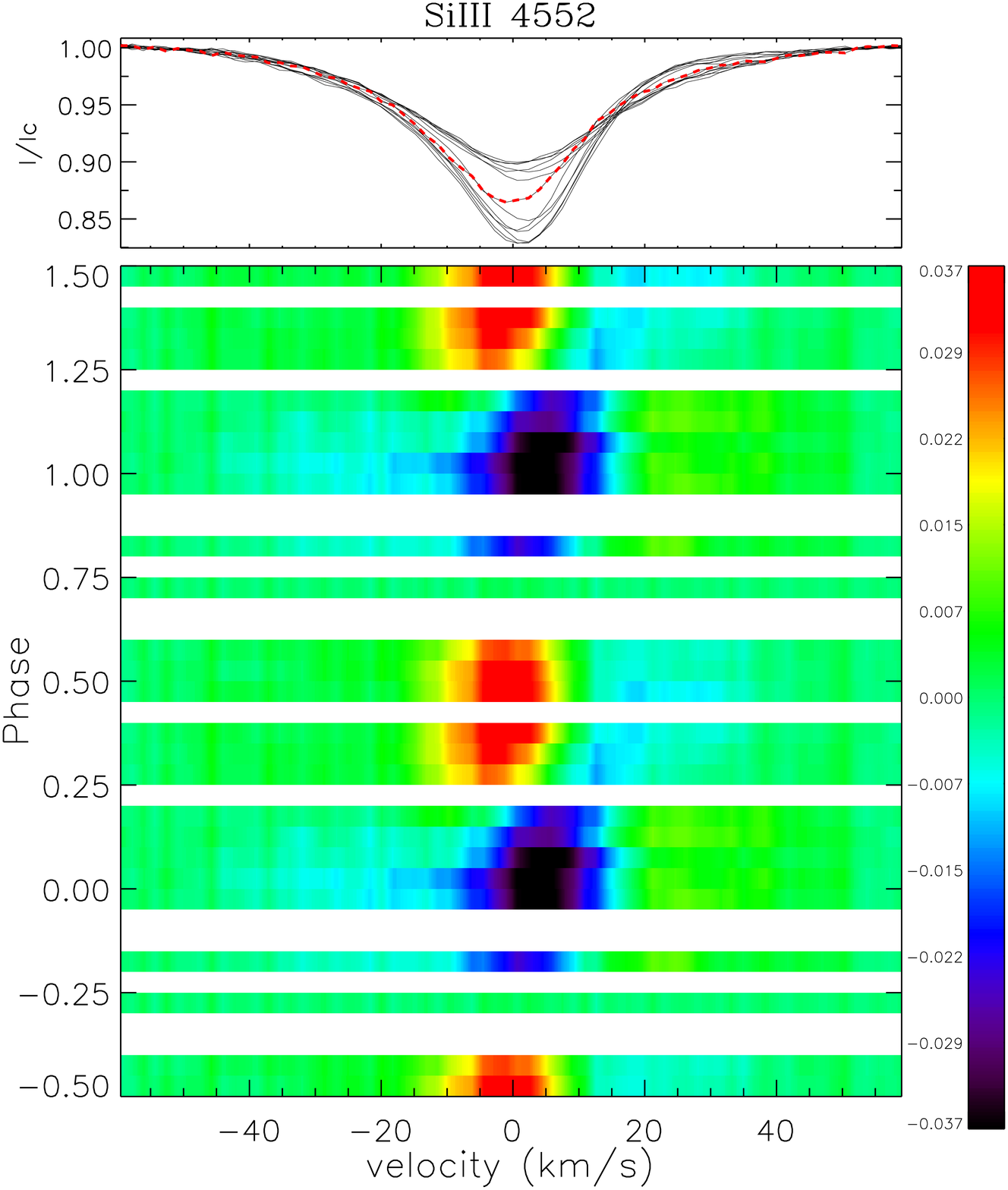}
\includegraphics[width=2.3in]{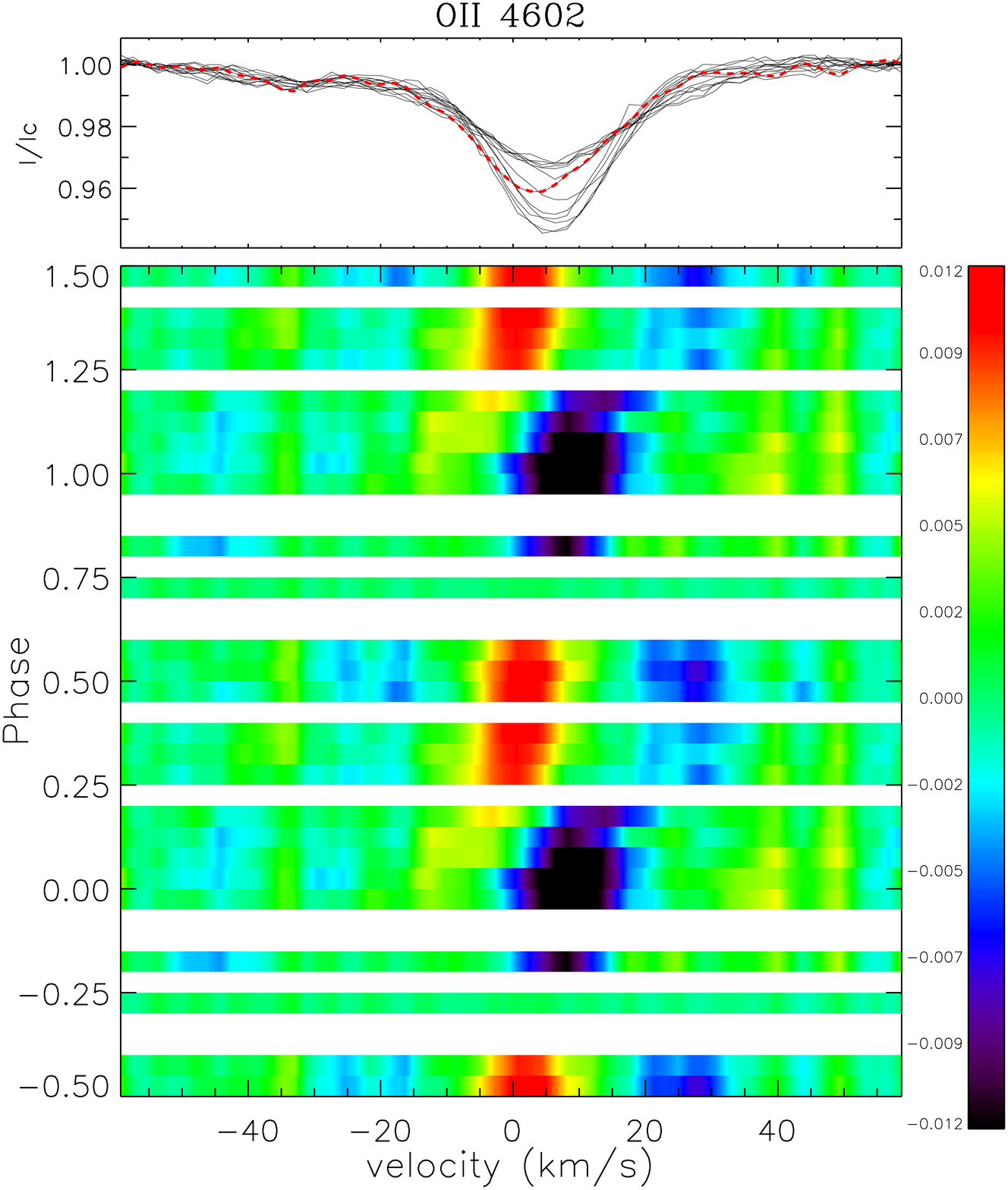}
\includegraphics[width=2.3in]{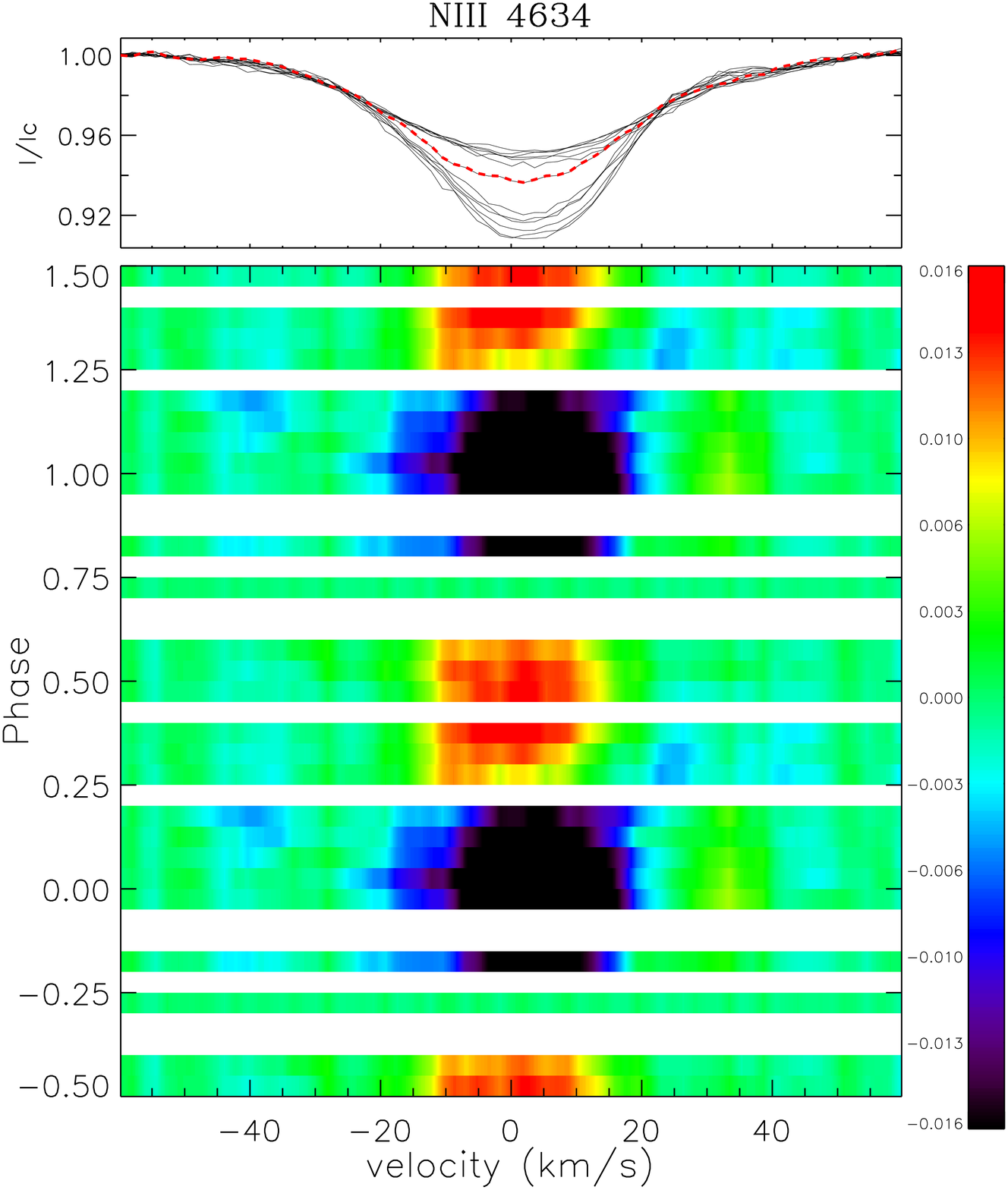}
\caption{Phased variations in selected spectral lines with sinusoidal equivalent width variations. Plotted is the difference between the observed profiles and the profile obtained on 24 December 2010 (dashed-red, top panel), which occurred at phase $\sim$0.75, corresponding to when the magnetic equator crosses the line-of-sight.}
\label{sin_dyn_fig}
\end{figure*}

\begin{figure*}
\centering
\includegraphics[width=2.3in]{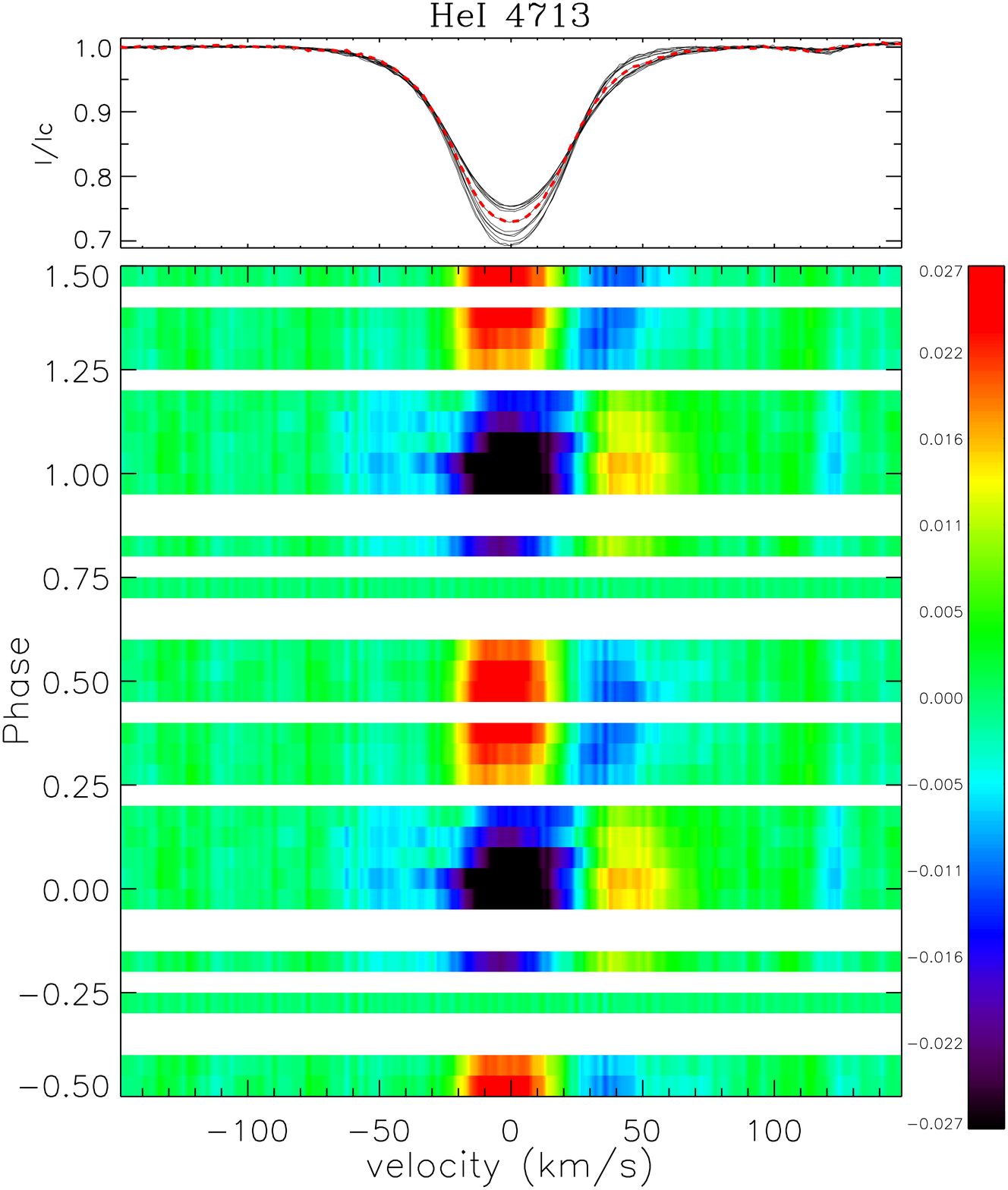}
\includegraphics[width=2.3in]{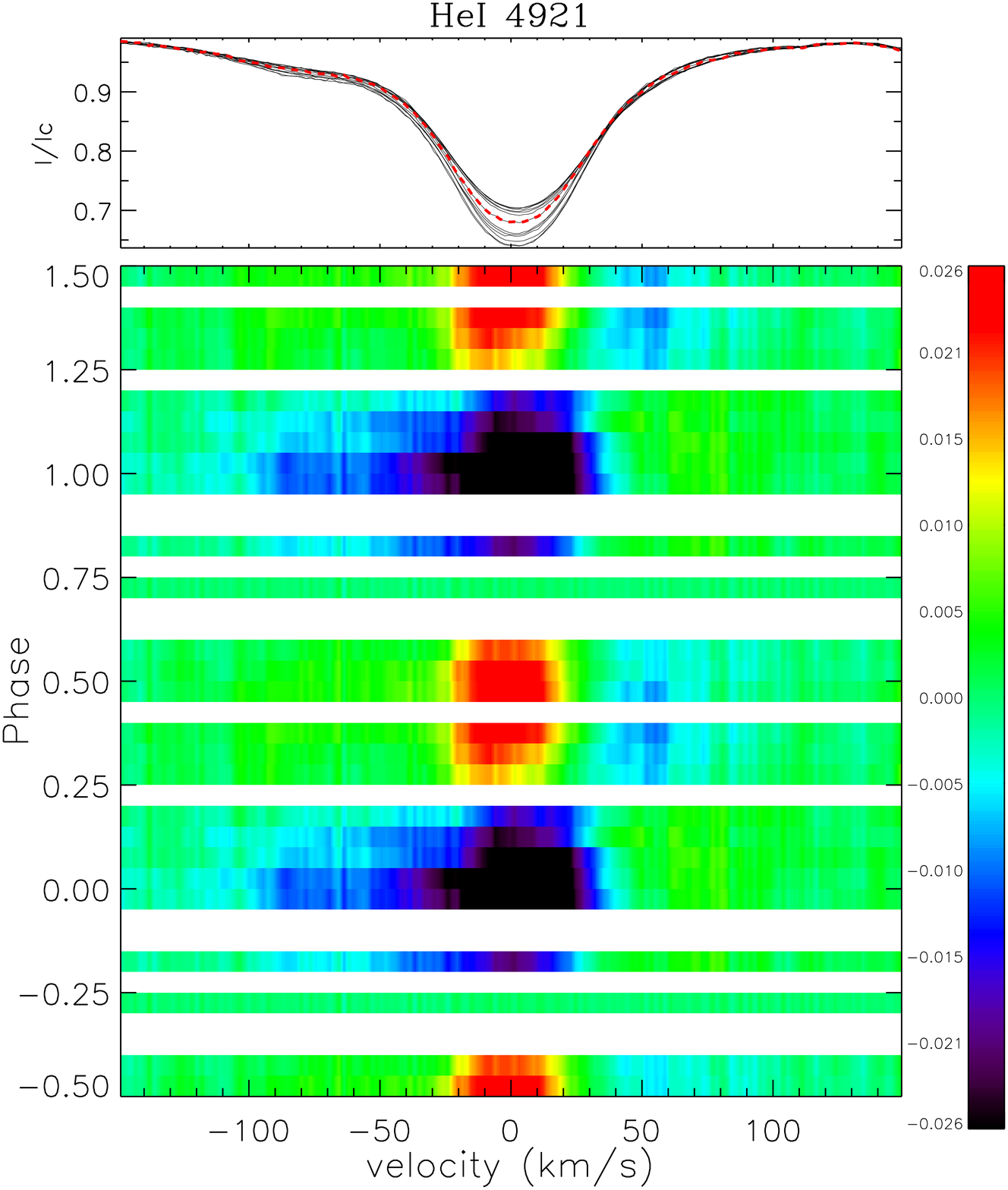}
\includegraphics[width=2.3in]{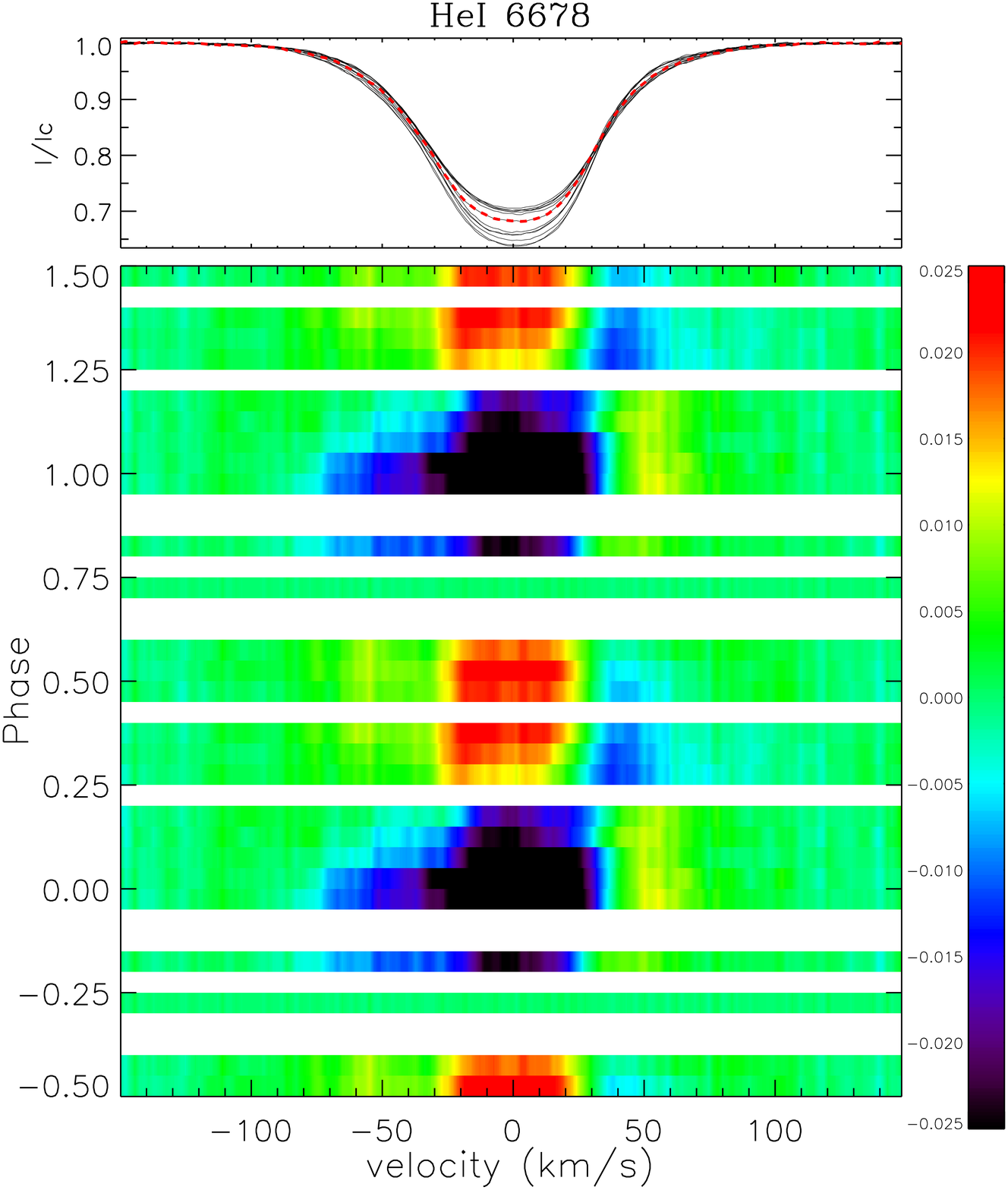}
\contcaption{}
\end{figure*}

\begin{figure*}
\centering
\includegraphics[width=2.3in]{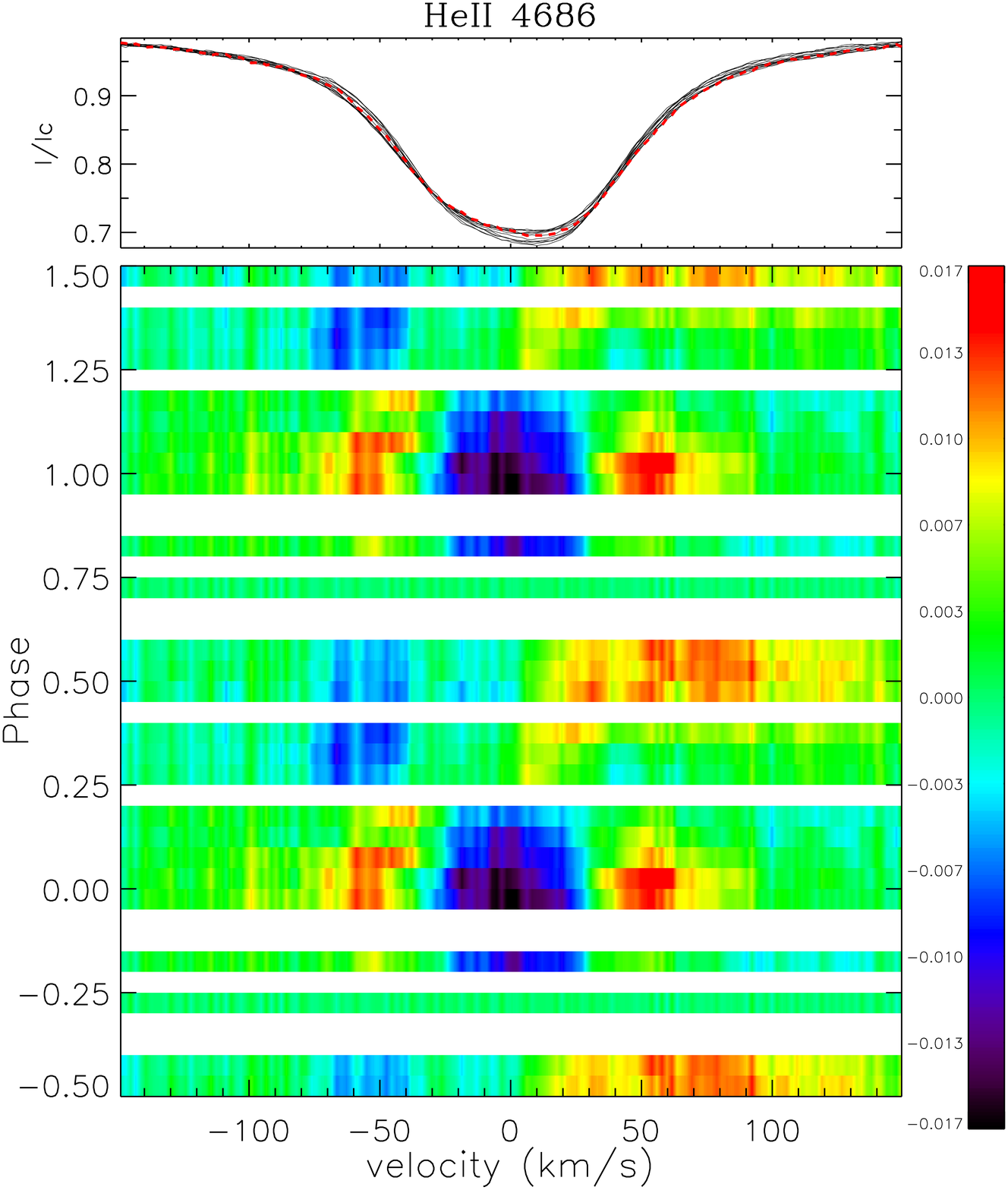}
\includegraphics[width=2.3in]{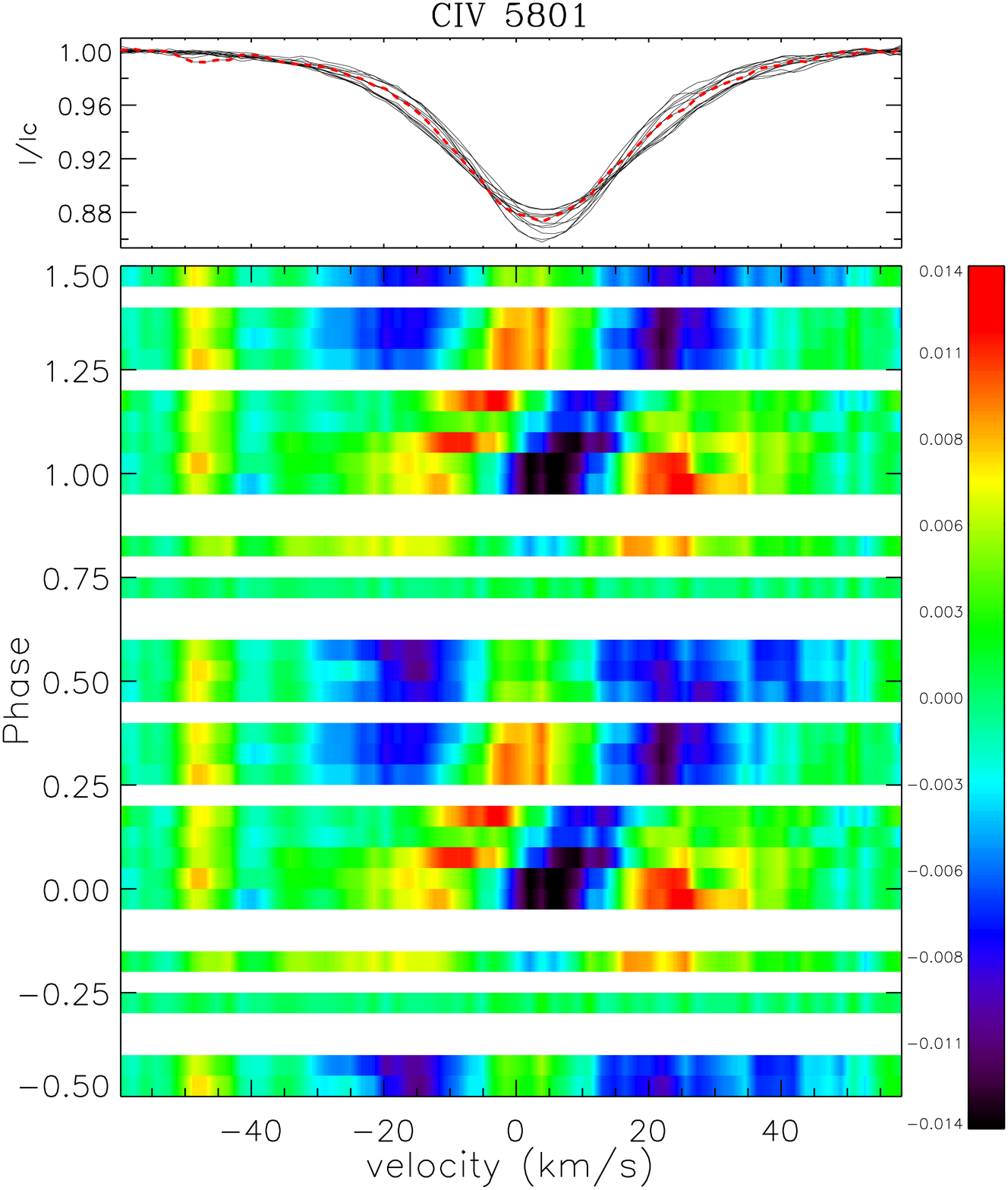}
\includegraphics[width=2.3in]{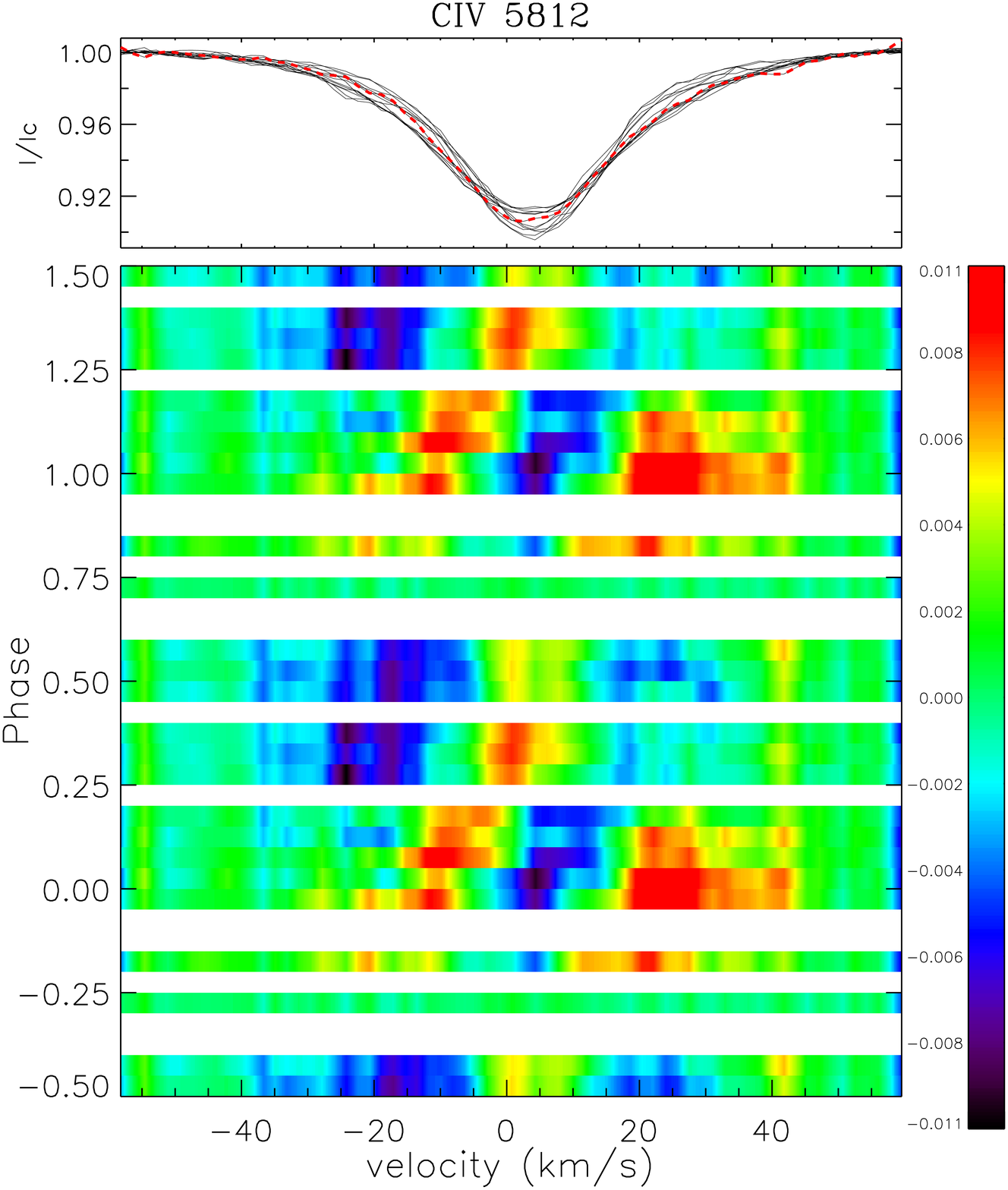}
\caption{Same as Fig.~\ref{sin_dyn_fig} but for selected spectral lines with clearly non-sinusoidal EW variations or EW variations that appear anti-phased with the other lines. Plotted is the difference between the observed profiles and the profile obtained on 24 December 2010 (dashed-red, top panel), which occurred at phase $\sim$0.75, corresponding to when the magnetic equator crosses the line-of-sight.}
\label{nonsin_dyn_fig}
\end{figure*}

\begin{figure*}
\centering
\includegraphics[width=2.3in]{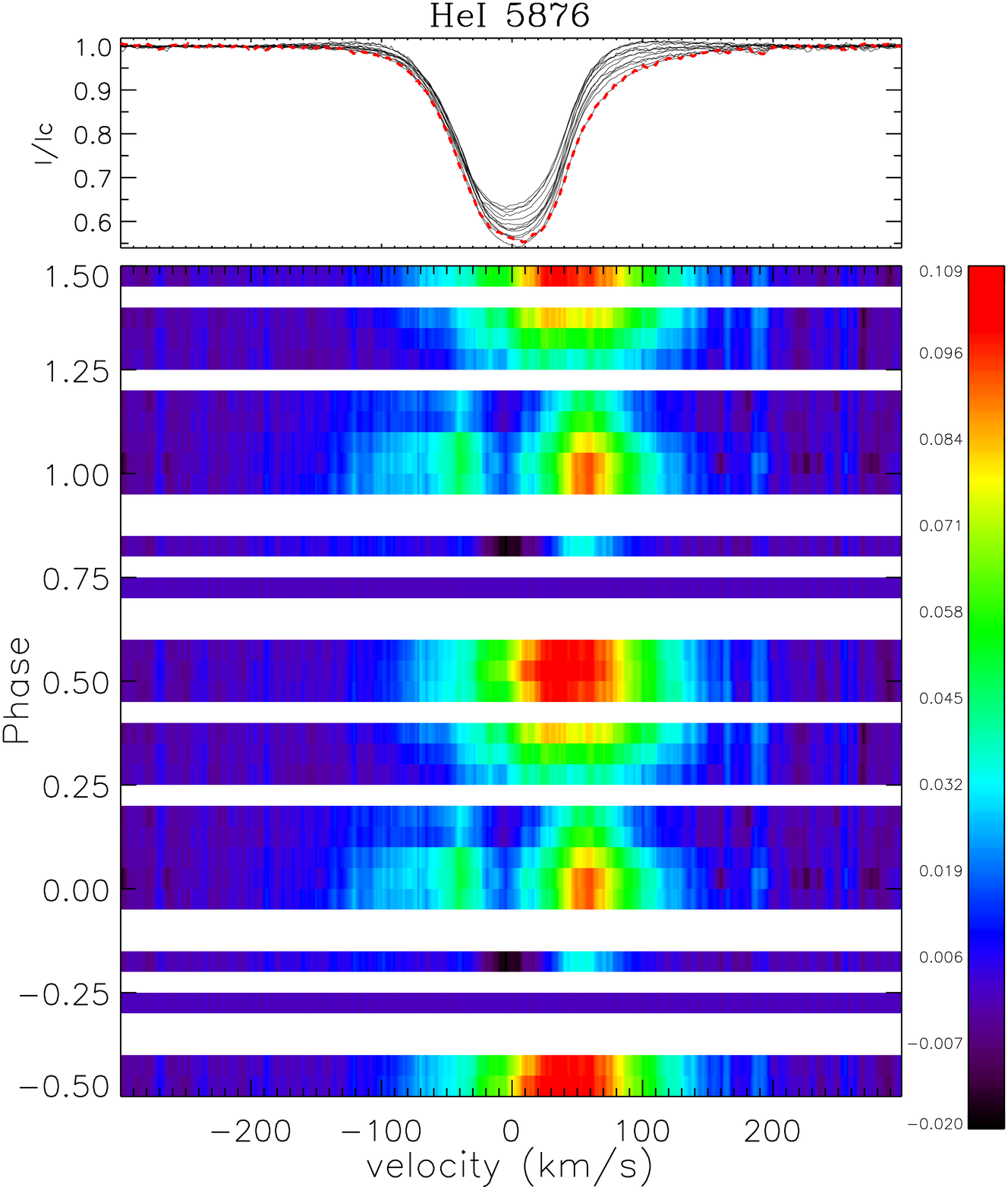}
\includegraphics[width=2.3in]{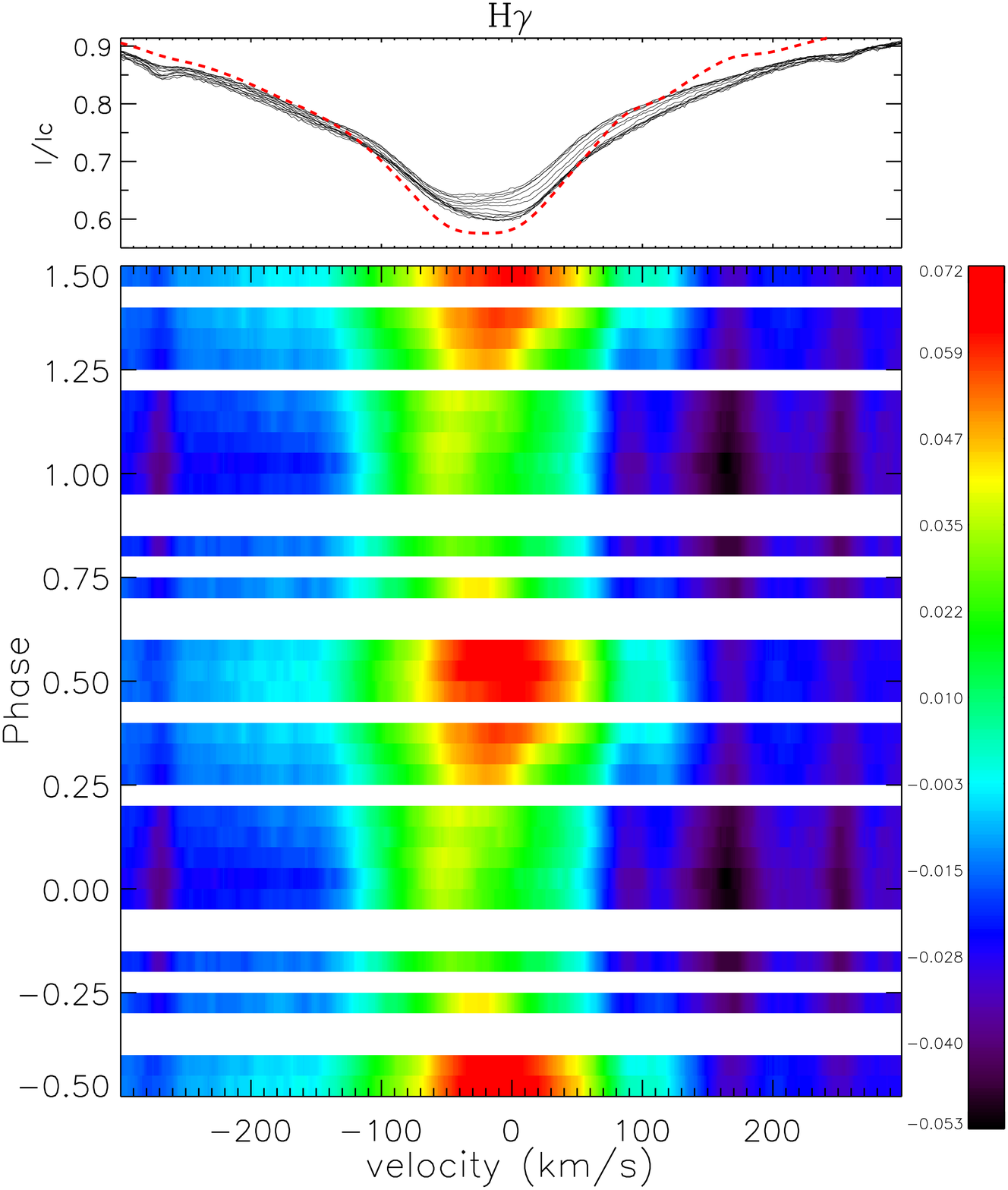}
\includegraphics[width=2.3in]{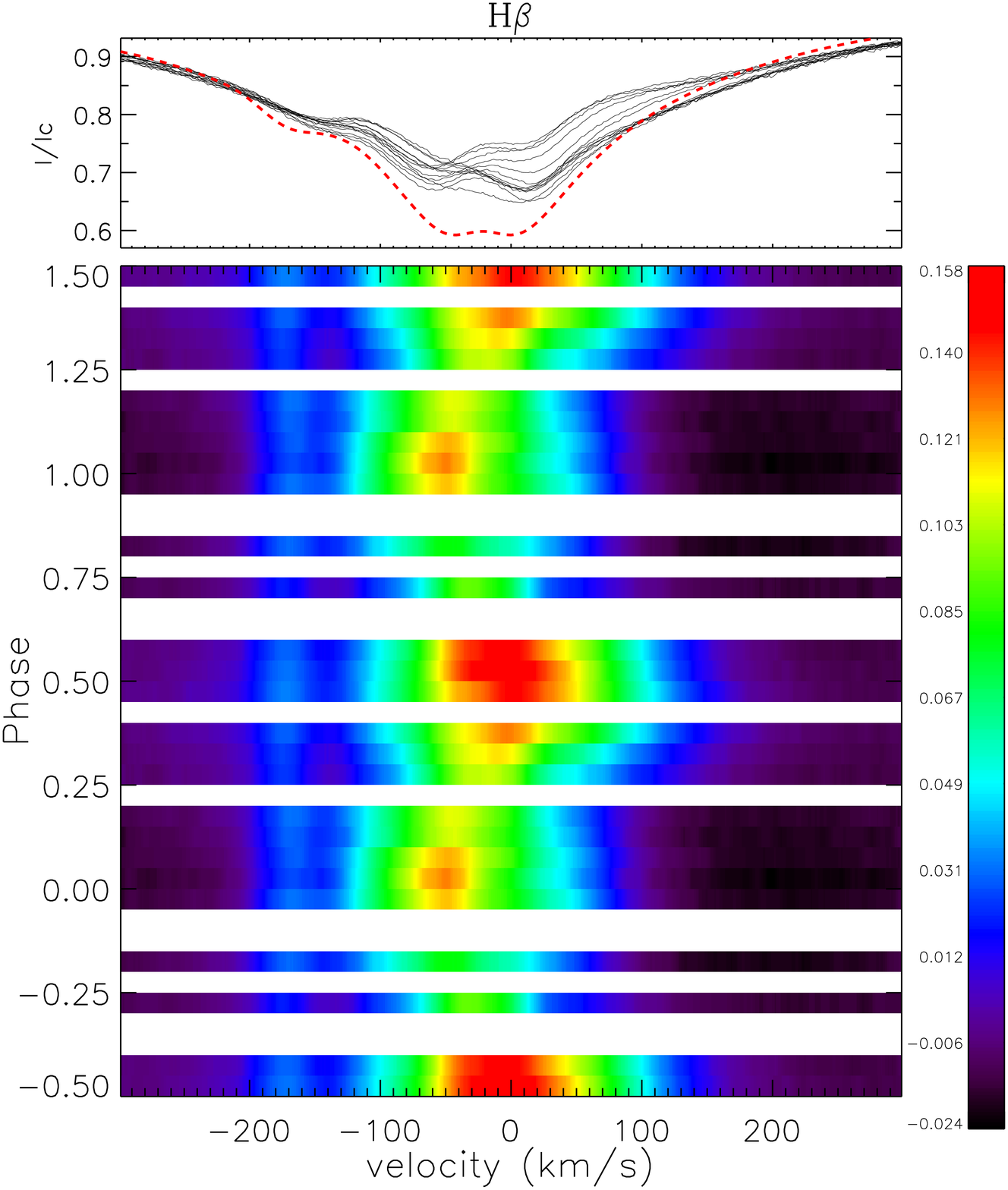}
\includegraphics[width=2.3in]{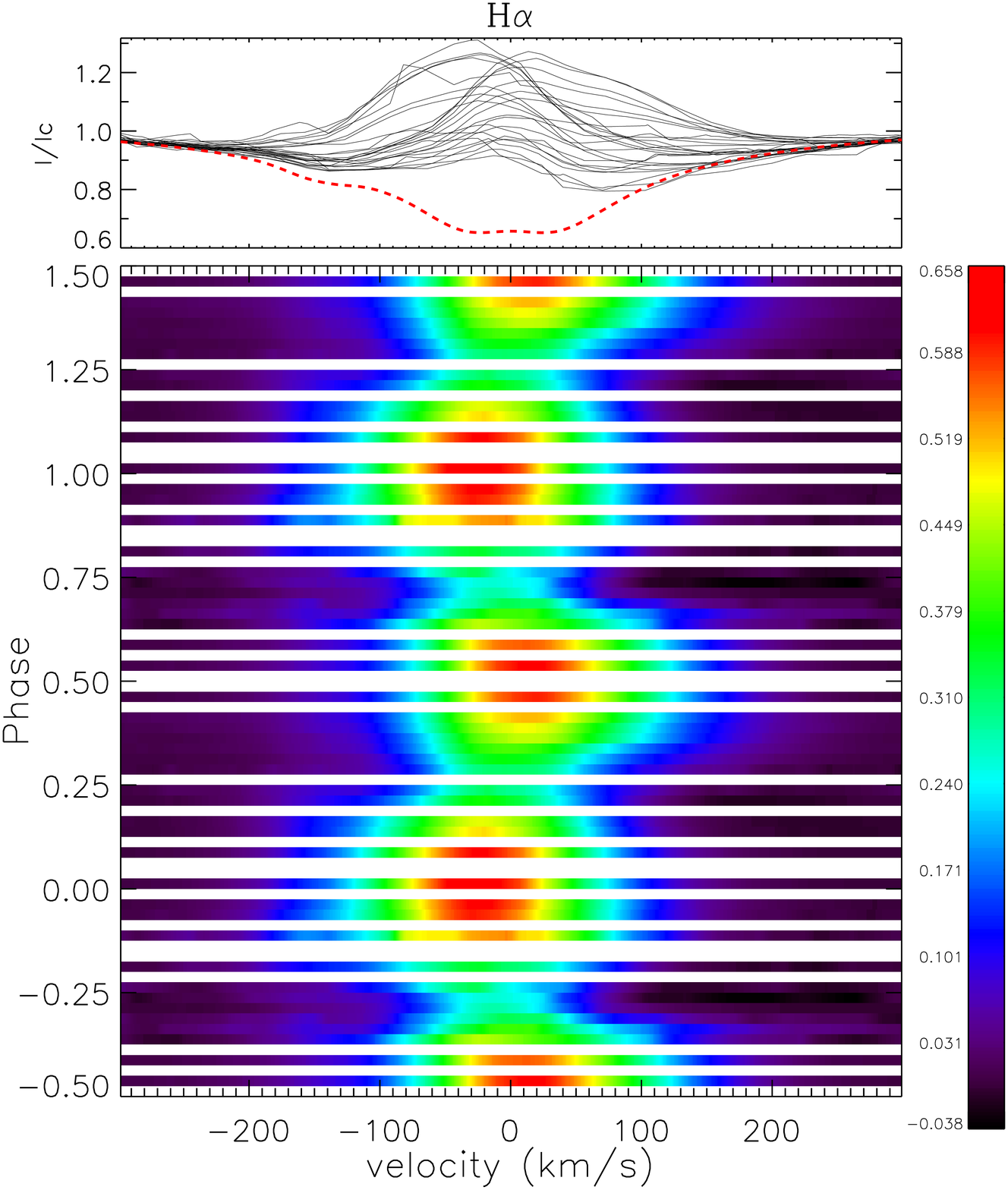}
\caption{Same as Fig.~\ref{sin_dyn_fig} but for spectral lines that show evidence of two emission features per rotation cycle. The Balmer lines are plotted as the difference between the observed profiles and a NLTE {\sc tlusty} model profile (top panel, dashed-red) to highlight the circumstellar emission. The He\,{\sc i} $\lambda$5876 lines is plotted as the difference between the observed profiles and the profile obtained on 24 December 2010 (dashed-red, top panel).}
\label{halpha_dyn_fig}
\end{figure*}

\section{Circumstellar Environment}\label{magneto_sect}
\subsection{Magnetosphere}
\begin{figure*}
\centering
\includegraphics[width=1.7in]{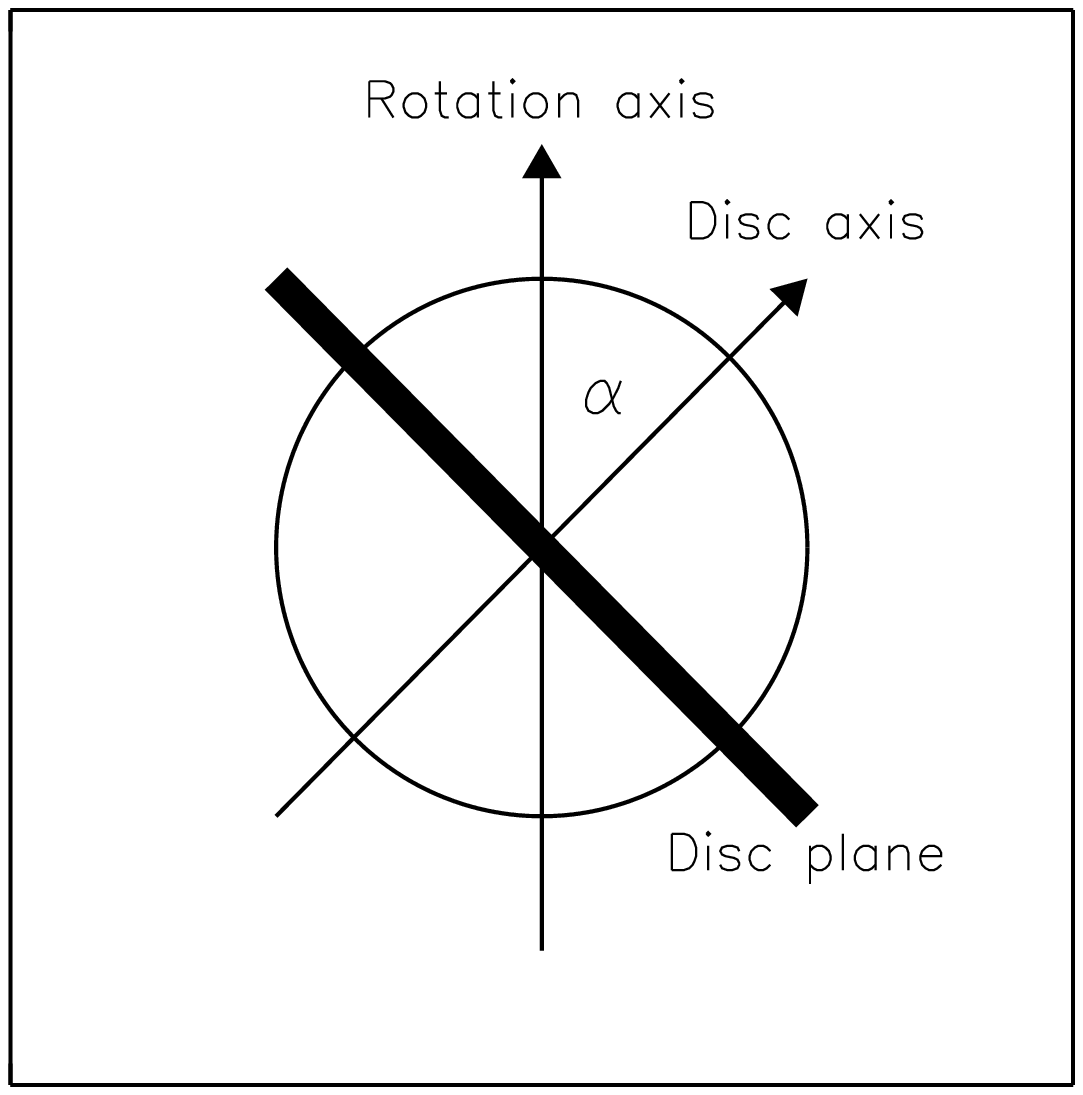}
\includegraphics[width=1.7in]{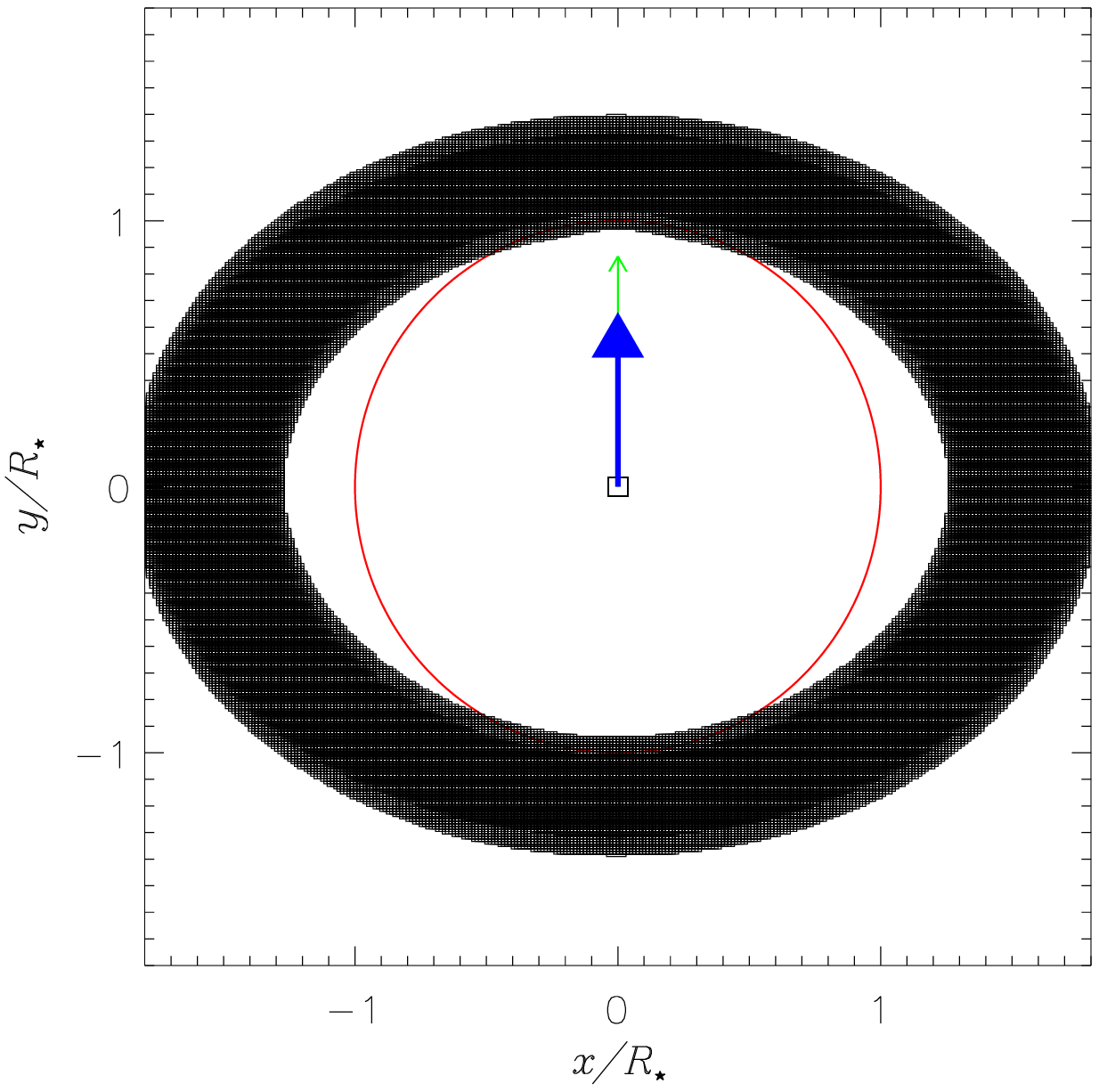}
\includegraphics[width=1.7in]{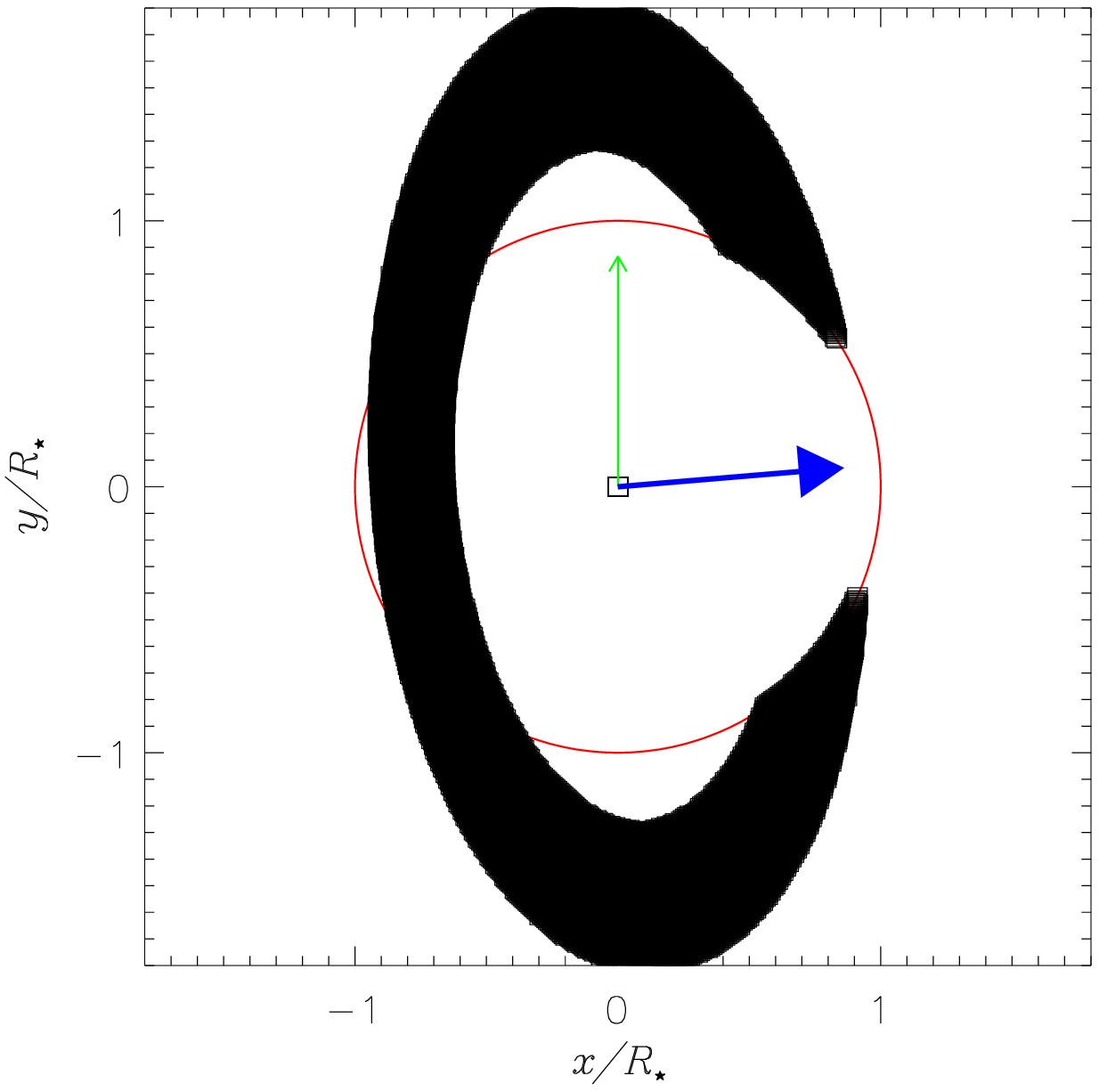}
\includegraphics[width=1.7in]{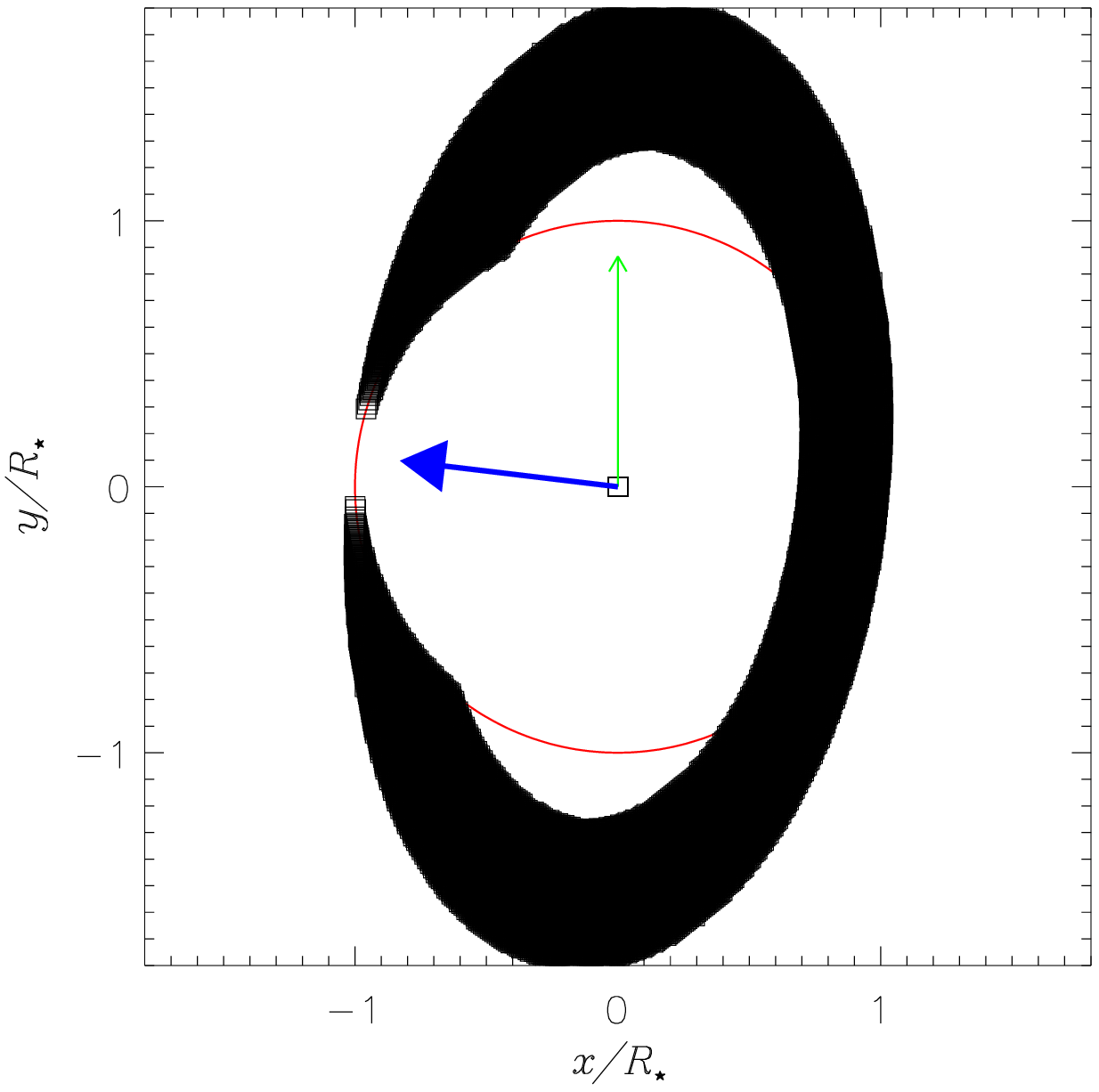}
\caption{Illustration of our ``Toy" model for the H$\alpha$ emission disc. The left panel provides an example schematic diagram showing the orientation of the plane of the disc for a given disc inclination $\alpha$ relative to the rotation axis. The other panels represent projections of the disc (solid black) and central star (solid red) onto our line-of-sight during phases 0.0, 0.33, and 0.66. The thin green arrow in these panels represents the projected rotation axis relative to our line-of-sight (taken to be 60$\degr$), while the thick blue arrow represents the disc axis of 79$\degr$ relative to our line-of-sight.}\label{toy_fig}
\end{figure*}

As discussed in the previous section, the H$\alpha$ emission variation is often qualitatively attributed to the variable projection of a flattened distribution of magnetospheric plasma trapped in closed loops near the magnetic equatorial plane. To this end, we explored the potential of using a ``Toy" model, similar to that described by \citet{how07}, to fit the observed H$\alpha$ EW variations. This model assumes that the H$\alpha$ emission is formed in a centred, tilted, infinitely thin, optically thick disc, in which the relative emission is only a function of the projected area of the disc, taking into account occultation by the star. We are only able to model the relative emission, but our model indicates that if we assume the inclination angle between the disc axis and the rotation axis ($\alpha$) is the same as the magnetic obliquity ($\beta$) then only subtle variations are predicted in the EW curve for different inclinations (assuming the relationship between $i$ and $\beta$ is constrained by the ORM), which are indistinguishable with the precision of our current measurements. We also find that higher $\alpha$ values provide much better fits and therefore we adopt the $\alpha=\beta=79\degr$ as derived from the $B_\ell$ measurements as our disc inclination. Using this value, we first attempted to constrain the disc radii but realised that the inner ($R_{\rm in}$) and outer ($R_{\rm out}$) radii of the disc are poorly constrained by the precision of our observations. However, a best-fit $R_{\rm in}=1.3\pm0.3$\,$R_\star$ and $R_{\rm out}=1.8^{+0.3}_{-0.2}$\,$R_\star$ (as measured from the centre of the star, i.e. a distance of $0.3\pm0.3$\,$R_\star$ and $0.8^{+0.3}_{-0.2}$\,$R_\star$ from the stellar surface) are found, where the uncertainties represent the 1$\sigma$ limits. An illustration of this model is presented in Fig.~\ref{toy_fig}.

\begin{figure}
\centering
\includegraphics[width=3.2in]{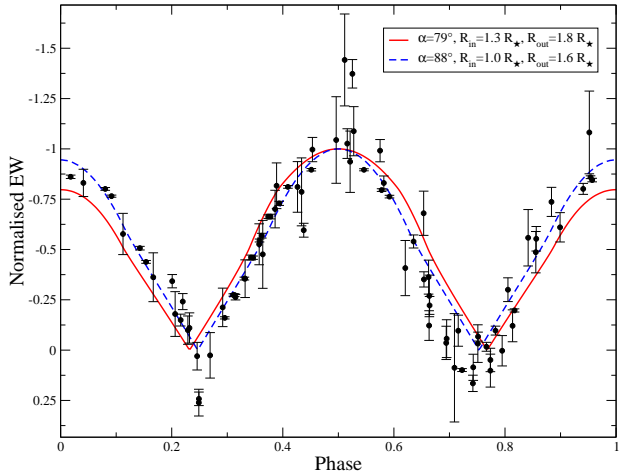}
\caption{``Toy" model compared to phased H$\alpha$ equivalent width measurements (black circles). The solid red curve represents a model with an $\alpha=79\degr$ between the rotation axis and the disc axis, as inferred from the fits to the $B_\ell$ variations, while the dashed blue curve represents the best-fit model with an $\alpha=88\degr$. Both models assume $i=60\degr$.}
\label{disc_comp}
\end{figure}

A comparison between the predicted EW variations from our ``Toy" model and the observed EW variations is shown in Fig.~\ref{disc_comp}. As illustrated, the model is able to qualitatively reproduce the general features of the observed EW variations including the sharp emission minima, the smooth variation during maximum emission, and the double peaked nature of the variability, suggesting that the H$\alpha$ emission does form in an optically thick, flattened distribution of circumstellar plasma with a disc-like structure. However, our model predicts a lower emission level between phases 0.0 to 0.25 and 0.75 to 1.0 and a higher emission level between phases 0.25 to 0.4 and 0.6 to 0.75 than observed, and there is a visible offset between the phases of emission minima; our model predicts the minima to occur at phases 0.23 and 0.77, not 0.25 and 0.75 as observed. To investigate these discrepancies, we conducted another search allowing the disc axis inclination to vary. Our results show that $\alpha=88\pm1\degr$ (and a corresponding $R_{\rm in}=1.0\,R_{\star}$ and $R_{\rm out}=1.6\,R_\star$) provides a better overall fit to the observations, but is inconsistent with our derived value of $\beta$. This disc orientation is better able to reproduce the emission level throughout the rotation cycle and correctly predicts the phases of minimum emission as illustrated in Fig.~\ref{disc_comp}.

\begin{figure}
\centering
\includegraphics[width=3.1in]{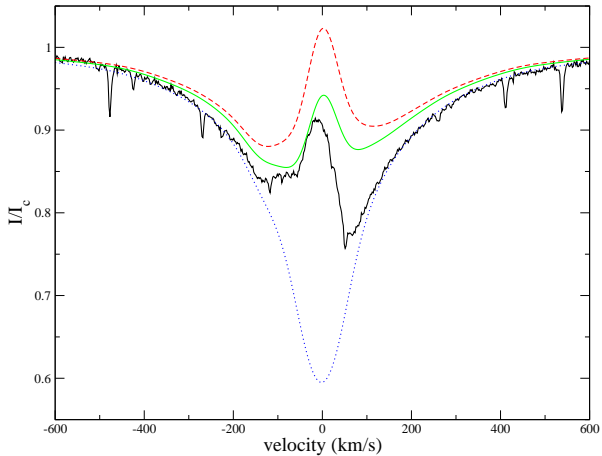}
\caption{Comparison between observed H$\alpha$ profile (solid black) from 2010 December 24, corresponding to a phase of 0.72, and synthetic profiles following the procedure of \citet{sund12} at this same phase. The model profiles were created with an adopted mass-loss rate as determined by \citet{grun09} ($10^{-8.85}$; dotted blue), a mass-loss rate of 17 (solid green) and 21 times this value (dashed red.)}
\label{mdot_comp_fig}
\end{figure}

In addition to investigating the magnetospheric properties via the ``Toy" model, we also used the method of \citet{sund12} to study the H$\alpha$ variability. \citet{sund12} showed that in a slow rotator such as HD\,57682, the transient suspension of wind material in closed magnetic loops, which is constantly fed by the quasi-steady wind outflow, leads to a statistically overdense, low-velocity region in the vicinity of the magnetic equator that causes persistent, periodic variations of Balmer lines. This method utilises the 2D MHD wind simulations of \citet{uddoula08} and the magnetic parameters determined from the $B_\ell$ measurements to determine the properties of the magnetically confined plasma. We note however that the original MHD models were computed for an O-type supergiant. Therefore, in the H$\alpha$ synthesis here, we have simply re-scaled the H$\alpha$ scaling invariant $Q=(\dot{M}\times f^2_{cl})/(v_\infty\times R_\star)^{3/2}$ \citep{puls08} from the original simulation to the value obtained for HD\,57682 using the parameters of \citet{grun09}. These properties are then used to synthesise the expected H$\alpha$ line profile as viewed from our line-of-sight as the star rotates. Unlike the ``Toy" model, these simulations are sensitive to the adopted mass-loss rate. We characterise the mass-loss rate as the product of the stellar mass-loss rate and the square-root of a wind clumping factor, which we assume to be about 4-10 as suggested by several recent studies \citep[e.g.][]{bouret03, najarro11, sund11b}. We attempted to constrain the wind properties by fitting the observed H$\alpha$ profile during phases that have the lowest contributions from the disc emission. As already noted by \citet{grun09}, we cannot fit the H$\alpha$ profile using a mass-loss rate as derived from the UV lines ($\log \dot{M}\sim-8.85$\,$M_\odot$\,yr$^{-1}$). The additional emission resulting from the wind is too low at phases 0.25 or 0.75 (minimum emission), which results in a line profile that is mainly in absorption, contrary to what is actually observed (see Fig.~\ref{mdot_comp_fig}). 

\begin{figure}
\centering
\includegraphics[width=3.1in]{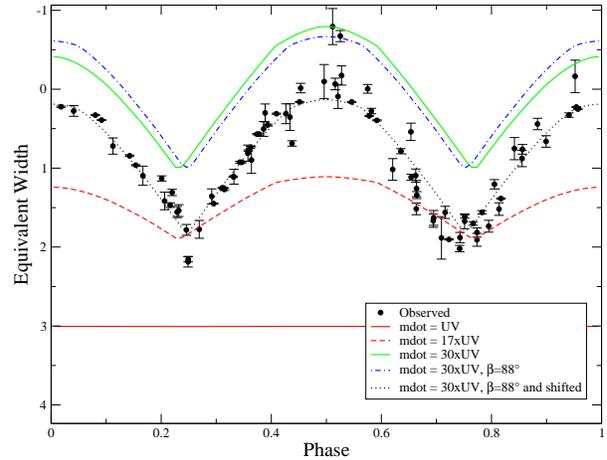}
\caption{Comparison between the measured EW variations of HD\,57682 (black circles) and the theoretical EW variations of our MHD model for different mass-loss rates as labelled (UV corresponds to the UV mass-loss rate derived by \citet{grun09}). While the absolute EW variations of the $\dot M/M_\odot=30\times$ UV (dash-dotted blue) model does not fit the observed EW curve, the amplitude of the predicted variations from this model does provide a good fit, as illustrated for the same model but vertically shifted to better match the variations implied by the observations (black dotted line).}
\label{halpha_synth_eqw_fig}
\end{figure}

\begin{figure*}
\centering
\includegraphics[width=2.3in]{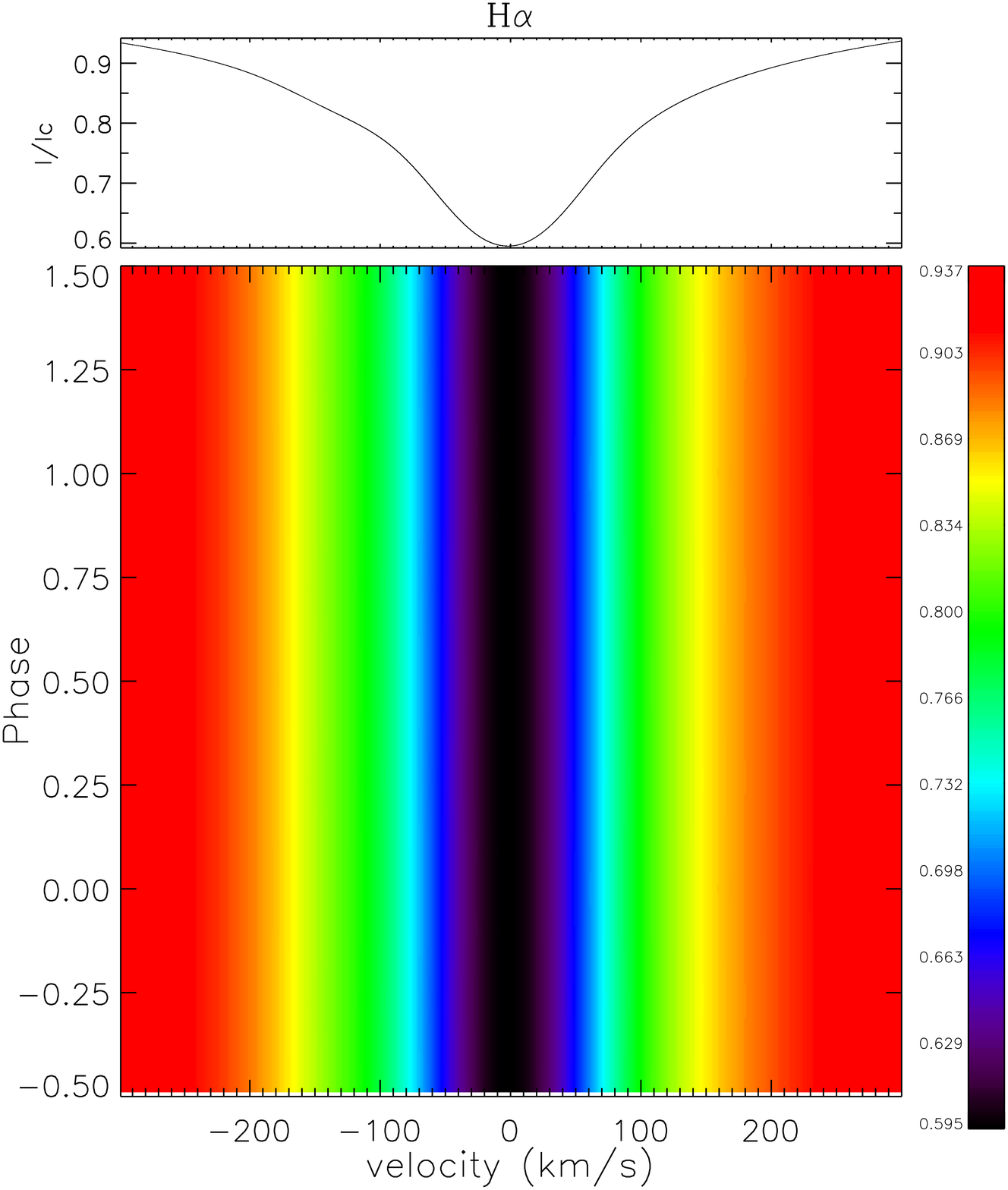}
\includegraphics[width=2.3in]{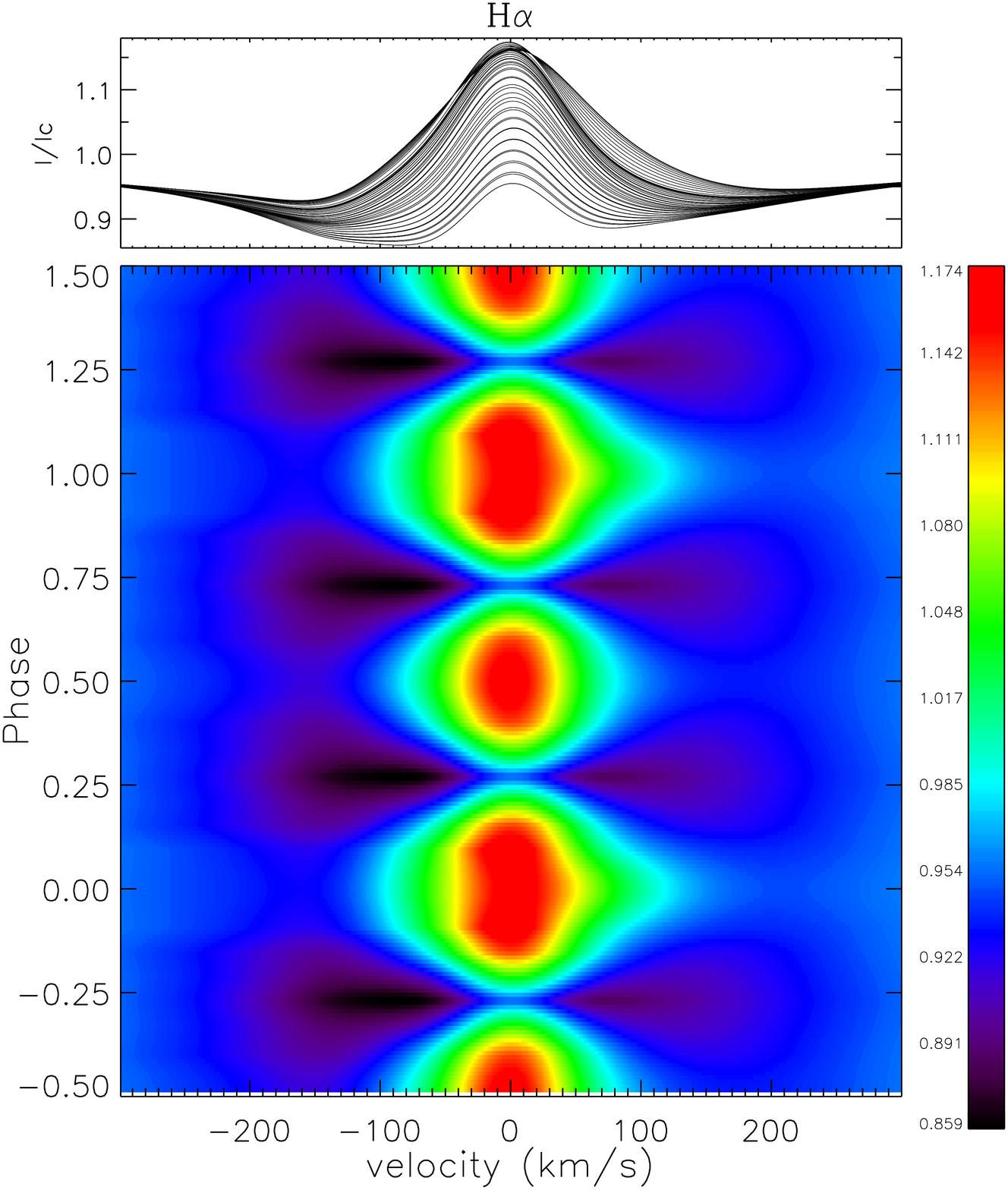}
\includegraphics[width=2.3in]{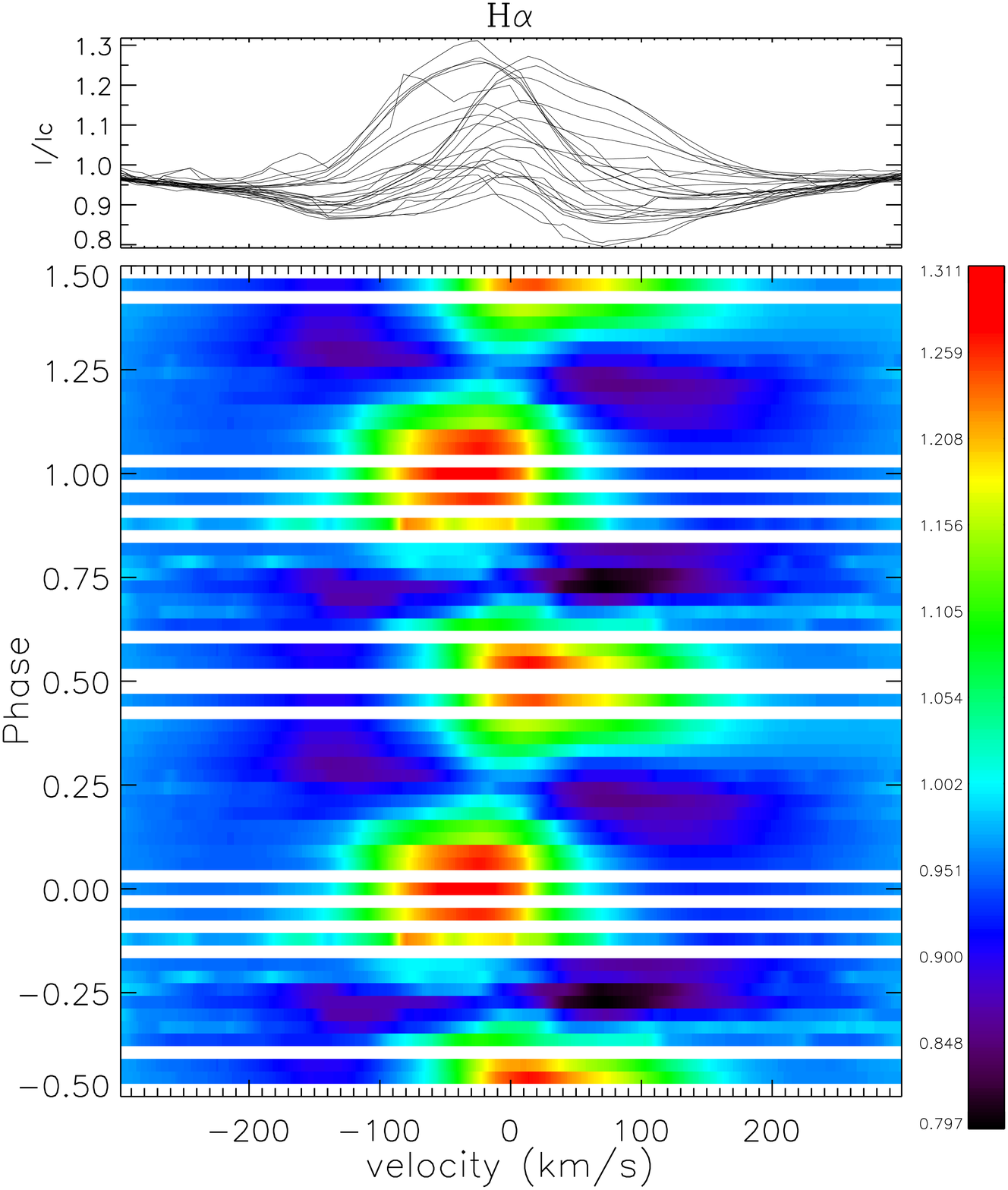}
\caption{Rotationally phased variations in observed (right panel) and modelled H$\alpha$ profiles. The model in the left panel shows the predicted variations assuming the mass-loss rate of \citet{grun09}, while the central panel uses an adopted mass-loss rate of 21 times this value. Note the lack of predicted variations of the model adopting the mass-loss rate of \citet{grun09} and the predicted emission symmetry.}
\label{synth_dyn_fig}
\end{figure*}

Our results (based on fits to the H$\alpha$ profile) suggest that the lack of emission is either due to an underestimated stellar mass-loss rate (by a factor of 10 to 30), or that the wind is very clumpy (note that clumping was not taken into account in the study of \citet{grun09}). This is consistent with many recent studies that find large discrepancies between UV and optical mass-loss rates (see \citet{sund11a} and references therein). Furthermore, according to Figs.~\ref{halpha_synth_eqw_fig} and \ref{synth_dyn_fig}, we find that the model using the UV derived mass-loss rate produces little to no emission with any significant variability. In fact, the amplitude and the offset of the EW variations shown in Fig.~\ref{halpha_synth_eqw_fig} are very sensitive to the adopted mass-loss rate. If we choose a mass-loss rate that best fits the profiles at minimum emission phases, we obtain a mass-loss rate of $\sim$15 times the suggested UV mass-loss rate (see Fig.~\ref{mdot_comp_fig}). However, while the predicted minimum emission of the EW variations is well-fit, the maximum emission is underestimated. A mass-loss rate of $\sim$20 times the previous estimate provides a better fit to the EW variations, by underestimating the maximum emission and over-estimating the minimum emission. A mass-loss rate of $\sim$30 times the previously estimated value is needed to provide the correct amplitude of the EW variation, but, as the wind emission is predicted to be significantly higher, there exists a significant offset between the predicted and observed EW values. In any case, our MHD simulations indicate that a higher mass-loss rate than derived by \citet{grun09} is necessary in order to observe significant variability, with similar characteristics to the observed H$\alpha$ dynamic spectrum (Fig.~\ref{synth_dyn_fig}). The characteristics of the MHD EW curves are similar to the predictions of our ``Toy" model; the phasing of minimum emission are not at phases 0.25 and 0.75 and the relative amplitude of the emission is slightly lower when using $\alpha=79\degr$ (note that the pseudo-disc naturally forms along the magnetic equator in these MHD simulations, i.e. $\alpha=\beta$). Just as was found with the ``Toy" model, a higher $\beta=88\degr$ provides a better agreement with observations. Of particular interest, is the fact that the synthesised emission variation in Fig.~\ref{synth_dyn_fig} is predicted to be symmetric about the systemic velocity of HD\,57682 with the adopted high $\beta$. However, a lower $\beta$ does provide increased occultation of the disc and therefore necessarily results in a more asymmetric line profile variation at some phases, similar to the observations.

\subsection{Greater circumstellar region}\label{grtr_circ}
We continued our investigation of the circumstellar environment to larger distances as probed at infrared wavelengths. Our original intention was to investigate the infrared properties of HD\,57682's SED to address the uncertain classification of HD\,57682 as a classical Oe star and to potentially probe the extent of the magnetosphere.

\begin{figure}
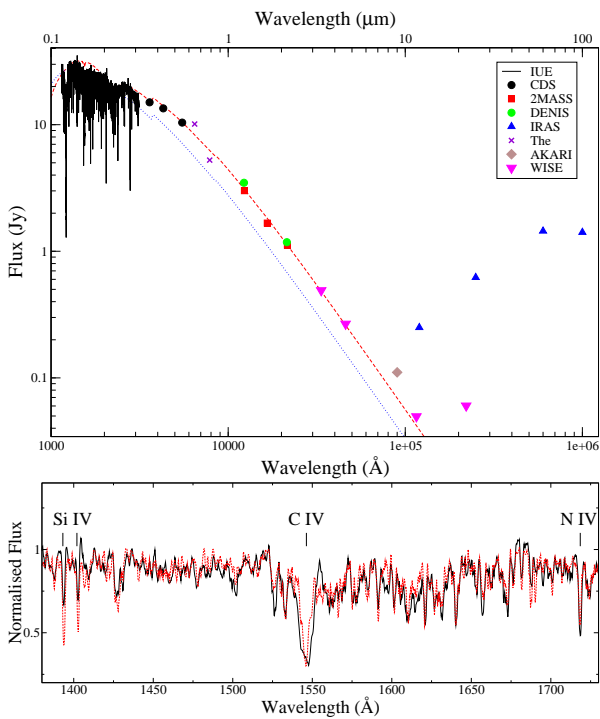

\centering
\includegraphics[width=3.1in]{fig18a.eps}\\
\includegraphics[width=3.1in]{fig18b.eps}
\caption{{\bf Upper:} Observed {\it IUE} observations and infrared flux measurements from the indicated sources. Also included are stellar models corresponding to the parameters derived by \citet{grun09} ($d=1.3$\,kpc and $E(B-V)=0.07$; dashed grey) and the revised model presented here ($d=1.0$\,kpc and $E(B-V)=0.12$; dashed red). The larger flux of the IRAS data points compared to the WISE measurements reflects the significantly larger aperture and therefore increased contribution from the surrounding emission. {\bf Lower:} Comparison between the observed {\it IUE} spectrum (solid black) and the {\sc cmfgen} model with the parameters inferred in this work (dashed red) in the wavelength range of 1380-1730\,\AA. The {\sc cmfgen} model still adopts the lower mass-loss rate of \citet{grun09} and not the higher value needed to fit H$\alpha$ (Fig.~\ref{mdot_comp_fig}; see Sect.~\ref{magneto_sect} for further discussion). Both the model and observed spectrum were smoothed for display purposes.}
\label{ir_fig}
\end{figure}

In Fig.~\ref{ir_fig}, we show the SED of HD\,57682 from the ultraviolet to infrared wavelengths. The data were collected from the {\it IUE} database and the available photometric archives discussed in Sect.~\ref{obs_sect}. Also included in this figure are theoretical SEDs based on the original stellar parameters of \citet{grun09} and updated values presented here. A comparison between our {\sc cmfgen} \citep{hillier98} model with the newly inferred parameters and the {\it IUE} data is also presented in the lower panel of this figure, illustrating the agreement between the observations and the model. We note that we have still adopted the lower UV mass-loss rate of \citet{grun09} in this {\sc cmfgen} model. Our new model has revised the distance and reddening parameters of \citet[][$d\sim1.3$\,kpc, $E(B-V)=0.07$]{grun09}, which were only derived from the {\it IUE} data and do not provide a good fit to the photometric measurements blue-ward of 50\,000\,\AA\ (5\,$\mu$m). Our revised parameters are $d\sim1.0$\,kpc and $E(B-V)=0.12$. The SED also shows a large disagreement between the IRAS and WISE data points, but this reflects the variation in the aperture sizes of the different instruments; the IRAS aperture is about five times larger and therefore includes more contribution of circumstellar emission (as further discussed below), while the WISE measurements represent more of the true ``stellar" flux, and appears consistent with the other infrared measurements. The observed SED does show a large disagreement with the theoretical SED for measurements red-ward of $1\times10^5$\,\AA\ (10 $\mu$m). The observed infrared emission appears to be increasing with increasing wavelength and not decreasing as predicted by contributions from pure stellar flux, and therefore represents an additional source of emission.

In order to better understand the characteristics of the infrared excess we utilised the procedure of \citet{ignace04} to model the free-free emission that would be emitted from a spherical cloud; Infrared excess is common amongst Be/Oe stars and is caused by free-free emission from the circumstellar disc. Using this procedure we find that we can qualitatively match the characteristics of the infrared excess by varying the properties of the cloud. A brief investigation of this model shows that the location of the infrared bump is mainly controlled by the temperature of the cloud (we find $T\sim20000$\,K to be sufficient), but the size of the bump is controlled by the relative sizes of the cloud and star. For simplicity we assume that the cloud is of constant density, which then requires a cloud of approximately 100\,$R_\star$ to reproduce the observations. If we instead adopt a cloud with many large clumps of different densities and at different radii \citep[as is done in ][]{ignace04} we can achieve similar results without the need for a 100\,$R_\star$ cloud. However, if we assume that the cloud is not highly clumped and results from stellar mass lost by HD\,57682, this would imply an unreasonable mass for the cloud, suggesting that HD\,57682 is in fact not a classical Oe star.

\begin{figure}
\centering
\includegraphics[width=1.3in]{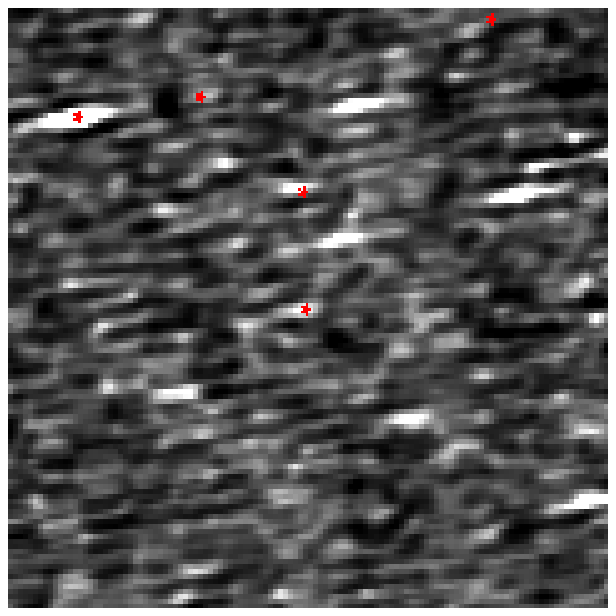}
\includegraphics[width=1.3in]{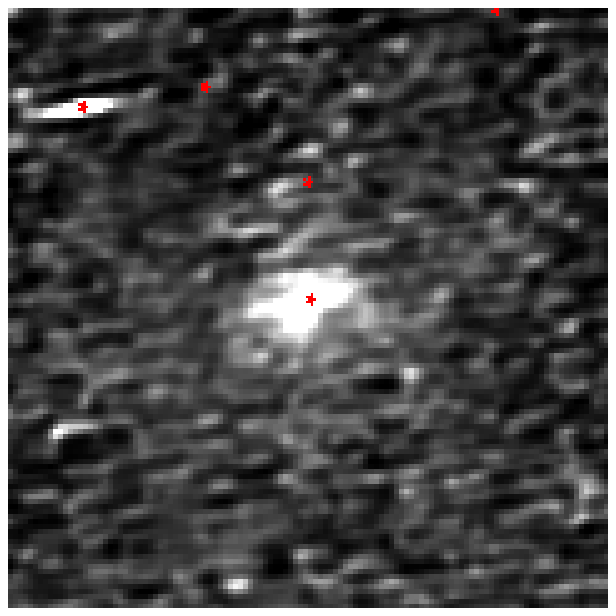}\\
\includegraphics[width=1.3in]{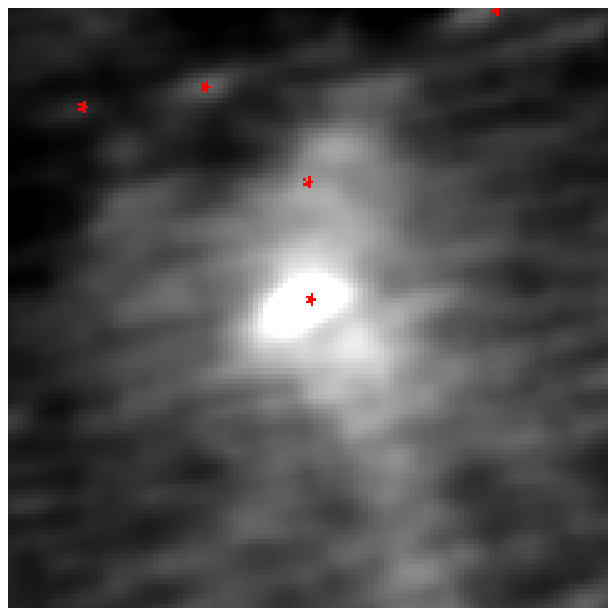}
\includegraphics[width=1.3in]{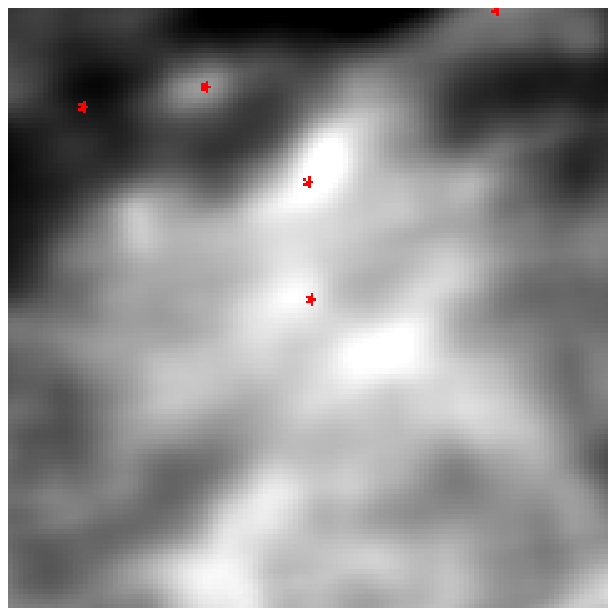}
\caption{Infrared emission as detected by IRAS/HIRES at different wavelengths (top left - 12 $\mu$m, top right - 25 $\mu$m, bottom left - 60 $\mu$m, bottom right - 100 $\mu$m). Each image represents a $30\times30$ arcminute scale. Point sources are indicated by the red circles and the field of view is centred on HD\,57682. Note that the elliptical shape of the point sources is a result of the scanning direction of the IRAS detector.}
\label{hires_map}
\end{figure}

In light of this conclusion we further investigated the IRAS photometry. In the end we suspect that the additional emission is a result of HD\,57682's illumination of the IC\,2177 nebula \citep{halb93}. A look at the HIRES/IRAS maps, as shown in Fig.~\ref{hires_map}, shows significant infrared emission centred on HD\,57682 at 25 $\mu$m and 60 $\mu$m, with the size of the emission cloud larger at 60 $\mu$m. Emission is also evident at 12 $\mu$m, but does not reach the same spatial extent as found at 25 or 60 $\mu$m. At 100 $\mu$m the cloud overwhelms the field of view ($30\times30$\,arcmin), but appears to be a blend of several clouds. It is also evident from these images that the majority of the point sources in the field of view are illuminating the nebula at these wavelengths, further corroborating our suggestion that the infrared emission does not result from a circumstellar disc or even the extended magnetosphere of HD\,57682.

\section{Discussion \& Conclusions}\label{disc_sect}
In this paper we report on our ongoing work towards understanding the magnetic properties and variability of the magnetic O9 subgiant star HD\,57682. Our current analysis is based on an extensive dataset spanning almost 15 years, including archival spectra from ESO's CES, UVES and FEROS spectrographs, and from the echelle spectrograph at LCO. In addition numerous low-resolution spectra were acquired and accessed from the BeSS database. Furthermore, we also present 13 newly-obtained, high-resolution spectropolarimetric observations acquired with ESPaDOnS at the CFHT within the context of the MiMeS project.

A period analysis performed on the spectroscopic H$\alpha$ EW variations and the polarimetric $B_\ell$ measurements resulted in a single, consistent period between the polarimetric and H$\alpha$ variations of $63.5708\pm0.0057$\,d, which we infer to be the rotational period of this star. This long rotational period suggests that the $v\sin i$ inferred by \citet{grun09} is too high by a factor of $\sim$2, which led us to attempt to remeasure this value. We found that the C\,{\sc iv} $\lambda$5801 photospheric line was the least susceptible to additional macroturbulence in the line profile and therefore were able to estimate a new $v \sin i=4.6\pm0.6$\,km\,s$^{-1}$, which is now consistent with the $\sim$63\,d period. This period also implies that the inclination of the rotation axis $i>30\degr$ with a best-fit value $i\sim56^{+34}_{-26}\degr$.

From fits to the rotationally phased $B_\ell$ measurements we were able to constrain the magnetic dipole parameters of HD\,57682, assuming that the magnetic field is well characterised by the ORM. Taking $i=60$\degr, the data imply that $B_d=880\pm50$\,G and $\beta=79\pm4\degr$. However, direct modelling of the individual mean LSD profiles suggests a somewhat weaker dipole field strength of $\sim$700-900\,G and a smaller obliquity angle of $\sim$60-68$\degr$. However, the results are strongly dependent on the adopted depth and width of the modelled LSD Stokes $I$ and $V$ profiles, as discussed in Sect.~\ref{mag_field_sect}. Therefore, it is likely that improved modelling of the observed line profiles that correctly takes into account the profile variations could reduce the disagreement and we therefore suggest that the real magnetic dipole parameters are closer to the parameters implied by the $B_\ell$ variation. 

The profile variations of nearly all observable lines in the high-resolution ESPaDonS spectra were also investigated (Figs.~\ref{sin_ew} and \ref{nonsin_ew}). Our analysis suggests that we can essentially separate all lines into one of two categories as determined by the rotationally-phased EW variations of the spectral line. There are those lines that show clear evidence of double-peaked variations similar to H$\alpha$, while the rest show single-peaked, sinusoidal variations. The LPV of these lines, as described in Sect.~\ref{spec_var_sect}, show rotationally phased velocity variations of the pseudo-absorption or pseudo-emission components, and the pattern of this variability is generally consistent amongst all the lines. 

In Sect.~\ref{magneto_sect}, we clearly showed that the double-peaked variations of the H$\alpha$ EW is well explained by the presence of the variable projection of a flattened distribution of magnetospheric plasma trapped in closed loops near the magnetic equatorial plane. This flattened distribution is also predicted by MHD simulations using the stellar and magnetic properties as determined here and by \citet{grun09}, although we find that a mass-loss rate that is 20-30 times greater than the value inferred by \citet{grun09} from the UV lines is necessary to produce the level of emission that is observed. However, at present we are not able to model UV resonance lines using 
our MHD wind simulation and therefore must resort to modelling these lines using a spherically symmetric wind model. The predicted UV spectrum, as modelled using {\sc cmfgen}, indicates the presence of several intense spectral emission lines that are observed mainly in absorption in the {\it IUE} spectrum (e.g. Si\,{\sc iv} $\lambda1400$ and N\,{\sc iv} $\lambda$1718) or that the observed lines are too weak (e.g. N\,{\sc v} $\lambda1240$ and C\,{\sc iv} $\lambda1550$). However, the {\it IUE} spectrum is well reproduced when adopting the lower mass-loss rate, as already demonstrated by \citet{grun09} and illustrated in the lower panel of Fig.~\ref{ir_fig}. Note that {\sc cmfgen} assumes a spherically symmetric wind, whereas the MHD wind simulations predict a highly non-spherical wind with important consequences for spectral line diagnostics \citep{uddoula02, sund12}. Furthermore, the UV lines are very sensitive to the ionisation structure of the wind, which may also be largely affected by a non-spherically symmetric wind. This is not the case for H$\alpha$ as it is a recombination-based optical emission line, which is insensitive to the exact modelling of the ionisation structure \citep[e.g.][]{sund12}. Therefore it is not surprising that there is a disagreement between the mass-loss rates inferred by our optical and UV analyses. Future MHD UV line modelling will be necessary to resolve this discrepancy.

The MHD model provides not only an independent analysis of the stellar mass-loss rate, but also of the magnetic geometry, which is constrained by the characteristics of the H$\alpha$ EW variations. Using the higher mass-loss rate shows that the H$\alpha$ emission variation can be reasonably modelled, however there are several inconsistencies between the models and observed data that reflect our uncertainty in the mass-loss rate and potentially the magnetic geometry. Our results show that the inclination of the disc relative to the rotation axis must be greater than the inferred magnetic obliquity (as measured from our polarimetric data), since a higher disc inclination provides a much better fit to the observed EW variations. This may also be attributed to warping of the disc, or may even reflect the fact that the disc does not necessarily lie in the magnetic equatorial plane as predicted by MHD simulations. 

As with the LPV for the sinusoidally varying lines, there also appears to be rotationally phased velocity variations in the position of the central emission component in the H$\alpha$ line profile that is not predicted by the MHD simulations. We discuss a few potential explanations. If the disc structure is not vertically symmetric along the rotation axis (i.e. not symmetric about the equatorial rotation plane) the resulting line profiles would be asymmetric and would result in velocity shifting of the disc emission. Likewise, enhanced emission from a non-uniform velocity flow where the velocity flow occurs at high magnetic latitudes could also explain our observations. For example, during phases where the disc is viewed nearly face on, the emission could appear systematically offset from the systemic velocity if there is an additional outflow or inflow, for instance, due to the stellar wind or circumstellar plasma falling back onto the star. This would result in no velocity offset during phases where the disc is viewed nearly edge on as the additional outflow or inflow would appear perpendicular to our line-of-sight. However, it is difficult to understand a scenario where there would be a preference for outflow or inflow of plasma at one magnetic hemisphere compared to the other. This is required to explain the velocity shifting of the emission that occurs as our line-of-sight rotates from one magnetic hemisphere to the other. Another possible solution is to introduce an offset of the dipole relative to the centre of the star. Our preliminary tests suggest that a small offset of $\sim$0.2\,$R_\star$ along the rotation axis could explain the velocity shifting of the emission that is observed in H$\alpha$, although MHD models implementing this offset are necessary to confirm this phenomenon, which will be the subject of future studies. 

In addition to the LPV, we also measure RV variability in nearly all spectral lines and find that all lines with sinusoidally varying EW measurements show approximately sinusoidally varying RV measurements. \citet{turner08} investigated the binarity of this star and found no evidence of a companion in either their $I$-band adaptive optics or RV measurements. If we assume that the RV variations are due to a binary companion, a least-squares analysis of the RV measurements obtained from the mean LSD $I$ profile would suggest a mass function $f(M/M_\odot)=2.2\pm0.9\times10^{-6}$ for the companion star. For reasonable orbital inclinations, this would require the companion to be a very low-mass (and presumably a low-luminosity) star, and therefore should not have a significant effect on the observed line profile, which we clearly do observe. Taking this into account, it is therefore unlikely that the RV variations are due to a binary companion since we do not observe clear shifts of the entire spectroscopic lines. We propose that the RV variations are likely a result of the variable asymmetry of the line profiles and simply reflects the varying location of the centre-of-gravity in the profiles. This is clearly evident in the lines that show a double-wave EW curve since the RV measurements also show a more complex double-wave pattern. Furthermore, there also appears to be more of a direct relationship between between the RV and EW measurements in these lines, while the majority of the sinusoidally varying lines show a $\sim$0.25 phase offset between the RV and EW variations.

The root cause of the LPV in the sinusoidally varying lines is still an outstanding issue. It is highly unlikely that these variations are caused by surface features (e.g. chemical ``spots") as the observed variability between multiple ions of a given species does not behave as predicted by LTE or NLTE spectrum synthesis. For example, our observations demonstrate that the weak N\,{\sc iv} $\lambda$4058 line exhibits the same relative EW variation as the strong N\,{\sc iii} $\lambda$4523 line, but spectral modelling of these lines show that the weaker line should show 30 percent stronger variation if due to abundance changes.  Furthermore, the strong Si\,{\sc iv} $\lambda$4115 line and the weak Si\,{\sc iv} $\lambda$4403 line show nearly the same level of EW modulation, but the weak Si\,{\sc iv} line should show a 60 percent larger amplitude in the EW variation if resulting from abundance changes. The weak Si\,{\sc iv} $\lambda$4667 line is also predicted to show the same relative EW variation as the $\lambda$4403 line, but our observations show a significantly greater level of variability (about 50 percent larger) in the $\lambda$4667 line. Moreover, abundance spots require that all spectral lines of a given species vary in phase. This is clearly not the case for the C\,{\sc iv} lines (5801\,\AA, 5811\,\AA) that vary in antiphase with the C\,{\sc iii} lines (as shown in Fig.~\ref{sin_ew}), or the N\,{\sc iv} $\lambda$4058 and N,{\sc iii} $\lambda$4523 lines, which appear shifted relative to one another by a quarter of a cycle. The interpretation of the variability in terms of surface spots would also imply that all elements (except perhaps C) are concentrated at the same general region on the stellar surface, which would be a remarkable new phenomenon. Another possibility is that the variations may result from pulsations; however, pulsations should occur on much shorter time-scales \citep[e.g.][]{aerts10}. The rotationally-phased, double-peaked EW variations are well explained by the variable projection of a flattened distribution of magnetospheric plasma, and at first glance it is difficult to apply this same disc model to explain the variations in the other lines. Occultation or variable emission from a symmetric disc would result in double-peaked variations in all lines and not the single-peaked emission that we observe. However, the LPV of the higher Balmer lines (Fig.~\ref{halpha_dyn_fig}) likely provide the link between the double-peaked variability and single-peaked variability; a decrease in the strength of the emission feature at phase 0.25 and 0.75 compared to the emission feature at phase 0 is clearly observed in higher Balmer lines. Furthermore, this is also illustrated in the EW variations of the Balmer lines (Figs.~\ref{sin_ew} and \ref{nonsin_ew}), which show a clear transition from the double-peaked EW curve at H$\alpha$ to a single-peaked curve at H$\gamma$. Therefore, we tentatively conclude that all line profile variability observed in HD\,57682 is a result of variations of the flattened distribution of magnetospheric plasma. The details of the variable emission for different lines likely reflects the fact that each line probes a different region of what is likely an asymmetric spatial or velocity  distribution of the magnetospheric plasma and that the disc is not completely optically thick for all lines-of-sight, for all wavelengths. We suspect that full 3D MHD models and an extension of the line profile synthesis techniques of \citet{sund12} to lines other than H$\alpha$ are necessary to properly model these properties.

In summary, the narrow-lined photospheric spectrum and weak wind of HD\,57682 have allowed us to study its magnetic and magnetospheric properties at a level of detail not currently achievable for any other magnetic O-type star. The results of our analysis indicate a highly complex behaviour that can be used as a testbed for future more sophisticated 3D MHD simulations to better understand non-rotationally supported magnetospheres \citep[also referred to as dynamical magnetospheres][]{petit11,sund12} of hot stars.

\section*{ACKNOWLEDGEMENTS}
JHG acknowledges financial support in the form of an Alexander Graham Bell Canada Graduate Scholarship from the Natural Sciences and Engineering Research Council, and helpful discussions with C.P. Folsom. GAW acknowledges Discovery Grant support from the Natural Science and Engineering Research Council of Canada. JOS acknowledges support from NASA ATP grant NNX11AC40G. We thank Evgenya Shkolnik for reduction of the LCO observations. This work has made use of the BeSS database, operated at LESIA, Observatoire de Meudon, France: http://basebe.obspm.fr. The DENIS project has been partly funded by the SCIENCE and the HCM plans of the European Commission under grants CT920791 and CT940627. It is supported by INSU, MEN and CNRS in France, by the State of Baden-W\"urttemberg in Germany, by DGICYT in Spain, by CNR in Italy, by FFwFBWF in Austria, by FAPESP in Brazil, by OTKA grants F-4239 and F-013990 in Hungary, and by the ESO C\&EE grant A-04-046. This research has made use of the NASA/IPAC Infrared Science Archive, which is operated by the Jet Propulsion Laboratory, California Institute of Technology, under contract with the National Aeronautics and Space Administration.

\appendix
\section{H$\alpha$ Equivalent width measurements}\label{apndx1}
\begin{table}
\centering
\caption{Spectroscopic H$\alpha$ EW measurements. Included is the instrument or observatory name, the heliocentric Julian date, and the measured H$\alpha$ equivalent width (EW) and its corresponding 1$\sigma$ uncertainty.}
\begin{tabular}{cccc}
\hline
Instrument & HJD & H$\alpha$ EW\,(\AA) & $\sigma_{\rm EW}$\,(\AA) \\
\hline
CES	&	2450125.6216	&	1.270	&	0.014	\\
CES	&	2450126.6291	&	1.108	&	0.007	\\
CES	&	2450127.6296	&	0.926	&	0.008	\\
CES	&	2450128.6284	&	0.735	&	0.012	\\
CES	&	2450129.6287	&	0.569	&	0.011	\\
CES	&	2450130.6194	&	0.454	&	0.010	\\
CES	&	2450131.6155	&	0.311	&	0.007	\\
UVES	&	2452621.8553	&	0.278	&	0.035	\\
FEROS	&	2453042.6286	&	1.130	&	0.033	\\
FEROS	&	2453043.6016	&	1.468	&	0.029	\\
FEROS	&	2453045.6764	&	2.187	&	0.066	\\
FEROS	&	2453045.6873	&	2.154	&	0.034	\\
FEROS	&	2453361.6916	&	1.308	&	0.040	\\
LCO	&	2455111.8964	&	1.612	&	0.015	\\
LCO	&	2455112.8549	&	1.701	&	0.019	\\
LCO	&	2455113.8919	&	1.560	&	0.019	\\
BeSS	&	2454786.6621	&	0.785	&	0.032	\\
BeSS	&	2455233.4387	&	1.517	&	0.074	\\
BeSS	&	2455233.4676	&	1.259	&	0.103	\\
BeSS	&	2455233.4960	&	1.341	&	0.045	\\
BeSS	&	2455235.3993	&	1.667	&	0.083	\\
BeSS	&	2455235.4449	&	1.630	&	0.094	\\
BeSS	&	2455238.4500	&	2.019	&	0.041	\\
BeSS	&	2455238.4842	&	1.880	&	0.065	\\
BeSS	&	2455240.4484	&	1.816	&	0.058	\\
BeSS	&	2455240.4492	&	1.908	&	0.082	\\
BeSS	&	2455242.4600	&	1.205	&	0.059	\\
BeSS	&	2455247.4463	&	0.441	&	0.072	\\
BeSS	&	2455248.4281	&	0.662	&	0.073	\\
BeSS	&	2455257.4440	&	0.276	&	0.068	\\
BeSS	&	2455269.3994	&	1.556	&	0.070	\\
BeSS	&	2455270.4613	&	1.783	&	0.069	\\
BeSS	&	2455273.3852	&	1.358	&	0.096	\\
BeSS	&	2455279.3785	&	0.503	&	0.093	\\
BeSS	&	2455282.4116	&	0.354	&	0.169	\\
BeSS	&	2455286.3735	&	-0.097	&	0.215	\\
BeSS	&	2455287.3461	&	-0.794	&	0.228	\\
BeSS	&	2455288.3862	&	-0.174	&	0.122	\\
BeSS	&	2455291.3937	&	-0.004	&	0.055	\\
BeSS	&	2455296.3850	&	0.540	&	0.110	\\
BeSS	&	2455296.4176	&	1.115	&	0.038	\\
BeSS	&	2455300.3406	&	1.561	&	0.077	\\
BeSS	&	2455305.3715	&	1.736	&	0.075	\\
BeSS	&	2455308.3416	&	0.753	&	0.140	\\
BeSS	&	2455315.3491	&	-0.163	&	0.206	\\
BeSS	&	2455519.7252	&	1.096	&	0.121	\\
BeSS	&	2455542.5156	&	-0.673	&	0.071	\\
BeSS	&	2455548.5674	&	1.016	&	0.137	\\
BeSS	&	2455563.5060	&	0.878	&	0.103	\\
BeSS	&	2455563.5549	&	0.762	&	0.061	\\
BeSS	&	2455587.4412	&	1.537	&	0.075	\\
BeSS	&	2455595.4170	&	0.810	&	0.102	\\
BeSS	&	2455597.4102	&	0.301	&	0.114	\\
\hline
\end{tabular}
\label{online_tab2}
\end{table}

\begin{table}
\centering
\contcaption{}
\begin{tabular}{cccc}
\hline
Instrument & HJD & H$\alpha$ EW\,(\AA) & $\sigma_{\rm EW}$\,(\AA) \\
\hline
BeSS	&	2455600.5138	&	0.687	&	0.034	\\
BeSS	&	2455601.5064	&	-0.014	&	0.060	\\
BeSS	&	2455605.4999	&	-0.066	&	0.074	\\
BeSS	&	2455620.4429	&	1.673	&	0.094	\\
BeSS	&	2455624.4170	&	1.519	&	0.079	\\
BeSS	&	2455643.4021	&	0.720	&	0.102	\\
BeSS	&	2455649.3697	&	1.416	&	0.111	\\
BeSS	&	2455653.3678	&	1.775	&	0.113	\\
BeSS	&	2455657.3714	&	1.109	&	0.094	\\
BeSS	&	2455659.3947	&	0.897	&	0.169	\\
BeSS	&	2455663.3730	&	0.311	&	0.126	\\
BeSS	&	2455669.3974	&	0.092	&	0.152	\\
BeSS	&	2455678.3725	&	1.095	&	0.084	\\
BeSS	&	2455681.3408	&	1.884	&	0.269	\\
ESPaDOnS	&	2454806.0798	&	0.328	&	0.027	\\
ESPaDOnS	&	2454807.1081	&	0.251	&	0.009	\\
ESPaDOnS	&	2454955.7675	&	1.450	&	0.005	\\
ESPaDOnS	&	2454956.7498	&	1.251	&	0.006	\\
ESPaDOnS	&	2454958.7805	&	0.924	&	0.012	\\
ESPaDOnS	&	2454959.7489	&	0.777	&	0.007	\\
ESPaDOnS	&	2454960.7480	&	0.569	&	0.009	\\
ESPaDOnS	&	2455197.0957	&	0.391	&	0.005	\\
ESPaDOnS	&	2455201.0427	&	0.963	&	0.007	\\
ESPaDOnS	&	2455219.9544	&	0.162	&	0.006	\\
ESPaDOnS	&	2455225.9835	&	0.161	&	0.005	\\
ESPaDOnS	&	2455228.0015	&	0.338	&	0.007	\\
ESPaDOnS	&	2455228.9031	&	0.395	&	0.007	\\
ESPaDOnS	&	2455251.9131	&	0.226	&	0.007	\\
ESPaDOnS	&	2455255.9893	&	0.223	&	0.008	\\
ESPaDOnS	&	2455259.9422	&	0.327	&	0.007	\\
ESPaDOnS	&	2455263.9098	&	0.843	&	0.009	\\
ESPaDOnS	&	2455529.0354	&	1.251	&	0.007	\\
ESPaDOnS	&	2455555.0640	&	1.904	&	0.007	\\
ESPaDOnS	&	2455561.0733	&	1.387	&	0.008	\\
\hline
\end{tabular}
\end{table}
\label{lastpage}
\end{document}